\UseRawInputEncoding 
\pdfoutput=1
\documentclass[onecolumn,aps,prd,superscriptaddress,nofootinbib,amsmath,amssymb,floats,floatfix,showkeys,notitlepage,longbibliography]{revtex4-1}

\usepackage{comment}
\usepackage{lipsum}
\usepackage{graphicx}
\usepackage{subfigure}
\usepackage{palatino}
\usepackage{sans}
\usepackage{hyperref}
\hypersetup{colorlinks=true,linkcolor=blue,urlcolor=blue,citecolor=blue}
\usepackage[toc,page]{appendix}
\usepackage[normalem]{ulem}
\usepackage{adjustbox}
\usepackage{latexsym}
\usepackage{amsmath}
\usepackage{amssymb}
\usepackage{amsfonts}
\usepackage{dcolumn}
\usepackage{bm}
\usepackage{tikz}
\usepackage{bigints}
\usepackage{array,tabularx,multirow,booktabs}
\usepackage[tracking=true]{microtype}
\SetTracking{}{500}
\SetTracking{encoding={*}, shape=sc}{40}

\allowdisplaybreaks

\begin{document} \sloppy

\title{Constraints on charged Symmergent black hole from shadow and lensing}

\author{Beyhan Puli\c{c}e
}
\email{beyhan.pulice@sabanciuniv.edu (Corresponding Author)}
\affiliation{Faculty of Engineering and Natural Sciences, Sabanc{\i} University, 34956 Tuzla, \.{I}stanbul, Turkey}

\author{Reggie C. Pantig
}
\email{rcpantig@mapua.edu.ph}
\affiliation{Physics Department, Map\'ua University, 658 Muralla St., Intramuros, Manila 1002, Philippines}

\author{Ali \"Ovg\"un
}
\email{ali.ovgun@emu.edu.tr}
\affiliation{Physics Department, Eastern Mediterranean
University, Famagusta, 99628 North Cyprus, via Mersin 10, Turkey}

\author{Durmu\c{s}~Demir}
\email{durmus.demir@sabanciuniv.edu}
\affiliation{Faculty of Engineering and Natural Sciences, Sabanc{\i} University, 34956 Tuzla, \.{I}stanbul, Turkey}

\begin{abstract}
In this paper, we report on exact charged black hole solutions in symmergent gravity with Maxwell field. Symmergent gravity induces the gravitational constant $G$, quadratic curvature coefficient $c_{\rm O}$, and the vacuum energy $V_{\rm O}$ from the flat spacetime matter loops. In the limit in which all fields are degenerate in mass, the vacuum energy $V_{\rm O}$ can be expressed in terms of $G$ and $c_{\rm O}$. We parametrize deviation from this limit by a parameter ${\hat \alpha}$ such that the black hole spacetime is dS for ${\hat \alpha} < 1$ and AdS for  ${\hat \alpha} > 1$. In our analysis, we study horizon formation, shadow cast and gravitational lensing as functions of the black hole charge, and find that there is an upper bound on the charge. At relatively low values of charge, applicable to astronomical black holes, we determine constraints on $c_{\rm O}$ and ${\hat \alpha}$ using the EHT data from Sgr. A* and M87*. We apply these constraints to reveal how the shadow radius behaves as the observer distance $r_O$ varies. It is revealed that black hole charge directly influences the shadow silhouette, but the symmergent parameters have a tenuous effect. We also explored the weak field regime by using the Gauss-Bonnet theorem to study the weak deflection angle caused by the M87* black hole. We have found that impact parameters comparable to the actual distance $D = 16.8$ Mpc show the potential detectability of such an angle through advanced astronomical telescopes. Overall, our results provide new insights into the behavior of charged black holes in the context of symmergent gravity and offer a new way to test these theories against observational data.
\end{abstract}

\maketitle


\section{Introduction} \label{sect1}
Modified gravity theories are modifications or extensions of Einstein's theory of general relativity. They are motivated by various considerations, including the need to explain the observed acceleration of the universe's expansion, the desire to test the foundations of general relativity, and the possibility of solving various problems in cosmology and astrophysics. Modified gravity theories offer the possibility of new insights into the fundamental nature of gravity and the structure of the universe, and they are an active area of research in cosmology and astrophysics \cite{Lambiase:2022ucu,Lambiase:2020iul,Berti:2015itd,DeFelice:2010aj,Nojiri:2017ncd,Cardoso:2016oxy,Sharif:2012zzd,Sharif:2014fla,Barcelo:2005fc,Jacobson:2002ye}. 
In view of the difficulties with quantizing gravity and reconciling quantum fields with classical gravity, emergent gravity theories stand out as an important alternative. In this emergent approach, gravity is not a fundamental force but an emergent interaction arising from higher-energy dynamics. It is motivated by the observation that the behavior of gravity at large scales, as described by general relativity, is very different from the behavior of the other fundamental forces of nature. The idea of emergent gravity has been studied in various works \cite{Sakharov:1967pk,Verlinde:2010hp,VanRaamsdonk:2010pw,Liberati:2017jnr,Jacobson:1995ab,Padmanabhan:2009vy,Visser:2002ew}. While the emergent gravity approach is still a topic of active research and debate, it has the potential to provide a new perspective on the nature of gravity and the fundamental structure of the universe.
\cite{Nashed:2018efg,Nashed:2019tuk,Nashed:2018oaf}. Among various emergent gravity approaches, gravity theory emerging due to restoring gauge symmetries broken explicitly by the cutoff scale forms a special case. This approach, the so-called Symmergent gravity, differs from the others by its sensitivity only to the flat spacetime loops (the natural setup of quantum field theories (QFTs)) and by its ability to restore gauge symmetries, enabling the emergence of gravity holographically (via metric-affine gravity dynamics) and predict the existence of new particles beyond the known ones. Indeed, quantum loops generate effective QFTs with loop momenta cut at some UV scale $\Lambda$ such that scalar and gauge boson masses receive ${\mathcal{O}}(\Lambda^2)$ corrections, and vacuum energy gets corrected by ${\mathcal{O}}(\Lambda^4)$  and ${\mathcal{O}}(\Lambda^2)$  terms. All gauge symmetries are explicitly broken. The question of if gravity can emerge in a way restoring the explicitly broken gauge symmetries is answered affirmatively by forming a gauge symmetry-restoring emergent gravity model \cite{demir1,demir2,demir3}. This model, the Symmergent gravity, has been built by the observation that, in parallel with the introduction of the Higgs field to restore gauge symmetry for a massive vector boson (with Casimir invariant mass), spacetime affine curvature can be introduced to restore gauge symmetries for gauge bosons with loop-induced (Casimir non-invariant) masses proportional to the UV cutoff $\Lambda$ \cite{demir1,demir2,demir3}. Symmergent gravity is essentially emergent general relativity (GR) with a quadratic curvature term. It exhibits distinctive signatures, as revealed in recent works on static black hole spacetimes \cite{irfan,Symmergent-bh,Symmergent-bh2,Symmergentresults}. 

Charged black holes are important because they are a key test case for the predictions of general relativity, they provide a unique environment for studying the interaction between gravity and electromagnetism, they allow scientists to study the behavior of matter and energy in extreme conditions, and they provide a way to probe the structure of the universe at small scales. The no-hair theorem for black holes is a remarkable principle that suggests these cosmic entities can be uniquely characterized by only a few fundamental properties: their mass, angular momentum, and electric charge. This theorem emphasizes the simplicity and elegance with which black holes can be described, setting them apart from other astrophysical objects. However, modified gravity theories may challenge this notion, in some cases, can alter the behavior of charged black holes, potentially allowing for additional degrees of freedom or exotic properties beyond the scope of the classical no-hair theorem. Studying charged black holes in symmergent gravity offers the possibility of new insights into effective charges as the hair of the black holes and the nature of the principles that govern its behavior \cite{Chamblin:1999tk,Kubiznak:2012wp,Gregory:1994bj,Graham:2014mda}.

The main aim of the present paper is to build a comprehensive study of a new charged black hole solution in the Symmergent gravity - Maxwell framework and study its physical properties in detail. We show that the contribution of the quadratic curvature coefficient to the black hole solution has significant effects on the physical properties of the black hole.  

First, we study the shadow of the charged Symmergent black hole (CSBH). The shadow of a black hole is the region of space from which light cannot escape the black hole's gravitational pull. It appears as a region of darkness in the sky when light from a bright background is bent around the black hole and absorbed by it. The size and shape of the shadow are determined by the mass and spin of the black hole, as well as the distance of the observer from the black hole. Historically, the shadow through an accretion disk was first studied by Luminet \cite{Luminet:1979nyg}, and Synge pioneered the photon sphere that has a fundamental relation to the shadow \cite{Synge:1966okc}. The shadow is an important observational signature of black holes, and it has been observed by telescopes such as the Event Horizon Telescope \cite{EventHorizonTelescope:2019dse,EventHorizonTelescope:2022xnr}. Since then, the study of the shadow of black holes has become a key area of research in astrophysics, and it is expected to provide new insights into the behavior and properties of these objects. Many authors have considered the fingerprints of alternative theories of gravity through the shadow \cite{Contreras:2020kgy,Panotopoulos:2021tkk,Panotopoulos:2022bky,Pantig:2022gih,Ovgun:2019jdo,Ovgun:2020gjz,Okyay:2021nnh,Javed:2021arr,Cimdiker:2021cpz,Uniyal:2022vdu,Pantig:2022ely,Mustafa:2022xod,Pantig:2022qak,Kumaran:2022soh,Atamurotov:2022knb,Vagnozzi:2022moj,Chen:2022nbb,Dymnikova2019,Kuang:2022xjp,Kuang:2022ojj,Wei2019,Hou:2018avu,Tsukamoto:2017fxq,Kumar:2020hgm,Wang2017,Tsupko_2020,Konoplya2019,Belhaj:2020rdb,Belhaj:2020okh,Cunha:2018acu,Gralla:2019xty,Perlick:2015vta,Khodadi:2021gbc,Khodadi:2022pqh,Cunha:2016wzk,Shaikh:2019fpu,Allahyari:2019jqz,Cunha:2016bjh,Zakharov:2014lqa,Chakhchi:2022fl}, while others explored the effects of the astrophysical environment into the shadow \cite{Pantig:2020uhp,Pantig:2022toh,Pantig:2022whj,Xu:2020jpv,Xu:2021dkv,Konoplya:2021ube}. Through the black hole shadow, it also can penetrate through its quantum nature \cite{Devi:2021ctm,Xu:2021xgw,Lobos:2022,Anacleto:2021qoe,Hu:2020usx, Pantig:2022sjb,Pantig:2023yer, Ovgun:2023ego}.

Lastly, we study its deflection angle in weak field limits using the Gauss-Bonnet theorem, which is a mathematical result that relates the curvature of a surface to its topological properties. Research on weak gravitational lensing by black holes is an active area of study in astrophysics and cosmology. Weak gravitational lensing is a phenomenon in which the path of light is slightly bent as it passes through a region of the gravitational field, which can cause a distortion of images of distant objects, such as galaxies and quasars, and it can result in the formation of multiple images of the same object. In 1919, Arthur Eddington led an expedition to verify Einstein's theory of relativity by observing the phenomenon of gravitational lensing. This method has since become an essential tool in astrophysics, as evidenced by numerous studies and papers \cite{Virbhadra:1999nm,Virbhadra:2002ju,Adler:2022qtb,Bozza:2001xd,Bozza:2002zj,Perlick:2003vg,He:2020eah}. In the field of astrophysics, determining the distances of objects is crucial in understanding their properties. However, Virbhadra demonstrated that by observing the relativistic images alone, without any information about the masses and distances, it is possible to accurately determine an upper bound on the compactness of massive dark objects \cite{Virbhadra:2022ybp}. Additionally, Virbhadra discovered a distortion parameter that causes the signed sum of all images of singular gravitational lensing to vanish (this has been tested using Schwarzschild lensing in both weak and strong gravitational fields, \cite{Virbhadra:2022iiy}). On the other hand, in 2008, Gibbons and Werner applied the Gauss-Bonnet theorem to optical geometries in asymptotically flat spacetimes, and calculated the weak deflection angle for the first time in the literature \cite{Gibbons:2008rj}. Since then, this method has been used to study a variety of phenomena \cite{Ovgun:2018fnk,Ovgun:2019wej,Ovgun:2018oxk,Javed:2019ynm,Werner2012,Ishihara:2016vdc,Ono:2017pie,Li:2020dln,Li:2020wvn,Belhaj:2022vte, Pantig:2022toh, Javed:2023iih, Javed:2023IJGMMP, Javed:2022fsn, Javed:2022gtz}.

The paper is directed as follows: Sect. \ref{sec2} briefly introduces the Symmergent gravity, and its charged version will be derived in Sect. \ref{sec3}. Its properties will be explored through the Hawking temperature in Sect. \ref{sec4}. Constraints to the Symmergent property will be sought in Sect. \ref{sec5} by analyzing its shadow properties in conjunction with the EHT data. Finally, in Sect. \ref{sec6}, we apply the results to the weak field regime by applying the Gauss-Bonnet theorem to obtain the weak deflection angle. We state conclusive remarks and research prospects in Sect. \ref{conc}. Throughout the paper, we used geometrized units as $G = c = 1$, and the metric signature $(-,+,+,+)$.

\section{Symmergent Gravity in Brief} \label{sec2}
Symmergent gravity is a quadratic curvature gravity theory with a finite cosmological constant. It is a special case of the general $f(R)$ gravity theories. It has been proposed in \cite{demir2,demir3}, with the latest refinements and improvements in \cite{demir1}. It has recently been briefly discussed in \cite{Symmergent-bh,Symmergent-bh2} regarding its implications for black hole properties like quasi-periodic oscillations, shadow radius, and weak lensing. These studies on Symmergent black holes already give the most relevant properties of the curvature sector of Symmergent gravity. It is governed by the action
\begin{eqnarray}
\label{curvature-sector}
S[g]= \int d^4x \sqrt{-g}\left\{
\frac{R}{16\pi G} - \frac{c_{\rm O}}{16} R^2 - V_{\rm O}  + {\mathcal{L}}_{matter}\right\},
\end{eqnarray}
in which $R$ is the curvature scalar and ${\mathcal{L}}_{matter}$ is  the matter Lagrangian involving both the known matter fields (quarks, leptons, gauge bosons, and the Higgs)  plus new fields needed to induce Newton's constant in the form
\begin{eqnarray}
\label{params-0}
\frac{1}{G} = \frac{{\rm str}\left[{\mathcal{M}}^2\right]}{8 \pi},
\end{eqnarray}
where ${\rm str}[\dots]$ stands for the graded trace  ${\rm str}[{\mathcal{M}}^2] = \sum_s (-1)^{2s+1}\, {\rm tr}[{\mathcal{M}}^2]_s$, with $s$ being the particle spin  and ${\mathcal{M}}^2$ the mass-squared matrix of the matter fields. Not only the Newton's constant $G$ but also the quadratic curvature coefficient $c_{\rm O}$ and the vacuum energy density $V_{\rm O}$  
\begin{eqnarray}
\label{params}
 c_{\rm O} = \frac{n_\text{B} - n_\text{F}}{128 \pi^2}\,,\; V_{\rm O} = \frac{{\rm str}\left[{\mathcal{M}}^4\right]}{64 \pi^2}
\end{eqnarray}
are loop-induced parameters such that $n_\text{B}$ ($n_\text{F}$) stands for the total number of bosons (fermions) in the underlying QFT. One keeps in mind that $ n_\text {B}$ bosons and $n_\text{F}$ fermions contain not only the known standard model particles but also the completely new particles (massive as well as massless) that do not have to couple to the known particles non-gravitationally. 

Before going any further, one notes that if there are equal numbers of bosonic and fermionic degrees of freedom in nature (namely, $n_\text{B} = n_\text{F}$), then $c_{\rm O}\equiv 0$. In this particular case, Symmergent gravity reduces to Einstein's general relativity with no higher-curvature terms (with non-minimal couplings to scalars $S$ in the theory). This Bose-Fermi symmetric structure is reminiscent of the supersymmetric theories in which all particles (known and new ones) are coupled with significant (standard model-sized) couplings. Interestingly, symmergence predicts the pure Einstein gravity when $n_\text{B} = n_\text{F}$ in nature, with the additional property that, unlike the supersymmetric theories, the new particles do not have to interact with the known particles. In this case, one is led to the usual asymptotically-flat Schwarzschild or Kerr black holes. 

In the case of general $n_\text{B}$ and $n_\text{F}$, the Symmergent gravity action (\ref{curvature-sector}) can be brought into the $f(R)$ gravity from
\begin{eqnarray}
S[g]=\frac{1}{16 \pi G} \int d^4 x \sqrt{-g} \left(f(R) - 2 \Lambda-\frac{1}{2} \hat{F}_{\mu \nu} \hat{F}^{\mu \nu} \right)
\label{fr-action1}
\end{eqnarray}
in which
\begin{align}
f(R) &= R+ \beta R^2 
\end{align}
with the quadratic curvature coefficient 
\begin{align}
\beta = - \pi G c_{\rm O}\,,
\end{align}
and the cosmological constant 
\begin{align}
\Lambda=8 \pi G V_{\rm O}
\end{align}
such that a detailed analysis of the vacuum energy $V_{\rm O}$ will be given in Sec. \ref{sec3} below starting from its definition in (\ref{params}). 

For the purpose of the present paper, from the matter sector, in the action (\ref{fr-action1}), we retain only the electromagnetic field tensor
\begin{align}
\hat{F}_{\mu \nu} = \partial_{\mu} \hat{A}_{\nu} -\partial_{\nu} \hat{A}_{\mu} 
\end{align}
with the  dimensionless electromagnetic potential $\hat{A}_\mu = A_\mu/\sqrt{8\pi G}$.

\section{Charged Symmergent Black Hole} \label{sec3}
In this section, we present the charged black hole solution of the Symmergent gravity plus Maxwell system in (\ref{fr-action1}). The gravitational field equations take the form  ($F(R)\equiv d f(R)/dR$)
\begin{align}
\label{Einstein-eqns}
E_{\mu\nu}\equiv R_{\mu \nu} F(R)-\frac{1}{2} g_{\mu \nu} f(R) + g_{\mu \nu} \Lambda + (g_{\mu \nu} \square -\nabla_{\mu} \nabla_{\nu}) F(R) - {\hat T}_{\mu \nu} = 0
\end{align} 
accompanied by the Maxwell field equations
\begin{align}
\label{Maxwell-eqn}
\partial_\mu (\sqrt{-g} \hat{F}^{\mu \nu}) &= 0
\end{align}
where ${\hat T}_{\mu\nu}$ in (\ref{Einstein-eqns}) is the energy-momentum tensor of the dimensionless Maxwell field and is given by
\begin{align}
{\hat T}_{\mu \nu}=g^{\alpha \beta} \hat{F}_{\alpha \mu}  \hat{F}_{\beta \nu}-\frac{1}{4}  g^{\gamma \alpha} g^{\rho \beta} \hat{F}_{\alpha \beta} \hat{F}_{\gamma \rho}.    
\end{align}

We now look for a static, spherically symmetric solution for the Symmergent gravity plus the Maxwell system. We, therefore, propose the metric  
\begin{align}
d s^{2} = -h(r) d t^{2}+\frac{1}{h(r)} d r^{2} + r^{2} (d \theta^{2} +  \sin ^{2} \theta d \phi^{2})
\label{metric-fR}, 
\end{align}
formed by the single metric potential $h(r)$ and the electromagnetic scalar potential
\begin{align}
\hat{A}_0 = \hat{q}(r)
\end{align}
with vanishing vector potential $\hat{A}_i=0$ ($i=1,2,3$).

Now, using the metric (\ref{metric-fR}) the curvature scalar is found to be 
\begin{align}
R = -h^{\prime \prime} - \frac{4}{r} h^{\prime} - \frac{2}{r^2} (h -1),
\end{align}
where primes stand for derivatives with respect to the radial coordinate $r$. With this expression for $R$, non-vanishing components of the Einstein field equations $E_{\mu\nu}$ in (\ref{Einstein-eqns}) take the following forms:
\begin{align}
\label{E00-eqn}
E_0{}^0 & =  \Lambda +\frac{h^{\prime}(r)}{r} +\frac{h(r)}{r^2}  -\frac{1}{r^2} + \frac{1}{2} \hat{q}^{\prime}(r)^2  \nonumber \\ &+ \beta \left( -2 h^{\prime \prime \prime \prime}(r) h(r) - h^{\prime \prime \prime}(r) h^{\prime}(r)+\frac{1}{2} h^{\prime \prime}(r)^2-\frac{12 h^{\prime \prime \prime}(r) h(r)}{r}- \frac{2 h^{\prime}(r) h^{\prime \prime}(r)}{r}-\frac{4 h(r) h^{\prime \prime}(r)}{r^2}  \right. \nonumber \\
&\left. +\frac{2 h^{\prime}(r)^2}{r^2}  
+\frac{8 h(r) h^{\prime}(r)}{r^3}-\frac{10 h(r)^2}{r^4}+\frac{12 h(r)}{r^4}-\frac{2}{r^4} \right),
\end{align}

\begin{align}
\label{E11-eqn}
E_1{}^1 & = \Lambda + \frac{h^{\prime}(r)}{r}+\frac{h(r)}{r^2} -\frac{1}{r^2} + \frac{1}{2} \hat{q}^{\prime}(r)^2   \nonumber \\
&+ \beta \left(-h^{\prime \prime \prime}(r) h^{\prime}(r) +\frac{1}{2} h^{\prime \prime}(r)^2 - \frac{4 h^{\prime \prime \prime}(r) h(r)}{r}-\frac{2 h^{\prime}(r) h^{\prime \prime}(r)}{r}-\frac{16 h(r) h^{\prime \prime}(r)}{r^2}+\frac{2 h^{\prime}(r)^2}{r^2} \right.  \nonumber \\
&\left. +\frac{8 h(r) h^{\prime}(r)}{r^3}+\frac{14 h(r)^2}{r^4}-\frac{12 h(r)}{r^4}-\frac{2}{r^4}\right),  
\end{align}

\begin{align}
\label{E22-eqn}
E_2{}^2 & = \Lambda + \frac{h^{\prime}(r)}{r} + \frac{h^{\prime\prime}(r)}{2} -\frac{1}{2} \hat{q}^{\prime}(r)^2  \nonumber \\ 
&+\beta \left(-2 h^{\prime\prime\prime\prime}(r) h(r)-2 h^{\prime\prime\prime}(r) h^{\prime}(r) -\frac{1}{2} h^{\prime\prime}(r)^2 - \frac{10 h^{\prime\prime\prime}(r) h(r)}{r}-\frac{10 h^{\prime}(r) h^{\prime\prime}(r)}{r}+\frac{4 h^{\prime}(r)^2}{r^2} \right. \nonumber \\ 
&\left. +\frac{4 h(r) h^{\prime\prime}(r)}{r^2}+\frac{16 h(r) h^{\prime}(r)}{r^3}-\frac{12 h^{\prime}(r)}{r^3}-\frac{14 h(r)^2}{r^4}+\frac{12 h(r)}{r^4}+\frac{2}{r^4}\right),
\end{align}
with the expected relationship $E_3{}^3=\sin^2\theta E_2{}^2$.

In parallel with the Einstein field equations, using the metric (\ref{metric-fR}), the Maxwell equations (\ref{Maxwell-eqn}) reduce to
\begin{align}
r^2 \hat{q}^{\prime\prime}(r)+2 r \hat{q}^{\prime}(r)=0,
\end{align}
which is solved by the electrostatic potential
\begin{align}
\hat{q}(r) = \frac{Q}{r},
\end{align}
where we discarded a homogeneous part knowing that the scalar potential should have a purely  Coulomb form. 

Our goal now is to solve the metric potential $h(r)$ and the Coulomb potential $q(r)$ self-consistently using the system of equations (\ref{E00-eqn}), (\ref{E11-eqn}) and (\ref{E22-eqn}) supplemented by $\hat{q}(r) = \frac{Q}{r}$. To this end, the subtraction of (\ref{E11-eqn}) from (\ref{E00-eqn}) leads to the equation 
\begin{align}
\beta \left(-2 h^{\prime \prime \prime \prime}(r) h(r)-\frac{8 h^{\prime \prime \prime}(r) h(r)}{r}+\frac{12 h(r) h^{\prime \prime}(r)}{r^2}-\frac{24 h(r)^2}{r^4}+\frac{24 h(r)}{r^4}\right) = 0,
\end{align}
which is a fourth-order ordinary differential equation. Its solution is not obvious. To able to find a solution, we observe that $\beta=0$ (pure general relativity limit) is a solution, and for $\beta=0$, the (\ref{E22-eqn}) equation above would have the nontrivial solution 
\begin{align}
\label{h0(r)-sol}
h_{0}(r) = c_3 - \frac{2MG}{r} + \frac{Q^2}{2 r^2} - \frac{\Lambda r^2}{3},
\end{align}
which represents a massive ($M\neq 0$), charged ($Q\neq 0$), dS/AdS ($\Lambda\neq 0$) static black hole solution.  

Now, to include the non-vanishing $\beta$ effects, we propose a general solution of the form 
\begin{align}
\label{h(r)-decomp}
h(r) = h_0(r) + h_1(r),    
\end{align}
where we expand $h_1(r)$ as 
\begin{align}
\label{h1(r)-exp}
 h_1(r) = k_0 + \frac{k_1}{r} + \frac{k_2}{r^2} + \frac{k_3}{r^3} + \frac{k_4}{r^4} + \frac{k_5}{r^5} + \frac{k_6}{r^6}
\end{align}
up to the sixth order to be as precise as possible. Then, putting the decomposition (\ref{h(r)-decomp}) into the Einstein field equations (\ref{E00-eqn}), (\ref{E11-eqn}) and (\ref{E22-eqn}) we get the following common solution
\begin{align}
\label{h1(r)-sol}
h_1(r) = 1 - c_3 -  \frac{4 \beta \Lambda Q^2}{(1 + 8 \beta \Lambda)r^2 }    
\end{align}
for which  $k_1 = k_3 = k_4 = k_5 = k_6 = 0$. (Our trials with higher order $1/r$ terms in (\ref{h1(r)-exp}) yield all vanishing components). As a result, the decomposition in (\ref{h(r)-decomp}) leads to the general ($\beta\neq 0$) metric potential
\begin{eqnarray}
h(r) &=& 1 - \frac{2MG}{r} + \frac{1}{(1+8 \beta \Lambda)}
\frac{Q^2}{2r^2} -\frac{\Lambda r^2}{3}\\
&=& 1 - \frac{2MG}{r} + \frac{1}{(1-64 \pi^2 c_{\rm O} G^2 V_{\rm O})}
\frac{Q^2}{2r^2} -\frac{8\pi}{3}G V_{\rm O} r^2,
\label{hr-full}
\end{eqnarray}
where in the second line, we reverted to the original Symmergent gravity parameters. This solution makes it clear that the sole effect of the quadratic curvature parameter (which always comes accompanied by the cosmological constant in the form $\beta \Lambda$) is to rescale the charge of the black hole by the factor $1/(1+8\beta \Lambda)$. This solution also agrees with the recent study \cite{Nashed:2018efg} in which the potential also has vector potential components besides the scalar one. 

The metric potential (\ref{hr-full}) can be furthered thanks to the knowledge in Symmergent gravity of the vacuum energy $V_{\rm O}$ in (\ref{params}). First of all, as a loop-induced quantity,  Newton's constant in (\ref{params-0}) involves a super-trace of $({\rm masses})^2$ of the QFT fields. This means that  $V_{\rm O}$, proportional to the super-trace of  $({\rm masses})^4$ of the QFT fields, might be expressible in terms of $G$. To see this, one can go to the mass degeneracy limit in which all bosons and fermions have equal masses ($m_b=m_f=M_0$, for all $b$ and $f$). In essence, $M_0$ is the characteristic scale of the QFT (or mean value of all the field masses). Under this degenerate mass spectrum, the potential $V_{\rm O}$ can be expressed as follows:
\begin{eqnarray}
\label{analyze-VO-1}
V_{\rm O} = \frac{1}{64\pi^2}\left(\sum_{\rm B} m^4_{\rm B}- \sum_{\rm F} m^4_{\rm F}\right)\xrightarrow{\rm mass\,  degeneracy}\frac{M_0^4}{64\pi^2}(n_\text{B} - n_\text{F})=\frac{M_0^2}{8\pi G}=\frac{1}{2(8\pi G)^2 c_{\rm O}},
\end{eqnarray}
where, at the last equality, we used the relation $M_0^2=8\pi/(G(n_\text{B} - n_\text{F}))$ form the $G$ formula in (\ref{params-0}) in the degenerate limit and also used the relation $n_\text{B} - n_\text{F}=128\pi^2 c_{\rm O}$ from the $c_{\rm O}$ formula in (\ref{params}). The problem is to consider the realistic cases of non-degenerate field masses. To this end, for a QFT with a characteristic scale $M_0$  with no detailed knowledge of the mass spectrum, one can represent realistic cases by introducing the parametrization  
\begin{eqnarray}
\label{analyze-VO-2}
V_{\rm O} &=& \frac{1-{\hat \alpha}}{(8\pi G)^2 c_{\rm O}},
\end{eqnarray}
in which the parameter ${\hat \alpha}$ measures deviations of the boson and fermion masses from the characteristic scale $M_0$. Clearly, ${\hat \alpha}=1/2$ corresponds to the degenerate case in (\ref{analyze-VO-1}). Alternatively,  ${\hat \alpha}=1$ corresponds to $\sum_{\rm B} m^4_{\rm B}= \sum_{\rm F} m^4_{\rm F}$ in (\ref{analyze-VO-1}). In general,  ${\hat \alpha}>1$ (${\hat \alpha}<1$) corresponds to the fermion (boson) dominance in terms of the trace $({\rm masses})^4$. Also, ${\hat \alpha}>1$ (${\hat \alpha}<1$) corresponds to  AdS (dS) spacetime.

Now, with the vacuum energy in (\ref{analyze-VO-2}), the metric potential $h(r)$ in (\ref{hr-full}) becomes
\begin{eqnarray}
h(r) = 1 - \frac{2MG}{r} + 
\frac{Q^2}{2\hat{\alpha}r^2} -\frac{(1-\hat{\alpha})}{24 \pi G c_{\rm O}} r^2 
\label{metric}
\end{eqnarray} 
and it is seen to reduce to the usual Reissner-Nordstrom-AdS/dS black hole when $\hat{Q}^2 =\frac{Q^2}{2\hat{\alpha}}$ and $\hat{\Lambda}=\frac{(1-\hat{\alpha})}{8 \pi G c_{\rm O}}$. Henceforth, we call the metric (\ref{metric-fR}) with the potential  (\ref{metric}) as charged symmergent black hole (CSBH) to distinguish it from others in the literature. 

\begin{figure*}[!ht] 
    \centering
    \includegraphics[width=0.48\textwidth]{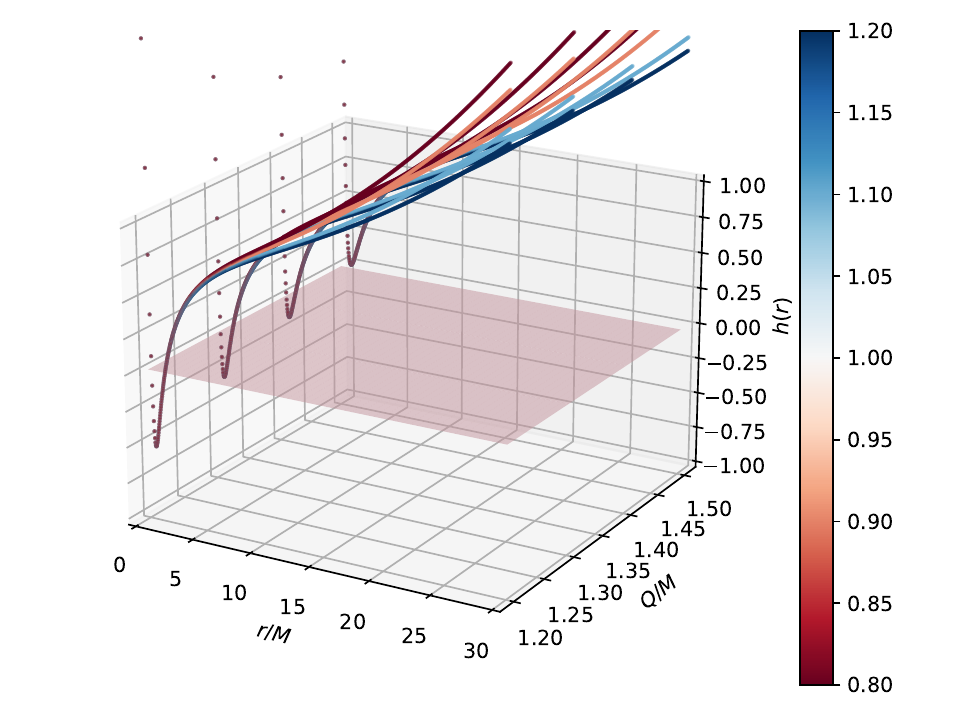}
    \includegraphics[width=0.48\textwidth]{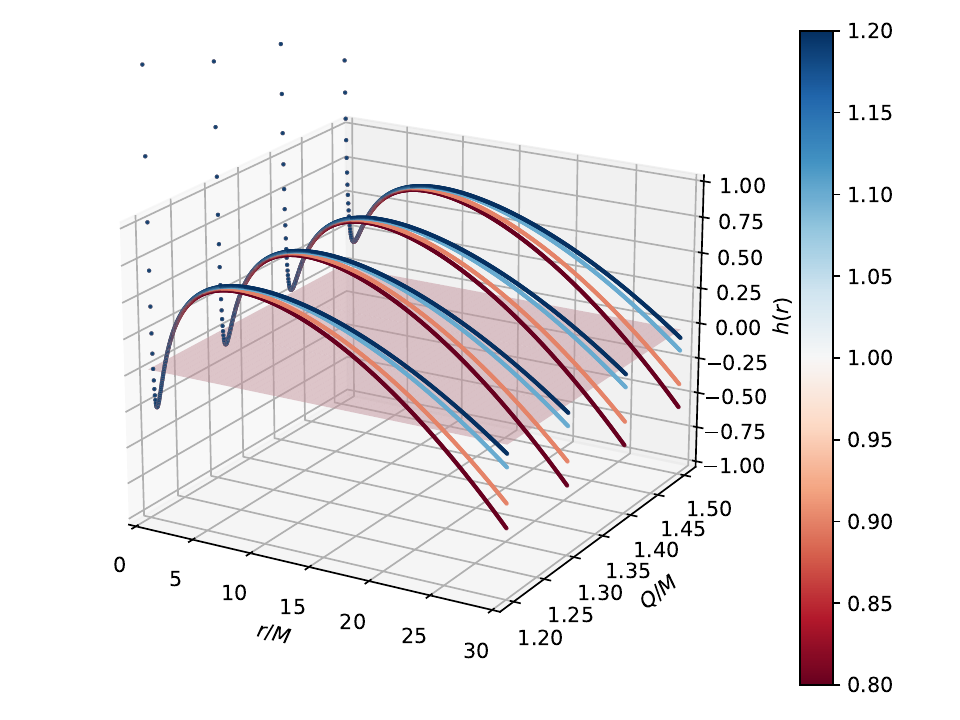}
    \caption{Variation of the metric potential $h(r)$ with the radial distance $r$, charge $Q$, and the quadratic curvature coefficient $c_\text{O}$ (in the colorbar) for $\hat{\alpha} = 1.10$ (left panel) and  $\hat{\alpha} = 0.90$ (right panel). The horizon $r=r_h$ is given by the points at which $h(r)$ vanishes ($r$-$Q$ plane at $h(r)=0$). The left (right) panel corresponds to the symmergent-AdS (symmergent-dS) spacetime. While the symmergent-AdS black hole has a single horizon (left panel), the symmergent-dS has two horizons (right panel). It is also clear that, in both panels, dependencies on $c_{\rm O}$ and $Q$ are mild (as $c_{\rm O}$ varies from $0.8$ to $1.2$, and $Q$ varies from $1.2$ to $1.55$).}
    \label{fig_hor}
\end{figure*}

\begin{figure*}[!ht] 
    \centering
    \includegraphics[width=0.48\textwidth]{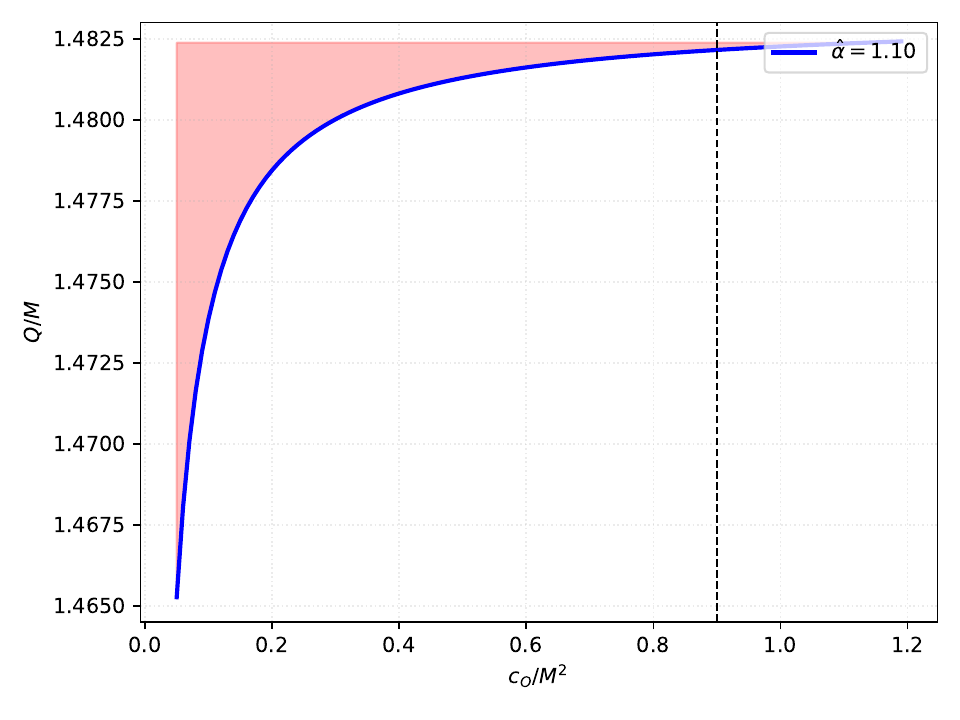}
    \includegraphics[width=0.48\textwidth]{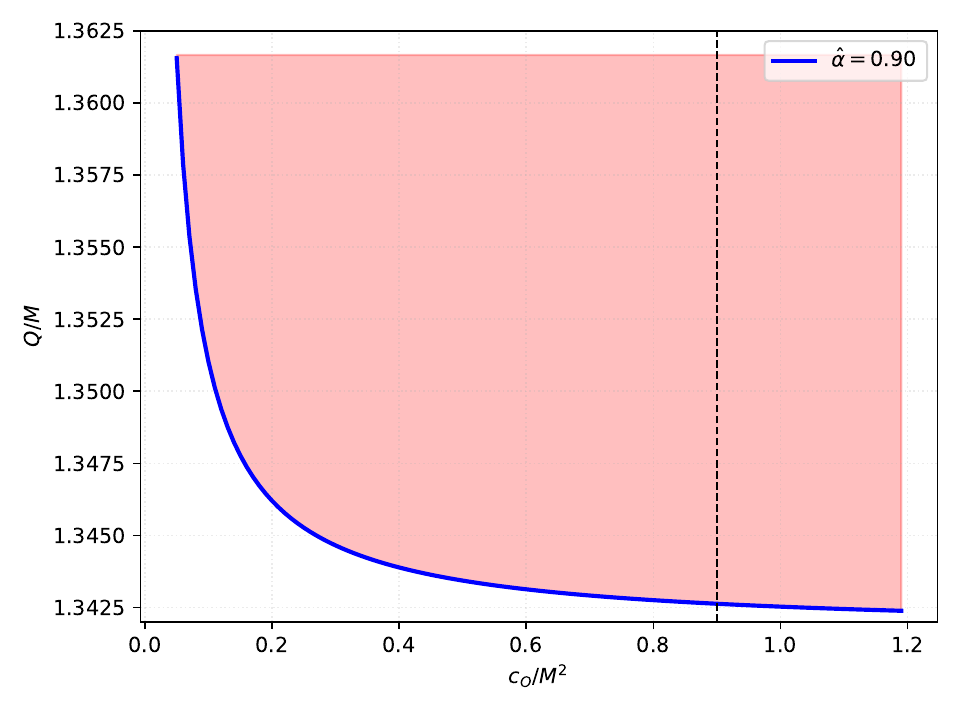}
    \caption{Variation of the upper bound on $Q$ with the symmergent parameter $c_{\rm O}$ for ${\hat \alpha}=1.1$ (left panel) and ${\hat \alpha}=0.9$ (right panel). The blue curve corresponds to points at which the minima of $h(r)$ satisfies $h(r) = 0$. In both panels, red region is forbidden since no horizon forms for  $Q$ values in these regions. The left (right) panel corresponds to the symmergent-AdS (symmergent-dS) case. The dashed vertical line is used to track the upper bound on $Q$ for a given value of the symmergent parameter $c_\text{O}$.}
    \label{fig_Q}
\end{figure*}

By definition, the radius $r=r_h$ at which $h(r_h)=0$ is the event horizon (for $r\geq 2 M$  the Schwarzschild horizon). In general, depending on the parameter values, $h(r_h)=0$ can have more than one solution (like an inner horizon $r_H=r_h^-$ and outer horizon $r=r_h^{+}$). Depicted in Fig.\ref{fig_hor} is the 3D plot of the metric potential $h(r)$ in which we explore how $h(r)$ varies with the radial distance $r$, charge $Q$, and the quadratic curvature coefficient $c_\text{O}$.
It is clear that one single horizon is formed in the symmergent-AdS case. In the symmergent-dS case, however, two horizons are formed (the second being far from the black hole). An important aspect we also notice is that for some values of $Q$, the minima rise to a point above the $r$--$Q$ plane at $h(r) = 0$, implying that no horizon is formed in the symmergent-AdS case. Only the outermost horizon is left for the symmergent-dS case. The upper bound on $Q$ is hit when the minima of $h(r)$ coincides with $h(r) = 0$ plane. We plot the results for both the symmergent-AdS/dS cases in Fig. \ref{fig_Q}.
Here, we notice that the symmergent-AdS case permits higher values for the upper bound of $Q$ than the symmergent-dS case. Let us suppose that we pick  $c_\text{O} = 0.90$, the upper bound for $Q$ in the symmergent-AdS case is $Q_\text{h} = 1.48M$, while that of the symmergent-dS case is $Q_\text{h} = 1.34M$ where we use the label $Q_\text{h}$ for $Q$ upper bound derived from the horizon formation. In addition, we also notice that the rate at which $Q$ increases relative to $c_\text{O}$ for the symmergent-AdS case seems to level off as $c_\text{O}$ gets larger. Similar observation holds also for the symmergent-AdS case as the rate at which $Q$ decreases relative to $c_\text{O}$.

\section{CSBH Hawking Temperature in Jacobi Metric formalism} \label{sec4}
\begin{figure}[ht!]
\centering
\includegraphics[width=0.48\textwidth]{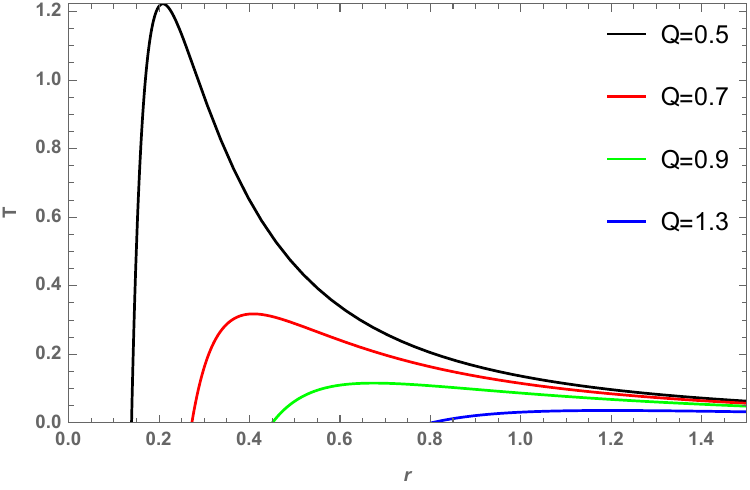}
    \caption{Hawking temperature $T\equiv (T_H)_h$ as a function of the event horizon radius $r\equiv r_h$ for $c_{\rm O}=0.5$, $\hat{\alpha} =0.9$ and different values of the black hole charge $Q$.}
    \label{fig:temp}
\end{figure}

In this section, we employ the Jacobi metric $j_{ij}$ corresponding to the covariant metrics in four dimensions. We aim to determine the Hawking temperature of the CSBH via the particle's tunneling probability through its horizon. To carry out this semi-classical analysis, we employ the WKB method, with the wavefunction $\psi=e^{(i/\hbar)S}$ for a particle with action $S$. The Jacobi, in relation to the metric in equation (\ref{metric-fR}), takes the form \cite{Gibbons:2015qja,Chanda:2016aph,Das:2016opi,Bera:2019oxg}
	\begin{equation}\label{10}
	ds^2 = j_{ij} dx^{i}dx^{j}= \Big(E^2-m^2 h(r)\Big) \Big(\frac{dr^2}{h^2(r)} + \frac{r^2}{h(r)}(d\theta^2 + \sin^2\theta d\phi^2)\Big),
	\end{equation}
where $E$ and $m$ are the energy and mass of the particle, respectively. The action of the particle is the integration of the Jacobi invariant distance
 \begin{equation}\label{11}
	S = -\int \sqrt{j_{ij}\frac{dx^i}{ds}\frac{dx^j}{ds}} ds~,
	\end{equation}
where 
	\begin{equation}\label{12}
	\sqrt{j_{ij}\frac{dx^i}{ds}\frac{dx^j}{ds}} = \pm\left(E^2- m^2 h(r)\right)^{1/2} \frac{1}{h(r)} \frac{dr}{ds}
	\end{equation}
as follows from (\ref{10}). The radial momentum of the particle using equation (\ref{12}) in equation (\ref{11}) is found to be
\begin{equation}\label{13}
p_r=\partial_r S= \mp \left(E^2-m^2h(r)\right)^{1/2}  \frac{1}{h(r)}~.
\end{equation}
In view of our tunneling approach, the particle is located inside the horizon and hence $h(r)<0$ \cite{Bera:2019oxg}. The radial momentum of the particle is $p_r=\partial_r S$, and the outgoing/incoming particle has a positive/negative momentum. Therefore, since $p_r$ becomes positive in our equation due to  $h(r)<0$, it corresponds to the particle going outwards (similar to the conventional tunneling approaches in \cite{Srinivasan:1998ty}). Tunneling occurs near the horizon at which metric gets effectively mapped to $(1+1)$-dimensions. Since only the radial movement counts \cite{Iso:2006wa} one can expand $h(r)$ around the horizon radius $r=r_h$ as
\begin{eqnarray}
h(r) = h(r_h) + h'(r_h)(r-r_h) + \mathcal{O}\left[(r-r_h)^2\right] \equiv 2\kappa(r-r_h) + \mathcal{O}(r-r_h)^2,
\label{expand}
\end{eqnarray}
in which 
\begin{eqnarray}
 \kappa=\frac{1}{2} h'(r_h)   
\end{eqnarray}
is the symmergent black hole's surface gravity. Now, substituting the expansion (\ref{expand}) in equation (\ref{12}) one obtains the near-horizon action for radial motion
		\begin{equation}\label{17}
	S = \mp \frac{E}{2\kappa} \int_{r_h-\epsilon}^{r_h+\epsilon} \frac{1}{(r-r_h)} dr \pm \frac{m^2}{2E}\int_{r_h-\epsilon}^{r_h+\epsilon} dr \mp \mathcal{O}(r-r_h), 
	\end{equation}
in which, for $\epsilon>0$, $r_h-\epsilon$ is close to the horizon and  $r_h+\epsilon$ is across the horizon. Redefining radial coordinate $r$ as $r-r_h = \epsilon e^{i\theta}$ in (\ref{17}), one gets $\int_{r_h-\epsilon}^{r_h+\epsilon} \frac{1}{(r-r_h)} dr = - i\pi$ for the first integral (residue theorem) and $\int_{r_h-\epsilon}^{r_h+\epsilon}=2\epsilon$ for the second integral. Then, the action (\ref{17}) takes the form
		\begin{equation}\label{19}
	S = \pm \frac{i\pi E}{2\kappa} + \textrm{real part}~,
	\end{equation}
in which $+$($-$) sign stands for outgoing (incoming) tunneling particles. Then, the WKB wavefunction becomes $\psi_{out} =A e^{\frac{i}{\hbar}S_{out}}$ and  $\psi_{in} = Ae^{\frac{i}{\hbar}S_{in}}$ for outgoing and incoming particles, respectively. In this regard, one obtains 
\begin{equation} \label{22}
	P_{em} = |\psi_{out}|^2 = |A|^2 \left|e^{\frac{i}{\hbar}S_{out}}\right|^2 
	=|A|^2 e^{-\frac{\pi E}{\hbar \kappa}}~
	\end{equation}
for the emission probability, and
 \begin{equation} \label{23}
	P_{ab} = |\psi_{in}|^2 = |A|^2 \left |e^{\frac{i}{\hbar}S_{in}}\right|^2 	
	= |A|^2 e^{\frac{\pi E}{\hbar \kappa}}.
	\end{equation}
for the absorption probability. One notes that the real part of the action (\ref{19}) does not contribute at all. These emission and absorption probabilities lead to the tunneling rate
		\begin{equation}\label{24}
	\Gamma = \frac{P_{out}}{P_{in}} = e^{-\frac{2\pi E}{\hbar \kappa}} \equiv  e^{-\frac{E}{T_{H}}},
	\end{equation}
	which is identical to Boltzmann factor,  with a temperature 	
	\begin{equation}\label{25}
	 T_{H} = \frac{\hbar \kappa}{2\pi}
	\end{equation}
given by the Hawking temperature. Having this formula at hand, the Hawking temperature of the CSBH at the event horizon $r=r_{h}$ takes the form:
\begin{equation}(T_H)_{h}=\frac{\hat{\alpha}  r_{h}}{48 \pi ^2 c_O G}-\frac{r_{h}}{48 \pi ^2 c_O G}+\frac{G M}{2 \pi  r_{h}^2}-\frac{Q^2}{4 \pi  \hat{\alpha} r_{h}^3}~. \label{haw-temp}\end{equation}
In Fig. \eqref{fig:temp}, we plot this Hawking temperature formula as a function of the event horizon radius for different charge values. As the figure reveals, Hawking temperature decreases slowly with decreasing the black hole charge. 

As follows from the formula (\ref{haw-temp}), the CSBH Hawking temperature reduces to the Schwarzschild black hole temperature $(T)_{h}=\frac{1}{8M\pi}$ in the limit $\hat{\alpha}= 1$ and $Q=0$.

In parallel with the Hawking temperature, the black hole mass can be expressed as 
\begin{equation}
(M)_{h}=\frac{12 \pi  c_O G Q^2+24 \pi  \hat{\alpha}  c_O G r_h^2+\hat{\alpha}^2 r_h^4-\hat{\alpha}  r_h^4}{48 \pi  \alpha  c_O G^2 r_h}
\label{mass}
\end{equation}
by using $h(r=r_h)=0$. One here notes that the CSBH’s mass reduces to the Schwarszschild black hole mass in the limit in which  $\hat{\alpha}= 1$ and $Q=0$. In fact, the formula (\ref{mass}) above is nothing but the relation 
$(M)_{h}=r_{h}/2$. 

\section{CSBH Shadow Cast with EHT Constraints} \label{sec5}
In this section, we aim to study the CSBH shadow cast as a function of the symmergent parameters and the black hole charge. This way, we will determine constraints on $c_{\rm O}$ for different values of the charge    $Q$. After determining the constraints, we will explore how the shadow radius varies with the observer distance $r_\text{obs}$. To these aims, we begin the analysis with the null-geodesic Lagrangian
\begin{equation}
    \mathcal{L} = \frac{1}{2}\left( -h(r) \dot{t}^2 +h(r)^{-1} \dot{r}^2 + r^2 \dot{\phi}^2 \right)
\end{equation}
in the equatorial plane for which $\theta=\pi/2$ in (\ref{metric-fR}). The least action principle gives two constants of motion: The energy
\begin{equation} \label{econs}
    E = h(r)\frac{dt}{d\lambda}
\end{equation}
and the angular momentum
\begin{equation} \label{lcons}
    L = r^2\frac{d\phi}{d\lambda}.
\end{equation}
Their ratio gives the impact parameter $b$ for null geodesics near the CSBH:
\begin{equation} \label{eb}
    b \equiv \frac{L}{E} = \frac{r^2}{h(r)}\frac{d\phi}{dt}.
\end{equation}
The null geodesic $ds^2=0$ leads to the photon orbit equation \cite{Khodadi:2022pqh}
\begin{equation}
\label{r-phi}
\left(\frac{d r}{d \phi}\right)^{2}=V_{e f f}(r)
\end{equation}
with the effective potential
\begin{equation}
V_{e f f}=r^{4}\left(\frac{E^{2}}{L^{2}}-\frac{h(r)}{r^{2}}\right).
\end{equation}
By using the expression for $E$ in (\ref{econs}) and $L$ in (\ref{lcons}), the effective potential takes the new form
\begin{equation}
\label{effpot2}
V_{e f f}=r^2 h(r)\left(\frac{\overline{H}(r)^2}{b^2}-1\right)
\end{equation}
after introducing
\begin{equation} \label{eh(r)}
 \overline{H}^2(r) = \frac{r^2}{h(r)}\,.
\end{equation}
Here, the null geodesic $r=r(\phi)$ remains stable if two conditions are satisfied: First, one must have $dr/d\phi=0$ and this condition implies $V_{e f f}(r_\text{ps})=0$ or $\overline{H}(r_{\rm ps})=b$ as follows from (\ref{r-phi}) and (\ref{effpot2}). Second, one must ensure  $d^2r/d\phi^2=0$, and this constraint necessitates 
$V_{e f f}^{\prime}(r_\text{ps})=0$. This latter condition reduces to $\frac{d}{d r}\left(\frac{h(r)}{r^{2}}\right)=0$ at $r=r_{\rm ps}$ and this relation takes the explicit form
\begin{equation}
\frac{h^{\prime}\left(r_{p s}\right)}{h\left(r_{p s}\right)}-\frac{2}{r_{p s}}=0\,.
\end{equation}
 From this equality follows the photon sphere radius $r_{\rm ps}$
\begin{equation} \label{erph}
    r_\text{ps}= \frac{3 M \hat{\alpha} \pm \sqrt{9 M^{2} \hat{\alpha}^{2}-4 Q^{2} \hat{\alpha}}}{2 \hat{\alpha}}
\end{equation}
as a function only of the charge $Q$ and the potential energy parameter $\hat{\alpha}$ defined in (\ref{analyze-VO-2}). This CSBH photon sphere radius is highly interesting because, compared to the RN-AdS/dS black holes, which involve only $Q$, the CSBH involves both $\hat{\alpha}$ and $Q$. In other words, the photon sphere radius in RN-AdS/dS black holes involves only  $Q$, implying that such black holes are insensitive to the cosmological constant in the strong field limit. In symmergent gravity, however, the vacuum energy in (\ref{analyze-VO-2}) generates the cosmological constant, and the CSBH exhibits, therefore, direct sensitivity to the cosmological constant in the strong field limit. 

Having determined the photon sphere radius, for an observer situated at the position $(t_\text{obs},r_\text{obs},\theta_\text{obs} = \pi/2, \phi_\text{obs})$, the angular shadow radius takes the form  \cite{Perlick:2015vta,Perlick:2021aok}
\begin{equation}
    \tan(\alpha_{\text{sh}}) = \left(\frac{r^2}{h(r)^{-1}}\right)^{1/2} \frac{d\phi}{dr} \bigg|_{r=r_\text{obs}} = \frac{b_\text{crit}}{\sqrt{\overline{H}(r_\text{obs})^{2}-b_\text{crit}^{2}}}
    \label{eangrad}
\end{equation}
with the use of the orbit equation (\ref{r-phi}). The critical impact parameter $b_{crit}$ in this equation follows from the condition $dr^2/d^2\phi = 0$ and takes the form 
\begin{equation} \label{ebcrit}
    b_\text{crit}^2 = \frac{4r_\text{ps}^2}{r h'(r) |_{r=r_\text{ps}} + 2h(r_\text{ps})}
\end{equation}
 for any static and spherically symmetric spacetime \cite{Pantig:2022ely,Pantig:2022sjb}. 
 For the CSBH, it takes the form  
\begin{equation} \label{ebcrit2}
	b_\text{crit}^2=\frac{6 r_\text{ps}^{3}}{3 r_\text{ps}- \frac{(1-\hat{\alpha})}{4 \pi G c_{\rm O}} r_\text{ps}^{3}-3G M}
\end{equation}
and leads to the shadow radius
\begin{equation} \label{ersh}
	R_\text{sh}= b_\text{crit}\sqrt{h(r_\text{obs})}
\end{equation}
corresponding to the shadow angle in (\ref{eangrad}). 

\begin{figure*}[!ht]
    \centering
    \includegraphics[width=0.48\textwidth]{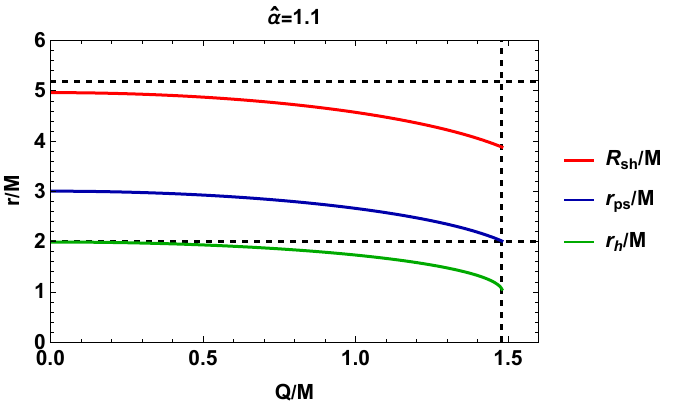}
    \includegraphics[width=0.48\textwidth]{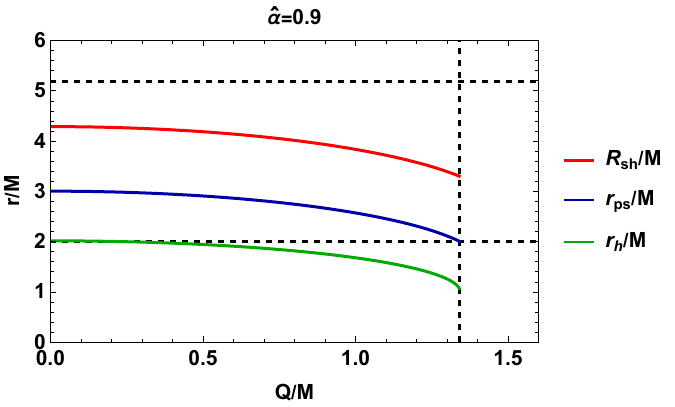}
    \caption{The horizon radius $r_h$, the photon sphere radius $r_\text{ps}$, and the shadow radius $R_\text{sh}$ (with $r_\text{obs}=10M$) as functions of the charge $Q$ for ${\hat \alpha}=1.1$ (symmergent-AdS case; left panel) and ${\hat \alpha}=0.9$ (symmergent-dS case; right panel), and $c_O = 0.9$ for both cases. The dashed horizontal lines are the photon sphere radius $r_{\text{ps}} = 2M$ in the limit case  set by the charge $Q_\text{h}$ and the shadow radius in the Schwarzschild case. The vertical dashed lines are the upper bounds on $Q_\text{h}$ for ${\hat \alpha}=1.1$ (left panel) and ${\hat \alpha}=0.9$ (right panel). It is clear that $r_h < r_\text{ps} < R_\text{sh}$ in the allowed range $0 < Q < Q_\text{h}$ of the black hole charge.}
    \label{fig_allradius}
\end{figure*}

Another important aspect of the $r_{\rm ps}$ solution in \eqref{erph} is the existence of an upper bound on $Q$. The upper bound  can be determined by requiring $r_{\rm ps}$ not to take any imaginary value. It can be denoted as $Q_\text{ps}$ to emphasize its photon sphere origin. In fact, it is given by the simple expression 
\begin{equation}
    Q_\text{ps} = \frac{3}{2}\sqrt{\hat{\alpha}} M
    \label{Q-upper-rps}
\end{equation}
as the maximal value of $Q$ (as a function of ${\hat \alpha}$) such that the $r_{\rm ps}$ in \eqref{erph} remains real. For the symmergent-AdS case with $\hat{\alpha} = 1.10$ the upper bound is $Q_\text{ps} \sim 1.57 M$. For the symmergent-dS case with $\hat{\alpha} = 0.90$, however, the bound is $Q_\text{ps} \sim 1.42 M$. To remark, we see that this behavior is similar to the upper bound $Q_\text{h}$ from the horizon radius. The only difference is that  while  $Q_\text{ps}$ involves only $\hat{\alpha}$ (as revealed by equation \eqref{Q-upper-rps} above) $Q_\text{h}$ involves also the symmergent parameter $c_\text{O}$ (see Fig. \ref{fig_Q} above). In Fig. \ref{fig_allradius} we plot the horizon radius $r_\text{h}$, photon sphere radius $r_\text{ps}$  and the shadow radius $R_\text{sh}$ (for $r_\text{obs}=10 M$) by taking into account the upper bound on $Q$ from the horizon formation ($Q<Q_\text{h}$) for ${\hat \alpha}=0.9$ (symmergent-dS) and ${\hat \alpha}=1.1$ (symmergent-AdS). The horizontal dashed lines at $r=2M$ correspond to the photon sphere radius value in the limit case set by the charge upper bound $Q_\text{h}$ from the horizon radius, and the points this dashed line intersects the $r_\text{ps}$ curves give the actual $Q_\text{h}$ values. The horizontal dashed lines at $r=3\sqrt{3}M$ correspond to the shadow radius in the Schwarzschild case, and the points it intersects the $R_\text{sh}$ curves give the $Q$ upper bound $Q_\text{h}$ from the horizon radius. The vertical dashed lines correspond to the upper bounds $Q_\text{h}$ for the given ${\hat \alpha}$ values. In general, the allowed range for the electric charge $Q$ is $0 < Q < Q_\text{h}$. As follows from Fig. \ref{fig_allradius}, in general, $r_h < r_\text{ps} < R_\text{sh}$ in the allowed range $0 < Q < Q_\text{h}$ of the black hole charge. This hierarchy is what is expected of the radii  $r_h$, $r_\text{ps}$, $R_\text{sh}$ on physical grounds. (As a side note, the shadow radius $R_\text{sh}$ is found to fall below the photon sphere radius $r_\text{ps}$ for $Q\lesssim M/2$ and $r_\text{obs}\lesssim 3 M$ (the Schwarzschild photon sphere radius).)

\begin{figure*}[!ht]
    \centering
    \includegraphics[width=0.48\textwidth]{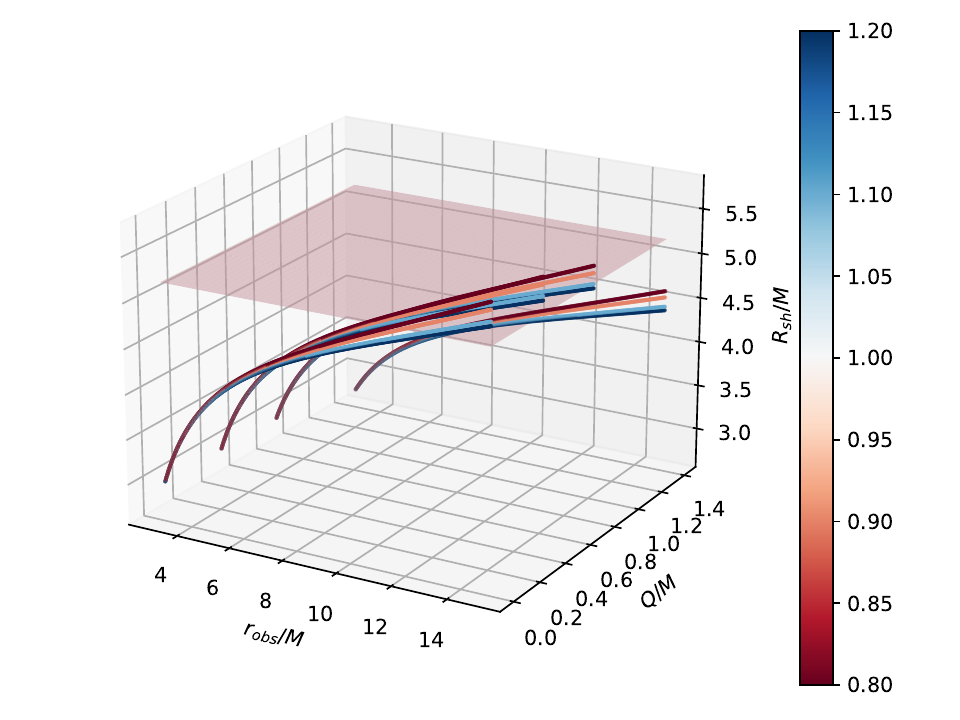}
    \includegraphics[width=0.48\textwidth]{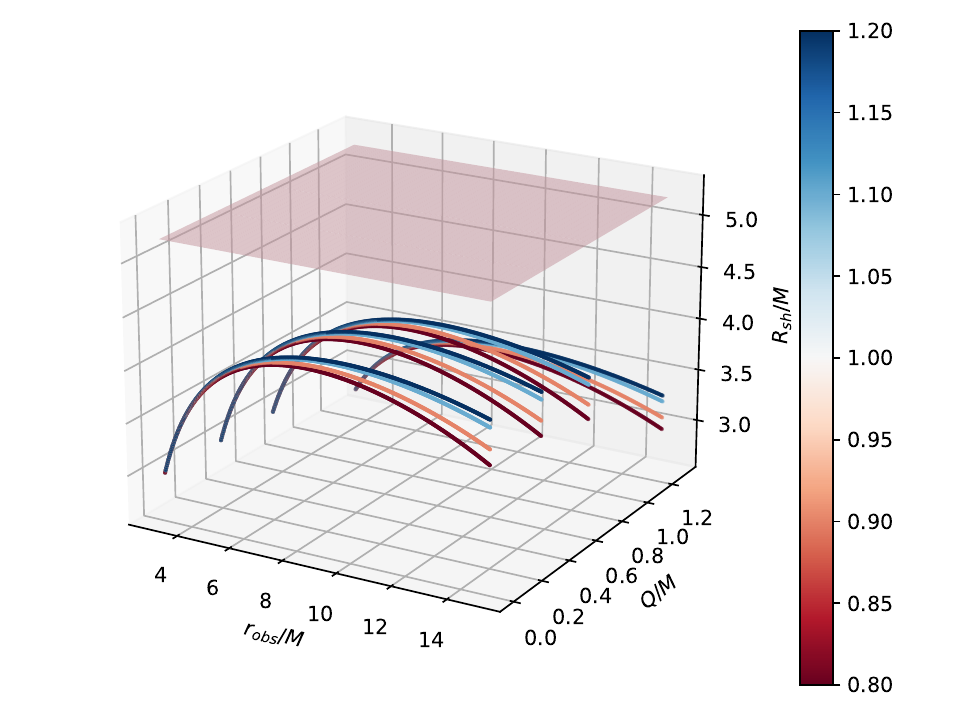}
    \caption{Variation of the shadow radius $R_\text{sh}$ with the observer position $r_\text{obs}$, black hole charge $Q$, and the quadratic curvature coupling $c_\text{O}$ (indicated by a colorbar). The left panel is for the symmergent-AdS case ($\hat{\alpha} = 1.1$), and the right panel is for the symmergent-dS case ($\hat{\alpha} = 0.90$). The shaded plane corresponds to the Schwarzschild radius of $3\sqrt{3}M$.}
    \label{fig_rsh}
\end{figure*}

In Fig. \ref{fig_rsh}, depicted is the variation of $R_\text{sh}$ in \eqref{ersh} with the observer position $r_\text{obs}$, charge $Q$, and the quadratic curvature parameter $c_\text{O}$ (given in color bar).  Here, each plane of the 3D plot gives information on a different aspect of $R_\text{sh}$. One notices a certain observer position in the symmergent-AdS case at which $R_\text{sh}$ coincides with the shadow radius of the Schwarzschild black hole (see Fig. \ref{fig_allradius}). As the observer position increases, manifestation of the AdS effect becomes stronger. In the symmergent-dS case, our 3D plot has already revealed possibility of forming another shadow radius at large $r_\text{obs}$ although, with the chosen parameter ranges, the shadow radius remains smaller than the standard Schwarzschild case of $3\sqrt{3}M$.


\subsection{Constraints through the black hole shadow} \label{sub}
Allowed values of this shadow radius under the EHT data on the supermassive black holes M87* and Sgr. A* put bounds on the symmergent parameter $c_{\rm O}$. We tabulate these observational data in Table \ref{tab1}, where the distance $r_\text{obs}$ of the observer from the SMBH (in ${\rm kpc}$) is also given. In addition to these data, we obtain the allowed  $1\sigma-$ bands for the Schwarzschild deviation \cite{EventHorizonTelescope:2019dse,EventHorizonTelescope:2022xnr,EventHorizonTelescope:2021dqv,Vagnozzi:2022moj}, which read $4.55M \leq R_\text{sh} \leq 5.22M$, and $ 4.31M \leq R_\text{sh} \leq 6.08M$ for Sgr. A* and M87*, respectively.
\begin{table}
    \centering
    \begin{tabular}{l|l|l|l}
\hline
\hline
{} &   Mass ($M_\odot$) & Angular diameter $2\alpha_\text{sh}$ ($\mu$as) &       Observer distance $r_\text{obs}$  (kpc) \\
\hline
Sgr. A* &  $4.3 \pm 0.013$x$10^6$ (VLTI) &   $48.7 \pm 7$ (EHT) &  $8.277 \pm 0.033$ \\
M87*    & $6.5 \pm 0.90$x$10^9$ &   $42 \pm 3$ &  $16800$ \\
\hline
\end{tabular}
    \caption{Observational constraints of various black hole parameters from the EHT data.}
    \label{tab1}
\end{table}
\begin{table}
    \centering
    \begin{tabular}{ll}
\hline
{$\hat{\alpha} = 0.90$} &  $1\sigma$(lower)  \\
\hline
\hline
$Q = 0.05M$ &    $38.56$ \\
$Q = 0.25M$ &    $38.58$ \\
$Q = 0.50M$ &    $38.59$ \\
$Q = 0.75M$ &    $38.62$ \\
\hline
\end{tabular}
\quad
    \begin{tabular}{ll}
\hline
{$\hat{\alpha} = 1.10$} &  $1\sigma$(upper) \\
\hline
\hline
$Q = 0.05M$ &    $39.97$ \\
$Q = 0.25M$ &    $39.74$ \\
$Q = 0.50M$ &    $39.39$ \\
$Q = 0.75M$ &    $39.09$ \\
\hline
\end{tabular}
    \caption{Allowed values of the symmergent parameter $c_{\rm O}$ according to the EHT data on Sgr. A*.}
    \label{tab_sgrA}
\end{table}
\begin{table}
        \centering
    \begin{tabular}{llll}
\hline
{$\hat{\alpha} = 0.90$} &  $1\sigma$(lower)  \\
\hline
\hline
$Q = 0.05M$ &    $45.05$  \\
$Q = 0.25M$ &    $45.06$  \\
$Q = 0.50M$ &    $45.07$  \\
$Q = 0.75M$ &    $45.10$  \\
\hline
\end{tabular}
\quad
    \begin{tabular}{llll}
\hline
{$\hat{\alpha} = 1.10$} &  $1\sigma$(upper)  \\
\hline
\hline
$Q = 0.05M$ &    $44.98$  \\
$Q = 0.25M$ &    $44.97$  \\
$Q = 0.50M$ &    $44.94$  \\
$Q = 0.75M$ &    $44.90$  \\
\hline
\end{tabular}
    \caption{Allowed values of the symmergent parameter $c_{\rm O}$ according to the EHT data on M87*.}
    \label{tab_m87}
\end{table}

\begin{figure*}[!ht]
    \centering
    \includegraphics[width=0.48\textwidth]{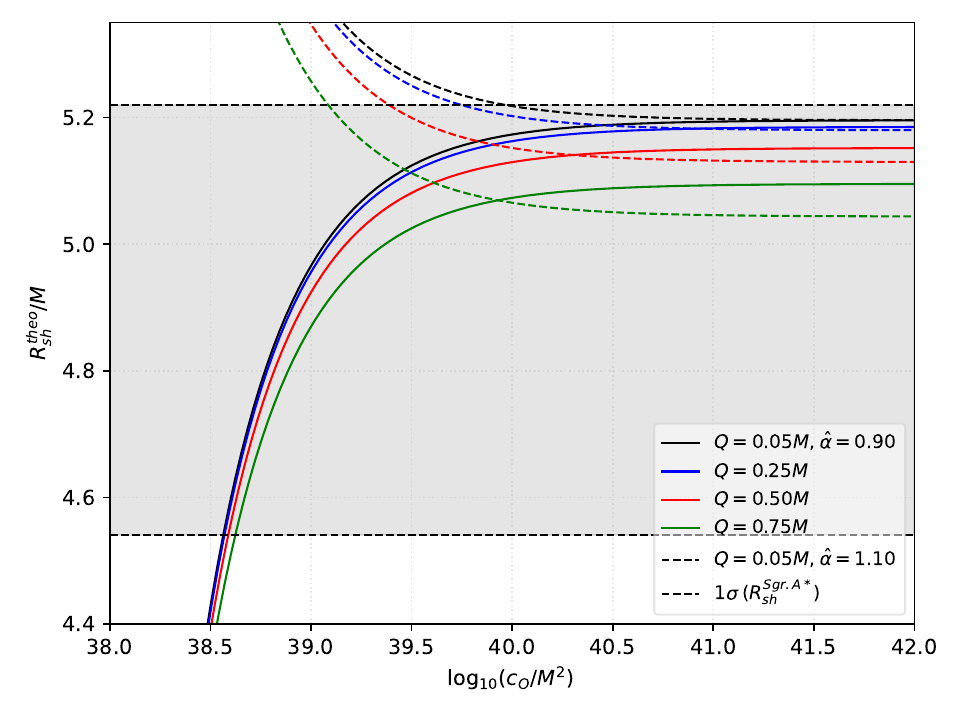}
    \includegraphics[width=0.48\textwidth]{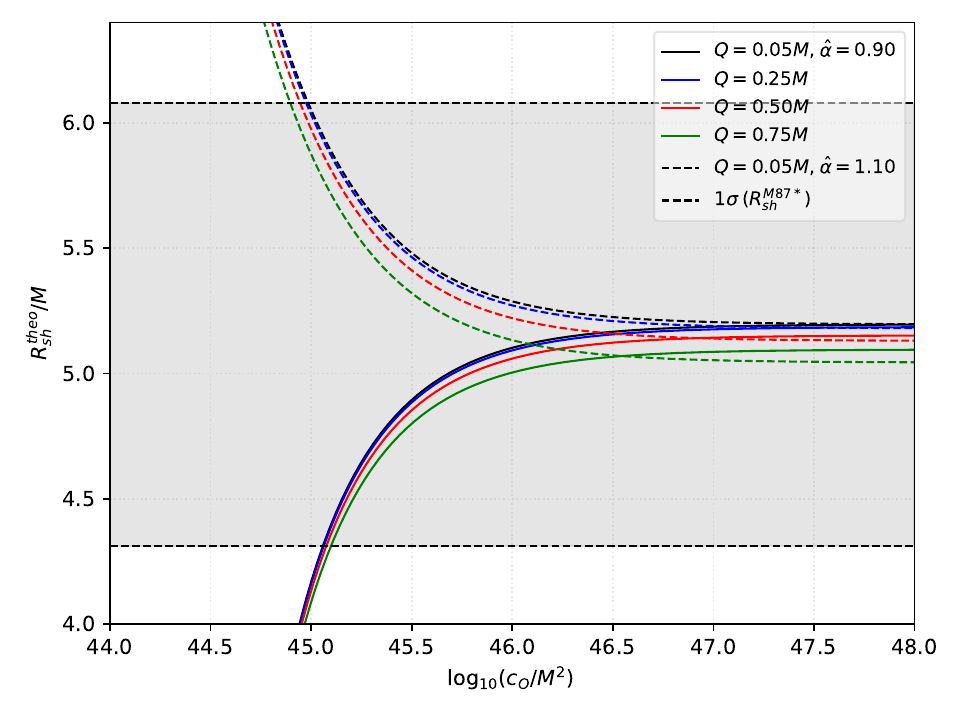}
    \caption{Allowed $1\sigma$ shadow radius bands ($68\%$ confidence level) as a function of the symmergent parameter $c_{\rm O}$ for Sgr. A* (left) and M87* (right). In each panel, with the given observer positions in Table \ref{tab1}, plotted are the shadow radius curves for certain values of the charge $Q$ and symmergent potential parameter ${\hat \alpha}$.}
    \label{fig_cons}
\end{figure*}

Fig. \ref{fig_cons} shows how the shadow radius varies with $c_{\rm O}$ at fixed observer position $r_\text{obs}$ (given in Table \ref{tab1}).
Each curve depicts a different value of the charge $Q$. The first crucial observation is that $Q$ makes the shadow radius smaller compared to the uncharged case regardless of whether the symmergent gravity mimics the dS ($\hat \alpha <1$) or AdS ($\hat \alpha > 1$) case. Somehow, the effect of dS and AdS cases becomes hard to distinguish as $c_\text{O}$ gets larger in size. This is the case such large values fall within the $1\sigma$ band as tabulated in Table \ref{tab_sgrA} and Table \ref{tab_m87}.

\begin{figure*}[!ht]
    \centering
    \includegraphics[width=0.50\textwidth]{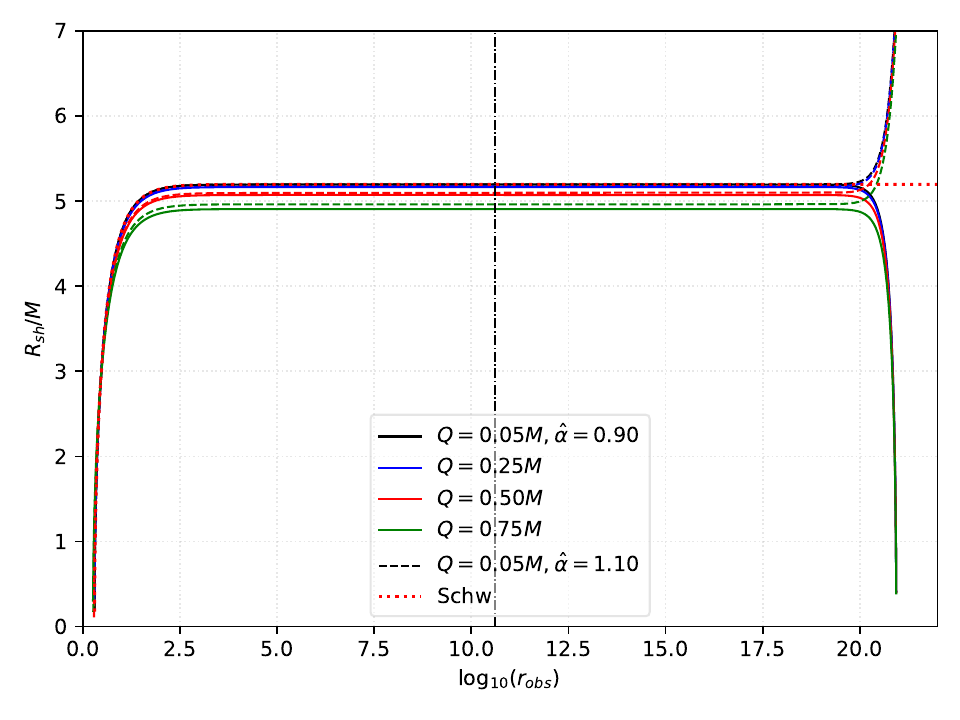}
    \caption{Variation of the shadow radius $R_\text{sh}$ with the position $r_\text{obs}$ of the observer for different values of the black hole charge $Q$. The solid line is for $\hat{\alpha} < 1$ (dS behavior) and dashed line for $\hat{\alpha} > 1$ (AdS behavior). The dotted horizontal line is for the Schwarzschild black hole. On the other hand, the dot-dashed vertical line designates the actual location of Earth from the black hole Sgr. A*.}
    \label{sha_obs}
\end{figure*}

At this stage, using the allowed parameter space in Fig. \ref{fig_cons}, it is possible to explore  how the shadow changes if the observer changes its position $r_\text{obs}$ relative to the black hole. We do this in Fig. \ref{sha_obs} by fixing the symmergent parameter $c_{\rm O}$ as $c_O/M^2 = 10^{39}$.
As the plot reveals, the spacetime of the symmergent gravity is not asymptotically-flat as the effect of $c_{\rm O}$ occurs far from the Earth's location. As the CSBH mimics the dS spacetime, the shadow radius tends to lower and lower values at a vast distance. On the other hand, when it mimics the AdS behavior, the shadow radius grows larger and larger values again at a vast distance. The conclusion is that the mere effect seen by the observer at $r_\text{obs} = 8277$ pc is due to the charge $Q$. Mainly, large $Q$ causes the shadow radius to decrease for both dS and AdS spacetimes, with a slightly higher value for AdS.

\section{Weak deflection angle of CSBH with finite distance method} \label{sec6}
Considering our location from Sgr. A* and M87*, results from the previous section reveal that the symmergent effects become significant only at far-off regions. In this regard, it becomes necessary to find some other black hole observable with a higher potential to reveal the symmergent effects. To this end, in this section, we study the weak deflection angle from the CSBH in four dimensions (following the methodology in \cite{Li:2020wvn} and the earlier works \cite{Ishihara:2016vdc,Ono:2017pie} based on the Gauss-Bonnet theorem \cite{Carmo2016,Klingenberg2013,Gibbons:2008rj,Werner2012}).

With the CSBH metric in \eqref{metric-fR} and \eqref{10}, specializing to the equatorial plane $\theta = \pi/2$, the Jacobi takes the form
\begin{equation} \label{eJac}
    dl^2=(E^2-m^2h(r))\left(\frac{1}{h(r)^2}dr^2+\frac{r^2}{h(r)}d\phi^2\right),
\end{equation}
where
\begin{equation} \label{en}
    E = \frac{m}{\sqrt{1-v^2}}
\end{equation} 
is the energy of a time-like particle with relativistic speed $v$.

The weak deflection angle at a finite distance is obtained by the integral \cite{Li:2020wvn}
\begin{equation} \label{eIshi}
    \Theta = \iint_{D}KdS + \phi_{\text{RS}},
\end{equation}
in which the integration domain $D$ is a quadrilateral specified by ${_{r_\text{ps}}^{R }\square _{r_\text{ps}}^{S}}$, where $r_\text{ps}$ is the photon sphere radius, and $S$ and $R$ are the locations of the photon source and some static observer (the receiver), respectively. Also, the differential surface area $dS$ is given by
\begin{equation}
    dS = \sqrt{g}drd\phi,
\end{equation}
in which 
\begin{equation}
    g=\frac{r^2}{h(r)^3}(E^2-m^2 h(r))^2
\end{equation}
is the determinant of the Jacobi metric, and 
\begin{align}
    K&=\frac{1}{\sqrt{g}}\left[\frac{\partial}{\partial\phi}\left(h(r) \sqrt{g}\Gamma_{rr}^{\phi}\right)-\frac{\partial}{\partial r}\left(h(r) \sqrt{g}\Gamma_{r\phi}^{\phi}\right)\right] 
    =-\frac{1}{\sqrt{g}}\left[\frac{\partial}{\partial r}\left(h(r) \sqrt{g}\Gamma_{r\phi}^{\phi}\right)\right]
\end{align}
is the Gaussian curvature.

In the definition of the deflection angle \eqref{eIshi}, the offset angle $\phi_\text{RS}$ is the difference between the angular coordinates of the receiver ($\phi_R$) and source ($\phi_S$). It is defined as $\phi_\text{RS} = \phi_\text{R}-\phi_\text{S}$. This angle can be determined by iteratively solving 
\begin{equation}
    \left(\frac{du}{d\phi}\right)^2 = \left(\frac{E}{J}\right)^2-h(u)\left(\frac{1}{J^2}+u^2\right) \equiv F(u),
    \label{motion-d}
\end{equation}
in which $u=1/r$ is the usual celestial coordinate, and 
\begin{equation}
    J = \frac{m v b}{\sqrt{1-v^2}}
\end{equation}
is the relativistic angular momentum (like the energy $E$ in \eqref{en}), for the impact parameter $b$. Bringing in $h(u)$ from \eqref{metric}, the function $F(u)$ in \eqref{motion-d} takes the form
\begin{align}
    &F(u) = \frac{E^2-1}{J^2}-u^2-u^2\left(\frac{1}{J^2}+u^2\right)
\hat{Q}^2 
    +\left(\frac{1}{J^2 u^2}+1\right)
    \frac{\hat{\Lambda}}{3} 
    + \left(\frac{1}{J^2}+u^2\right)2 M u
\end{align}
whose iterative solution gives the sought trajectory
\begin{equation} \label{orb}
    u(\phi) = \frac{\sin(\phi)}{b}+\frac{1+v^2\cos^2(\phi)}{b^2v^2}M - \frac{{\hat Q}^2}{2 v^2 b^3} + \frac{{\hat \Lambda}b}{6v^2},
\end{equation}
where one recalls that ${\hat Q}^2=Q^2/2{\hat \alpha}$ and ${\hat \Lambda} = (1-{\hat \alpha})/8\pi c_{\rm O}$ in the language of RN-dS or RN-AdS black holes. 

\begin{figure*}
    \centering
    \includegraphics[width=0.48\textwidth]{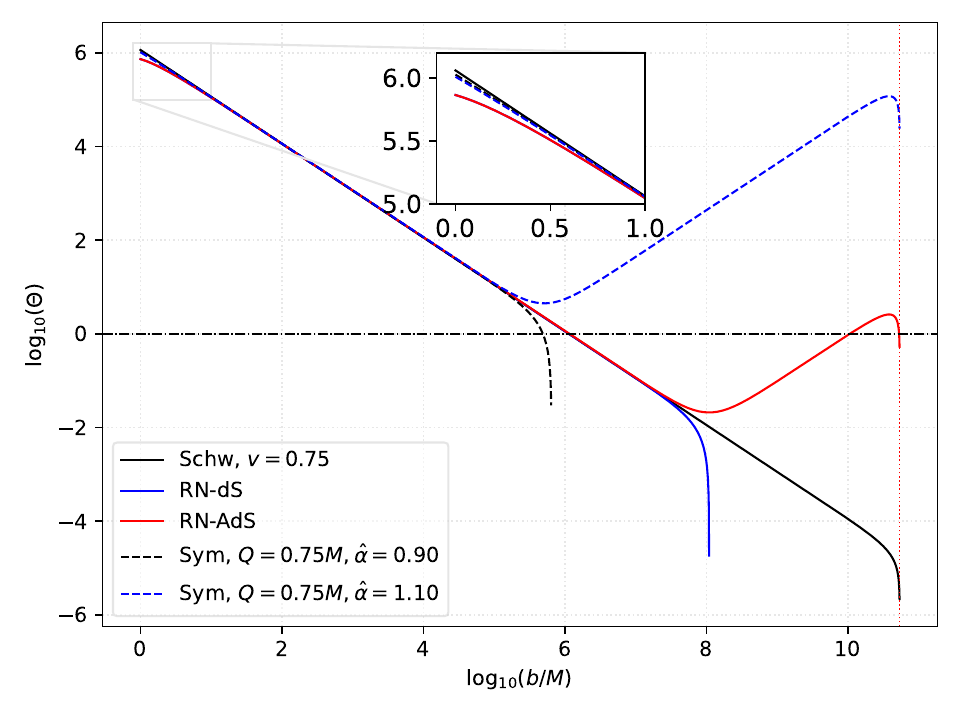}
    \includegraphics[width=0.48\textwidth]{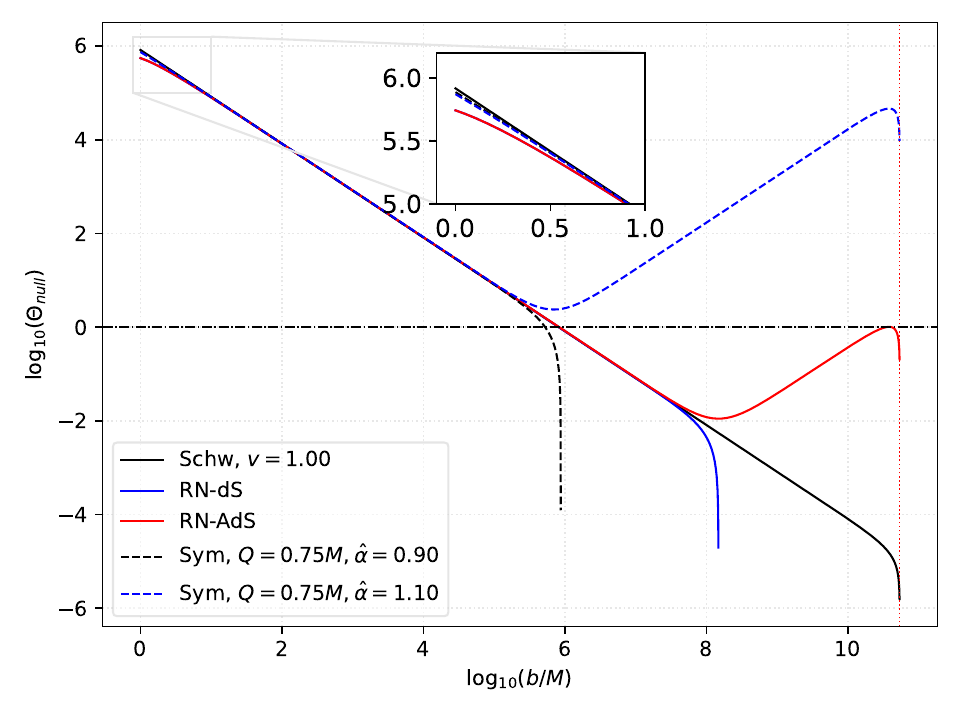}
    \caption{Weak deflection angle based on the M87* EHT data. The left panel is for time-like particles with speed $v = 0.75$. Conversely, the right panel is for null particles like photons ($v=1$). The figure contrasts CSBH results with the RN-AdS/dS black holes for $\hat{\alpha} = 0.90$ using $c_\text{O} = 45.10 M^2$, and $\hat{\alpha} = 1.10$ using $c_\text{O} = 44.90 M^2$, both at $D = 16.8$ Mpc and black hole charge $Q = 0.75 M$.}
    \label{fig_wda}
\end{figure*}

Going back to \eqref{eIshi}, the radial integration \cite{Li:2020wvn}
\begin{align} \label{gct}
    \int_{r_\text{ps}}^{r(\phi)} K\sqrt{g}dr &= -\frac{2rh(r)\left(E^{2}-h(r)\right)-r^2 E^{2}h(r)'}{2r\left(E^{2}-h(r)\right)\sqrt{h(r)}}\bigg|_{r = r(\phi)}\nonumber\\
 &= -\frac{\left(2E^{2}-1\right)M(\cos(\phi_\text{R})-\cos(\phi_\text{S}))}{\left(E^{2}-1\right)b} 
    -\frac{\left(3E^{2}-1\right)\hat{Q}^2\left[\phi_\text{RS}-\frac{(\sin(2\phi_\text{R})-\sin(2\phi_\text{S})}{2}\right]}{4\left(E^{2}-1\right)b^{2}} \nonumber\\
    &+\frac{\left(1+E^{2}\right)b^{2}\hat{\Lambda}(\cot(\phi_\text{R})-\cot(\phi_\text{S}))}{6\left(E^{2}-1\right)} -\phi_\text{RS} 
    + \mathcal{O}[M\hat{Q}^2,M\hat{\Lambda},\hat{\Lambda} \hat{Q}^2,M\hat{Q}^2\hat{\Lambda}]
\end{align}
sets the deflection angle $\alpha_{\rm def}$. The angular positions of the sources and receiver read as 
\begin{align} \label{s}
    \phi_\text{S} =\arcsin(bu)+\frac{M\left[v^{2}\left(b^{2}u^{2}-1\right]-1\right)}{bv^{2}\sqrt{1-b^{2}u^{2}}} 
    +\frac{\hat{Q}^2}{2b^{2}v^{2}\sqrt{1-b^{2}u^{2}}}-\frac{b^{2}\hat{\Lambda}}{3\sqrt{2}v^{2}\sqrt{2-2b^{2}u^{2}}} 
    + \mathcal{O}[M\hat{Q}^2,M\hat{\Lambda},\hat{\Lambda} \hat{Q}^2,M\hat{Q}^2\hat{\Lambda}],
\end{align}
and $\phi_\text{R}=\pi-\phi_\text{S}$. Now, since $\phi_\text{RS} = \pi - 2\phi_\text{S}$ and since $\sin(\pi-\phi_\text{S})=\sin(\phi_\text{S})$ the $\sin(\dots)$ terms in \eqref{gct} cancel out, and one is led to 
\begin{align} \label{ewda}
    \Theta &= \int^{\phi_\text{R}}_{\phi_\text{S}} \left[-\frac{2rh(r)\left(E^{2}-h(r)\right)-r^2 E^{2}h(r)'}{2r\left(E^{2}-h(r)\right)\sqrt{h(r)}}\bigg|_{r = r(\phi)}\right]d\phi + \phi_\text{RS}\nonumber\\
    &= \frac{M\left(v^{2}+1\right)}{bv^{2}}\left(\sqrt{1-b^{2}u_\text{R}^{2}}+\sqrt{1-b^{2}u_\text{S}^{2}}\right) 
    -\frac{\hat{Q}^2\left(v^{2}+2\right)}{4b^{2}v^{2}}\left[\pi-(\arcsin(bu_\text{R})+\arcsin(bu_\text{S}))\right] \nonumber\\
    &+\frac{b\hat{\Lambda}\left(v^{2}-2\right)}{6v^{2}}\left(\frac{\sqrt{1-b^{2}u_\text{R}^{2}}}{u_\text{R}}+\frac{\sqrt{1-b^{2}u_\text{S}^{2}}}{u_\text{S}}\right) 
    +\mathcal{O}[M\hat{Q}^2,M\hat{\Lambda},\hat{\Lambda} \hat{Q}^2,M\hat{Q}^2\hat{\Lambda}]
\end{align}
after expanding $\cos(\phi_\text{S})$ and $\cot(\phi_\text{S})$ via the source angle in \eqref{s}. This expression for the deflection angle, which involves the finite source distance $u_\text{S}$ and receiver distance $u_\text{R}$, can be approximated by going to large distances (small  $u$) so that  $b^2 u^2 \lll 1$ and one gets  the simple expression
\begin{align} \label{ewda_exact}
    \Theta = \frac{2M\left(v^{2}+1\right)}{bv^{2}}-\frac{Q^2 \pi\left(v^{2}+2\right)}{8{\hat \alpha}b^{2}v^{2}} 
    +\frac{b(1-\hat{\alpha})\left(v^{2}-2\right)}{48\pi c_{\rm O} v^{2}}\left(\frac{1}{u_\text{R}}+\frac{1}{u_\text{S}}\right) 
    +\mathcal{O}[M\hat{Q}^2,M\hat{\Lambda},\hat{\Lambda} \hat{Q}^2,M\hat{Q}^2\hat{\Lambda}],
\end{align}
where the original parameters $Q$, ${\hat \alpha}$ and $c_{\rm O}$ are restored. For null particles like the photon, one has
\begin{align} \label{wdafin}
      \Theta_{\rm null} = \frac{4M}{b}-\frac{3\pi Q^2}{8{\hat \alpha}b^{2}} 
    -\frac{b(1-\hat{\alpha})}{48\pi c_{\rm O}}\left(\frac{1}{u_\text{R}}+\frac{1}{u_\text{S}}\right) 
    +\mathcal{O}[M\hat{Q}^2,M\hat{\Lambda},\hat{\Lambda} \hat{Q}^2,M\hat{Q}^2\hat{\Lambda}]
\end{align}
in agreement with \cite{Ishihara:2016vdc}.
We now study the deflection angle in \eqref{ewda_exact} under the M87* constraints. In our analysis, we include the Schwarzschild black hole for comparison. In addition, we contrast our results to those of the RN-AdS and RN-dS spacetimes. Our results are depicted in Fig. \ref{fig_wda}. Its
left panel depicts the deflection angles of time-like particles in equation \eqref{ewda_exact}. Its right panel, on the other hand, depicts the deflection angle of the null particles in equation \eqref{wdafin}. It is clear from the figure that, at large impact parameters ($b\gtrsim 10^6 M$), the time-like deflection angle is significantly larger than the null deflection angle. To study the effect of charge $Q$, we set $Q =0.75 M$ (non-extremal) to enhance the theoretical result. To this end, our results show that $Q$ dominates for distances near the black hole. We can also say that the cosmological constant (parametrized by $\hat \alpha$) and the loop factor $c_{\rm O}$) have no discernible effect on $\Theta$ since $\Theta$ coincides with the Schwarzschild limit as the impact parameter $b$ gets smaller. At considerably large $b$, however, effects of the cosmological effect become discernible, and a distinction between the AdS and dS cases (both for $c_{\rm O}$ and ${\hat \alpha}$) becomes possible. The analysis in Fig. \ref{fig_wda} sets an example of deflection angle in asymptotically non-flat spacetimes like the symmergent one in (\ref{metric}).

\section{Conclusion and Future Prospects} \label{conc}

In the present work, we have performed a detailed study of the black hole solutions in symmergent gravity with the Maxwell field. As discussed in detail in Sec. II, symmergent gravity is an emergent gravity theory in which gravity emerges from quantum loops in a way restoring the gauge symmetries. It generates Newton's constant $G$,  quadratic curvature coefficient $c_{\rm O}$, and the vacuum energy $V_{\rm O}$ (parametrized by ${\hat \alpha}$) from the loops. We have constructed charged symmergent black holes and contrasted them with the Schwarzschild solution and the RN-AdS and RN-dS black hole solutions.

In our analysis, we studied various observable features of the CSBH. Firstly, we studied the CSBH metric potential $h(r)$ in a way revealing the combined effect of the charge $Q$, symmergent quadratic curvature parameter $c_\text{O}$, and the symmergent potential parameter $\hat{\alpha}$. In view of this parameter space, we found that there arises one horizon for the symmergent-AdS case and two horizons for the symmergent-dS case. There is also an upper bound on the charge $Q$ for both cases, where for any $Q$ greater than the said bound the horizon turns to imaginary. Secondly, we studied the CSBH photon sphere and the shadow radius. We found that photon sphere radius does not depend on the quadratic curvature parameter $c_\text{O}$. This independence allows for a $Q$ upper bound which is larger than the upper bound found by the horizon formation. Our analysis focuses on relatively low values of $Q$ in view of  the analyses of astrophysical black holes which restrict $Q$ to be nearly zero, far from the extremal limit $Q=M$ \cite{Zajacek:2018ycb}. In fact,  the highest recorded observational bound on the electric charge of Sgr A* is $\sim 3 \times 10^8 $ C (or $\sim 9.16 \times 10^{-9}$ m in geometrized units). Apart from the charge, one notes that exclusion of M87* spin parameter $a$ in the present study is justified by the analyses of \cite{Vagnozzi:2022moj}.

Weak field deflection provides a window into symmergent effects when the light rays scatter with very large impact parameters. In contrast, the charge $Q$ has no significant effect on the weak deflection angle. In this sense, shadow serves as a more sensitive probe of asymptotically non-flat spacetimes. Future astronomical devices can probe the symmergent parameter space. One such device would be the EHT reaching $10-15 \mu$as level within $345$ GHz. Another device would be the ESA GAIA, capable of resolving $7-20\mu$as \cite{Liu:2016nwt}. And yet another device would be the powerful VLBI RadioAstron which can reach an angular resolution as small as $1-10\mu$as \cite{Kardashev:2013cla}. As suggested by Fig. \ref{fig_wda}, symmergent gravity with RN-dS behavior becomes observable at smaller impact parameters, and what is needed are astronomical devices having smaller than $1\mu$as resolution. Conversely, weak gravitational lensing is a subtle effect and is difficult to accurately measure. However, advances in technology and observation techniques (in relation to weak lensing) can make it possible in the near future.

Studies of symmergent gravity in relation to black holes \cite{Symmergent-bh,Symmergent-bh2,Symmergentresults} have started a novel research direction. As the present work has shown, black holes can provide windows into the symmergent parameter space, and the few topics below can provide further windows into symmergence:
\begin{enumerate}

\item Investigation of the effects of symmergent gravity in other astrophysical objects: It could be possible to test the symmergence in other astrophysical objects such as neutron stars, boson stars, Proca stars, etc. \cite{stars} Such tests can be useful to the extent one has a precise knowledge of the density and pressure of the astrophysical object. Another important factor is that the quadratic curvature term (proportional to $c_{\rm O}$) gives cause to ghosts, and to avoid them, one treats the quadratic curvature term as an energy-momentum tensor of some exotic fields -- a new dynamics outside the existing gravitational framework.

\item Study of the effects of symmergent gravity in the strong field regime: It could be interesting to study strong field regimes and probe the symmergent gravity via gravitational waves, quasinormal modes, and such.

\item Investigation of the observational signatures of Symmergent gravity: It would be interesting to investigate other observational signatures of symmergent gravity, such as its implications for dark matter, dark photons, and the cosmic microwave background radiation.
\end{enumerate}

\acknowledgements
The authors thank the anonymous referees for their helpful comments that improved the quality of the manuscript. B. P., R. P. and A. {\"O}. would like to acknowledge networking support by the COST Action CA18108 - Quantum gravity phenomenology in the multi-messenger approach (QG-MM). B. P., A. {\"O}. and D. D.  would like to acknowledge networking support by the COST Action CA21106 - COSMIC WISPers in the Dark Universe: Theory, astrophysics and experiments (CosmicWISPers). The work of B.P. is supported by T{\"U}B{\.I}TAK B{\.I}DEB-2218 national postdoctoral fellowship program.

\bibliography{ref.bib}

\begin{thebibliography}{127}%
\makeatletter
\providecommand \@ifxundefined [1]{%
 \@ifx{#1\undefined}
}%
\providecommand \@ifnum [1]{%
 \ifnum #1\expandafter \@firstoftwo
 \else \expandafter \@secondoftwo
 \fi
}%
\providecommand \@ifx [1]{%
 \ifx #1\expandafter \@firstoftwo
 \else \expandafter \@secondoftwo
 \fi
}%
\providecommand \natexlab [1]{#1}%
\providecommand \enquote  [1]{``#1''}%
\providecommand \bibnamefont  [1]{#1}%
\providecommand \bibfnamefont [1]{#1}%
\providecommand \citenamefont [1]{#1}%
\providecommand \href@noop [0]{\@secondoftwo}%
\providecommand \href [0]{\begingroup \@sanitize@url \@href}%
\providecommand \@href[1]{\@@startlink{#1}\@@href}%
\providecommand \@@href[1]{\endgroup#1\@@endlink}%
\providecommand \@sanitize@url [0]{\catcode `\\12\catcode `\$12\catcode
  `\&12\catcode `\#12\catcode `\^12\catcode `\_12\catcode `\%12\relax}%
\providecommand \@@startlink[1]{}%
\providecommand \@@endlink[0]{}%
\providecommand \url  [0]{\begingroup\@sanitize@url \@url }%
\providecommand \@url [1]{\endgroup\@href {#1}{\urlprefix }}%
\providecommand \urlprefix  [0]{URL }%
\providecommand \Eprint [0]{\href }%
\providecommand \doibase [0]{http://dx.doi.org/}%
\providecommand \selectlanguage [0]{\@gobble}%
\providecommand \bibinfo  [0]{\@secondoftwo}%
\providecommand \bibfield  [0]{\@secondoftwo}%
\providecommand \translation [1]{[#1]}%
\providecommand \BibitemOpen [0]{}%
\providecommand \bibitemStop [0]{}%
\providecommand \bibitemNoStop [0]{.\EOS\space}%
\providecommand \EOS [0]{\spacefactor3000\relax}%
\providecommand \BibitemShut  [1]{\csname bibitem#1\endcsname}%
\let\auto@bib@innerbib\@empty
\bibitem [{\citenamefont {Lambiase}\ \emph {et~al.}(2023)\citenamefont
  {Lambiase}, \citenamefont {Mastrototaro},\ and\ \citenamefont
  {Visinelli}}]{Lambiase:2022ucu}%
  \BibitemOpen
  \bibfield  {author} {\bibinfo {author} {\bibfnamefont {Gaetano}\ \bibnamefont
  {Lambiase}}, \bibinfo {author} {\bibfnamefont {Leonardo}\ \bibnamefont
  {Mastrototaro}}, \ and\ \bibinfo {author} {\bibfnamefont {Luca}\ \bibnamefont
  {Visinelli}},\ }\bibfield  {title} {\enquote {\bibinfo {title}
  {{Gravitational waves and neutrino oscillations in Chern-Simons axion
  gravity}},}\ }\href {\doibase 10.1088/1475-7516/2023/01/011} {\bibfield
  {journal} {\bibinfo  {journal} {JCAP}\ }\textbf {\bibinfo {volume} {01}},\
  \bibinfo {pages} {011} (\bibinfo {year} {2023})},\ \Eprint
  {http://arxiv.org/abs/2207.08067} {arXiv:2207.08067 [hep-ph]} \BibitemShut
  {NoStop}%
\bibitem [{\citenamefont {Lambiase}\ and\ \citenamefont
  {Mastrototaro}(2020)}]{Lambiase:2020iul}%
  \BibitemOpen
  \bibfield  {author} {\bibinfo {author} {\bibfnamefont {Gaetano}\ \bibnamefont
  {Lambiase}}\ and\ \bibinfo {author} {\bibfnamefont {Leonardo}\ \bibnamefont
  {Mastrototaro}},\ }\bibfield  {title} {\enquote {\bibinfo {title} {{Effects
  of modified theories of gravity on neutrino pair annihilation energy
  deposition near neutron stars}},}\ }\href {\doibase 10.3847/1538-4357/abba2c}
  {\bibfield  {journal} {\bibinfo  {journal} {Astrophys. J.}\ }\textbf
  {\bibinfo {volume} {904}},\ \bibinfo {pages} {19} (\bibinfo {year} {2020})},\
  \Eprint {http://arxiv.org/abs/2009.08722} {arXiv:2009.08722 [astro-ph.HE]}
  \BibitemShut {NoStop}%
\bibitem [{\citenamefont {Berti}\ \emph {et~al.}(2015)\citenamefont {Berti}
  \emph {et~al.}}]{Berti:2015itd}%
  \BibitemOpen
  \bibfield  {author} {\bibinfo {author} {\bibfnamefont {Emanuele}\
  \bibnamefont {Berti}} \emph {et~al.},\ }\bibfield  {title} {\enquote
  {\bibinfo {title} {{Testing General Relativity with Present and Future
  Astrophysical Observations}},}\ }\href {\doibase
  10.1088/0264-9381/32/24/243001} {\bibfield  {journal} {\bibinfo  {journal}
  {Class. Quant. Grav.}\ }\textbf {\bibinfo {volume} {32}},\ \bibinfo {pages}
  {243001} (\bibinfo {year} {2015})},\ \Eprint
  {http://arxiv.org/abs/1501.07274} {arXiv:1501.07274 [gr-qc]} \BibitemShut
  {NoStop}%
\bibitem [{\citenamefont {De~Felice}\ and\ \citenamefont
  {Tsujikawa}(2010)}]{DeFelice:2010aj}%
  \BibitemOpen
  \bibfield  {author} {\bibinfo {author} {\bibfnamefont {Antonio}\ \bibnamefont
  {De~Felice}}\ and\ \bibinfo {author} {\bibfnamefont {Shinji}\ \bibnamefont
  {Tsujikawa}},\ }\bibfield  {title} {\enquote {\bibinfo {title} {{f(R)
  theories}},}\ }\href {\doibase 10.12942/lrr-2010-3} {\bibfield  {journal}
  {\bibinfo  {journal} {Living Rev. Rel.}\ }\textbf {\bibinfo {volume} {13}},\
  \bibinfo {pages} {3} (\bibinfo {year} {2010})},\ \Eprint
  {http://arxiv.org/abs/1002.4928} {arXiv:1002.4928 [gr-qc]} \BibitemShut
  {NoStop}%
\bibitem [{\citenamefont {Nojiri}\ \emph {et~al.}(2017)\citenamefont {Nojiri},
  \citenamefont {Odintsov},\ and\ \citenamefont {Oikonomou}}]{Nojiri:2017ncd}%
  \BibitemOpen
  \bibfield  {author} {\bibinfo {author} {\bibfnamefont {S.}~\bibnamefont
  {Nojiri}}, \bibinfo {author} {\bibfnamefont {S.~D.}\ \bibnamefont
  {Odintsov}}, \ and\ \bibinfo {author} {\bibfnamefont {V.~K.}\ \bibnamefont
  {Oikonomou}},\ }\bibfield  {title} {\enquote {\bibinfo {title} {{Modified
  Gravity Theories on a Nutshell: Inflation, Bounce and Late-time
  Evolution}},}\ }\href {\doibase 10.1016/j.physrep.2017.06.001} {\bibfield
  {journal} {\bibinfo  {journal} {Phys. Rept.}\ }\textbf {\bibinfo {volume}
  {692}},\ \bibinfo {pages} {1--104} (\bibinfo {year} {2017})},\ \Eprint
  {http://arxiv.org/abs/1705.11098} {arXiv:1705.11098 [gr-qc]} \BibitemShut
  {NoStop}%
\bibitem [{\citenamefont {Cardoso}\ \emph {et~al.}(2016)\citenamefont
  {Cardoso}, \citenamefont {Hopper}, \citenamefont {Macedo}, \citenamefont
  {Palenzuela},\ and\ \citenamefont {Pani}}]{Cardoso:2016oxy}%
  \BibitemOpen
  \bibfield  {author} {\bibinfo {author} {\bibfnamefont {Vitor}\ \bibnamefont
  {Cardoso}}, \bibinfo {author} {\bibfnamefont {Seth}\ \bibnamefont {Hopper}},
  \bibinfo {author} {\bibfnamefont {Caio F.~B.}\ \bibnamefont {Macedo}},
  \bibinfo {author} {\bibfnamefont {Carlos}\ \bibnamefont {Palenzuela}}, \ and\
  \bibinfo {author} {\bibfnamefont {Paolo}\ \bibnamefont {Pani}},\ }\bibfield
  {title} {\enquote {\bibinfo {title} {{Gravitational-wave signatures of exotic
  compact objects and of quantum corrections at the horizon scale}},}\ }\href
  {\doibase 10.1103/PhysRevD.94.084031} {\bibfield  {journal} {\bibinfo
  {journal} {Phys. Rev. D}\ }\textbf {\bibinfo {volume} {94}},\ \bibinfo
  {pages} {084031} (\bibinfo {year} {2016})},\ \Eprint
  {http://arxiv.org/abs/1608.08637} {arXiv:1608.08637 [gr-qc]} \BibitemShut
  {NoStop}%
\bibitem [{\citenamefont {Sharif}\ and\ \citenamefont
  {Zubair}(2012)}]{Sharif:2012zzd}%
  \BibitemOpen
  \bibfield  {author} {\bibinfo {author} {\bibfnamefont {M.}~\bibnamefont
  {Sharif}}\ and\ \bibinfo {author} {\bibfnamefont {M.}~\bibnamefont
  {Zubair}},\ }\bibfield  {title} {\enquote {\bibinfo {title} {{Thermodynamics
  in f(R,T) Theory of Gravity}},}\ }\href {\doibase
  10.1088/1475-7516/2012/03/028} {\bibfield  {journal} {\bibinfo  {journal}
  {JCAP}\ }\textbf {\bibinfo {volume} {03}},\ \bibinfo {pages} {028} (\bibinfo
  {year} {2012})},\ \bibinfo {note} {[Erratum: JCAP 05, E01 (2012)]},\ \Eprint
  {http://arxiv.org/abs/1204.0848} {arXiv:1204.0848 [gr-qc]} \BibitemShut
  {NoStop}%
\bibitem [{\citenamefont {Sharif}\ and\ \citenamefont
  {Shafique}(2014)}]{Sharif:2014fla}%
  \BibitemOpen
  \bibfield  {author} {\bibinfo {author} {\bibfnamefont {M.}~\bibnamefont
  {Sharif}}\ and\ \bibinfo {author} {\bibfnamefont {Imrana}\ \bibnamefont
  {Shafique}},\ }\bibfield  {title} {\enquote {\bibinfo {title} {{Noether
  symmetries in a modified scalar-tensor gravity}},}\ }\href {\doibase
  10.1103/PhysRevD.90.084033} {\bibfield  {journal} {\bibinfo  {journal} {Phys.
  Rev. D}\ }\textbf {\bibinfo {volume} {90}},\ \bibinfo {pages} {084033}
  (\bibinfo {year} {2014})}\BibitemShut {NoStop}%
\bibitem [{\citenamefont {Barcelo}\ \emph {et~al.}(2005)\citenamefont
  {Barcelo}, \citenamefont {Liberati},\ and\ \citenamefont
  {Visser}}]{Barcelo:2005fc}%
  \BibitemOpen
  \bibfield  {author} {\bibinfo {author} {\bibfnamefont {Carlos}\ \bibnamefont
  {Barcelo}}, \bibinfo {author} {\bibfnamefont {Stefano}\ \bibnamefont
  {Liberati}}, \ and\ \bibinfo {author} {\bibfnamefont {Matt}\ \bibnamefont
  {Visser}},\ }\bibfield  {title} {\enquote {\bibinfo {title} {{Analogue
  gravity}},}\ }\href {\doibase 10.12942/lrr-2005-12} {\bibfield  {journal}
  {\bibinfo  {journal} {Living Rev. Rel.}\ }\textbf {\bibinfo {volume} {8}},\
  \bibinfo {pages} {12} (\bibinfo {year} {2005})},\ \Eprint
  {http://arxiv.org/abs/gr-qc/0505065} {arXiv:gr-qc/0505065} \BibitemShut
  {NoStop}%
\bibitem [{\citenamefont {Jacobson}\ \emph {et~al.}(2003)\citenamefont
  {Jacobson}, \citenamefont {Liberati},\ and\ \citenamefont
  {Mattingly}}]{Jacobson:2002ye}%
  \BibitemOpen
  \bibfield  {author} {\bibinfo {author} {\bibfnamefont {T.}~\bibnamefont
  {Jacobson}}, \bibinfo {author} {\bibfnamefont {Stefano}\ \bibnamefont
  {Liberati}}, \ and\ \bibinfo {author} {\bibfnamefont {D.}~\bibnamefont
  {Mattingly}},\ }\bibfield  {title} {\enquote {\bibinfo {title} {{A Strong
  astrophysical constraint on the violation of special relativity by quantum
  gravity}},}\ }\href {\doibase 10.1038/nature01882} {\bibfield  {journal}
  {\bibinfo  {journal} {Nature}\ }\textbf {\bibinfo {volume} {424}},\ \bibinfo
  {pages} {1019--1021} (\bibinfo {year} {2003})},\ \Eprint
  {http://arxiv.org/abs/astro-ph/0212190} {arXiv:astro-ph/0212190} \BibitemShut
  {NoStop}%
\bibitem [{\citenamefont {Sakharov}(1967)}]{Sakharov:1967pk}%
  \BibitemOpen
  \bibfield  {author} {\bibinfo {author} {\bibfnamefont {A.~D.}\ \bibnamefont
  {Sakharov}},\ }\bibfield  {title} {\enquote {\bibinfo {title} {{Vacuum
  quantum fluctuations in curved space and the theory of gravitation}},}\
  }\href {\doibase 10.1070/PU1991v034n05ABEH002498} {\bibfield  {journal}
  {\bibinfo  {journal} {Dokl. Akad. Nauk Ser. Fiz.}\ }\textbf {\bibinfo
  {volume} {177}},\ \bibinfo {pages} {70--71} (\bibinfo {year}
  {1967})}\BibitemShut {NoStop}%
\bibitem [{\citenamefont {Verlinde}(2011)}]{Verlinde:2010hp}%
  \BibitemOpen
  \bibfield  {author} {\bibinfo {author} {\bibfnamefont {Erik~P.}\ \bibnamefont
  {Verlinde}},\ }\bibfield  {title} {\enquote {\bibinfo {title} {{On the Origin
  of Gravity and the Laws of Newton}},}\ }\href {\doibase
  10.1007/JHEP04(2011)029} {\bibfield  {journal} {\bibinfo  {journal} {JHEP}\
  }\textbf {\bibinfo {volume} {04}},\ \bibinfo {pages} {029} (\bibinfo {year}
  {2011})},\ \Eprint {http://arxiv.org/abs/1001.0785} {arXiv:1001.0785
  [hep-th]} \BibitemShut {NoStop}%
\bibitem [{\citenamefont {Van~Raamsdonk}(2010)}]{VanRaamsdonk:2010pw}%
  \BibitemOpen
  \bibfield  {author} {\bibinfo {author} {\bibfnamefont {Mark}\ \bibnamefont
  {Van~Raamsdonk}},\ }\bibfield  {title} {\enquote {\bibinfo {title} {{Building
  up spacetime with quantum entanglement}},}\ }\href {\doibase
  10.1142/S0218271810018529} {\bibfield  {journal} {\bibinfo  {journal} {Gen.
  Rel. Grav.}\ }\textbf {\bibinfo {volume} {42}},\ \bibinfo {pages}
  {2323--2329} (\bibinfo {year} {2010})},\ \Eprint
  {http://arxiv.org/abs/1005.3035} {arXiv:1005.3035 [hep-th]} \BibitemShut
  {NoStop}%
\bibitem [{\citenamefont {Liberati}(2017)}]{Liberati:2017jnr}%
  \BibitemOpen
  \bibfield  {author} {\bibinfo {author} {\bibfnamefont {Stefano}\ \bibnamefont
  {Liberati}},\ }\bibfield  {title} {\enquote {\bibinfo {title} {{Analogue
  gravity models of emergent gravity: lessons and pitfalls}},}\ }\href
  {\doibase 10.1088/1742-6596/880/1/012009} {\bibfield  {journal} {\bibinfo
  {journal} {J. Phys. Conf. Ser.}\ }\textbf {\bibinfo {volume} {880}},\
  \bibinfo {pages} {012009} (\bibinfo {year} {2017})}\BibitemShut {NoStop}%
\bibitem [{\citenamefont {Jacobson}(1995)}]{Jacobson:1995ab}%
  \BibitemOpen
  \bibfield  {author} {\bibinfo {author} {\bibfnamefont {Ted}\ \bibnamefont
  {Jacobson}},\ }\bibfield  {title} {\enquote {\bibinfo {title}
  {{Thermodynamics of space-time: The Einstein equation of state}},}\ }\href
  {\doibase 10.1103/PhysRevLett.75.1260} {\bibfield  {journal} {\bibinfo
  {journal} {Phys. Rev. Lett.}\ }\textbf {\bibinfo {volume} {75}},\ \bibinfo
  {pages} {1260--1263} (\bibinfo {year} {1995})},\ \Eprint
  {http://arxiv.org/abs/gr-qc/9504004} {arXiv:gr-qc/9504004} \BibitemShut
  {NoStop}%
\bibitem [{\citenamefont {Padmanabhan}(2010)}]{Padmanabhan:2009vy}%
  \BibitemOpen
  \bibfield  {author} {\bibinfo {author} {\bibfnamefont {T.}~\bibnamefont
  {Padmanabhan}},\ }\bibfield  {title} {\enquote {\bibinfo {title}
  {{Thermodynamical Aspects of Gravity: New insights}},}\ }\href {\doibase
  10.1088/0034-4885/73/4/046901} {\bibfield  {journal} {\bibinfo  {journal}
  {Rept. Prog. Phys.}\ }\textbf {\bibinfo {volume} {73}},\ \bibinfo {pages}
  {046901} (\bibinfo {year} {2010})},\ \Eprint {http://arxiv.org/abs/0911.5004}
  {arXiv:0911.5004 [gr-qc]} \BibitemShut {NoStop}%
\bibitem [{\citenamefont {Visser}(2002)}]{Visser:2002ew}%
  \BibitemOpen
  \bibfield  {author} {\bibinfo {author} {\bibfnamefont {Matt}\ \bibnamefont
  {Visser}},\ }\bibfield  {title} {\enquote {\bibinfo {title} {{Sakharov's
  induced gravity: A Modern perspective}},}\ }\href {\doibase
  10.1142/S0217732302006886} {\bibfield  {journal} {\bibinfo  {journal} {Mod.
  Phys. Lett. A}\ }\textbf {\bibinfo {volume} {17}},\ \bibinfo {pages}
  {977--992} (\bibinfo {year} {2002})},\ \Eprint
  {http://arxiv.org/abs/gr-qc/0204062} {arXiv:gr-qc/0204062} \BibitemShut
  {NoStop}%
\bibitem [{\citenamefont {Nashed}(2018{\natexlab{a}})}]{Nashed:2018efg}%
  \BibitemOpen
  \bibfield  {author} {\bibinfo {author} {\bibfnamefont {G.~G.~L.}\
  \bibnamefont {Nashed}},\ }\bibfield  {title} {\enquote {\bibinfo {title}
  {{Rotating charged black hole spacetimes in quadratic f(R) gravitational
  theories}},}\ }\href {\doibase 10.1142/S0218271818500748} {\bibfield
  {journal} {\bibinfo  {journal} {Int. J. Mod. Phys. D}\ }\textbf {\bibinfo
  {volume} {27}},\ \bibinfo {pages} {1850074} (\bibinfo {year}
  {2018}{\natexlab{a}})}\BibitemShut {NoStop}%
\bibitem [{\citenamefont {Nashed}\ and\ \citenamefont
  {Capozziello}(2019)}]{Nashed:2019tuk}%
  \BibitemOpen
  \bibfield  {author} {\bibinfo {author} {\bibfnamefont {Gamal G.~L.}\
  \bibnamefont {Nashed}}\ and\ \bibinfo {author} {\bibfnamefont {Salvatore}\
  \bibnamefont {Capozziello}},\ }\bibfield  {title} {\enquote {\bibinfo {title}
  {{Charged spherically symmetric black holes in $f(R)$ gravity and their
  stability analysis}},}\ }\href {\doibase 10.1103/PhysRevD.99.104018}
  {\bibfield  {journal} {\bibinfo  {journal} {Phys. Rev. D}\ }\textbf {\bibinfo
  {volume} {99}},\ \bibinfo {pages} {104018} (\bibinfo {year} {2019})},\
  \Eprint {http://arxiv.org/abs/1902.06783} {arXiv:1902.06783 [gr-qc]}
  \BibitemShut {NoStop}%
\bibitem [{\citenamefont {Nashed}(2018{\natexlab{b}})}]{Nashed:2018oaf}%
  \BibitemOpen
  \bibfield  {author} {\bibinfo {author} {\bibfnamefont {G.~G.~L.}\
  \bibnamefont {Nashed}},\ }\bibfield  {title} {\enquote {\bibinfo {title}
  {{Spherically symmetric charged black holes in f(R) gravitational
  theories}},}\ }\href {\doibase 10.1140/epjp/i2018-11849-7} {\bibfield
  {journal} {\bibinfo  {journal} {Eur. Phys. J. Plus}\ }\textbf {\bibinfo
  {volume} {133}},\ \bibinfo {pages} {18} (\bibinfo {year}
  {2018}{\natexlab{b}})}\BibitemShut {NoStop}%
\bibitem [{\citenamefont {Demir}(2021)}]{demir1}%
  \BibitemOpen
  \bibfield  {author} {\bibinfo {author} {\bibfnamefont {Durmus}\ \bibnamefont
  {Demir}},\ }\bibfield  {title} {\enquote {\bibinfo {title} {{Emergent Gravity
  as the Eraser of Anomalous Gauge Boson Masses, and QFT-GR Concord}},}\ }\href
  {\doibase 10.1007/s10714-021-02797-0} {\bibfield  {journal} {\bibinfo
  {journal} {Gen. Rel. Grav.}\ }\textbf {\bibinfo {volume} {53}},\ \bibinfo
  {pages} {22} (\bibinfo {year} {2021})},\ \Eprint
  {http://arxiv.org/abs/2101.12391} {arXiv:2101.12391 [gr-qc]} \BibitemShut
  {NoStop}%
\bibitem [{\citenamefont {Demir}(2019)}]{demir2}%
  \BibitemOpen
  \bibfield  {author} {\bibinfo {author} {\bibfnamefont {Durmus}\ \bibnamefont
  {Demir}},\ }\bibfield  {title} {\enquote {\bibinfo {title} {{Symmergent
  Gravity, Seesawic New Physics, and their Experimental Signatures}},}\ }\href
  {\doibase 10.1155/2019/4652048} {\bibfield  {journal} {\bibinfo  {journal}
  {Adv. High Energy Phys.}\ }\textbf {\bibinfo {volume} {2019}},\ \bibinfo
  {pages} {4652048} (\bibinfo {year} {2019})},\ \Eprint
  {http://arxiv.org/abs/1901.07244} {arXiv:1901.07244 [hep-ph]} \BibitemShut
  {NoStop}%
\bibitem [{\citenamefont {Demir}(2016)}]{demir3}%
  \BibitemOpen
  \bibfield  {author} {\bibinfo {author} {\bibfnamefont {Durmus~Ali}\
  \bibnamefont {Demir}},\ }\bibfield  {title} {\enquote {\bibinfo {title}
  {{Curvature-Restored Gauge Invariance and Ultraviolet Naturalness}},}\ }\href
  {\doibase 10.1155/2016/6727805} {\bibfield  {journal} {\bibinfo  {journal}
  {Adv. High Energy Phys.}\ }\textbf {\bibinfo {volume} {2016}},\ \bibinfo
  {pages} {6727805} (\bibinfo {year} {2016})},\ \Eprint
  {http://arxiv.org/abs/1605.00377} {arXiv:1605.00377 [hep-ph]} \BibitemShut
  {NoStop}%
\bibitem [{\citenamefont {\c{C}imdiker}(2020)}]{irfan}%
  \BibitemOpen
  \bibfield  {author} {\bibinfo {author} {\bibfnamefont {\.Ilim~\.Irfan}\
  \bibnamefont {\c{C}imdiker}},\ }\bibfield  {title} {\enquote {\bibinfo
  {title} {{Starobinsky inflation in emergent gravity}},}\ }\href {\doibase
  10.1016/j.dark.2020.100736} {\bibfield  {journal} {\bibinfo  {journal} {Phys.
  Dark Univ.}\ }\textbf {\bibinfo {volume} {30}},\ \bibinfo {pages} {100736}
  (\bibinfo {year} {2020})}\BibitemShut {NoStop}%
\bibitem [{\citenamefont {\c{C}imdiker}\ \emph
  {et~al.}(2021{\natexlab{a}})\citenamefont {\c{C}imdiker}, \citenamefont
  {Demir},\ and\ \citenamefont {\"Ovg\"un}}]{Symmergent-bh}%
  \BibitemOpen
  \bibfield  {author} {\bibinfo {author} {\bibfnamefont {\.Irfan}\ \bibnamefont
  {\c{C}imdiker}}, \bibinfo {author} {\bibfnamefont {Durmu\c{s}}\ \bibnamefont
  {Demir}}, \ and\ \bibinfo {author} {\bibfnamefont {Ali}\ \bibnamefont
  {\"Ovg\"un}},\ }\bibfield  {title} {\enquote {\bibinfo {title} {{Black hole
  shadow in symmergent gravity}},}\ }\href {\doibase
  10.1016/j.dark.2021.100900} {\bibfield  {journal} {\bibinfo  {journal} {Phys.
  Dark Univ.}\ }\textbf {\bibinfo {volume} {34}},\ \bibinfo {pages} {100900}
  (\bibinfo {year} {2021}{\natexlab{a}})},\ \Eprint
  {http://arxiv.org/abs/2110.11904} {arXiv:2110.11904 [gr-qc]} \BibitemShut
  {NoStop}%
\bibitem [{\citenamefont {Rayimbaev}\ \emph {et~al.}(2023)\citenamefont
  {Rayimbaev}, \citenamefont {Pantig}, \citenamefont {\"Ovg\"un}, \citenamefont
  {Abdujabbarov},\ and\ \citenamefont {Demir}}]{Symmergent-bh2}%
  \BibitemOpen
  \bibfield  {author} {\bibinfo {author} {\bibfnamefont {Javlon}\ \bibnamefont
  {Rayimbaev}}, \bibinfo {author} {\bibfnamefont {Reggie~C.}\ \bibnamefont
  {Pantig}}, \bibinfo {author} {\bibfnamefont {Ali}\ \bibnamefont {\"Ovg\"un}},
  \bibinfo {author} {\bibfnamefont {Ahmadjon}\ \bibnamefont {Abdujabbarov}}, \
  and\ \bibinfo {author} {\bibfnamefont {Durmu\c{s}}\ \bibnamefont {Demir}},\
  }\bibfield  {title} {\enquote {\bibinfo {title} {{Quasiperiodic oscillations,
  weak field lensing and shadow cast around black holes in Symmergent
  gravity}},}\ }\href {\doibase 10.1016/j.aop.2023.169335} {\bibfield
  {journal} {\bibinfo  {journal} {Annals Phys.}\ }\textbf {\bibinfo {volume}
  {454}},\ \bibinfo {pages} {169335} (\bibinfo {year} {2023})},\ \Eprint
  {http://arxiv.org/abs/2206.06599} {arXiv:2206.06599 [gr-qc]} \BibitemShut
  {NoStop}%
\bibitem [{\citenamefont {Ali}\ \emph {et~al.}(2023)\citenamefont {Ali},
  \citenamefont {Babar}, \citenamefont {Akhtar},\ and\ \citenamefont
  {\"Ovg\"un}}]{Symmergentresults}%
  \BibitemOpen
  \bibfield  {author} {\bibinfo {author} {\bibfnamefont {Riasat}\ \bibnamefont
  {Ali}}, \bibinfo {author} {\bibfnamefont {Rimsha}\ \bibnamefont {Babar}},
  \bibinfo {author} {\bibfnamefont {Zunaira}\ \bibnamefont {Akhtar}}, \ and\
  \bibinfo {author} {\bibfnamefont {Ali}\ \bibnamefont {\"Ovg\"un}},\
  }\bibfield  {title} {\enquote {\bibinfo {title} {{Thermodynamics and
  logarithmic corrections of symmergent black holes}},}\ }\href {\doibase
  10.1016/j.rinp.2023.106300} {\bibfield  {journal} {\bibinfo  {journal}
  {Results Phys.}\ }\textbf {\bibinfo {volume} {46}},\ \bibinfo {pages}
  {106300} (\bibinfo {year} {2023})},\ \Eprint
  {http://arxiv.org/abs/2302.12875} {arXiv:2302.12875 [gr-qc]} \BibitemShut
  {NoStop}%
\bibitem [{\citenamefont {Chamblin}\ \emph {et~al.}(1999)\citenamefont
  {Chamblin}, \citenamefont {Emparan}, \citenamefont {Johnson},\ and\
  \citenamefont {Myers}}]{Chamblin:1999tk}%
  \BibitemOpen
  \bibfield  {author} {\bibinfo {author} {\bibfnamefont {Andrew}\ \bibnamefont
  {Chamblin}}, \bibinfo {author} {\bibfnamefont {Roberto}\ \bibnamefont
  {Emparan}}, \bibinfo {author} {\bibfnamefont {Clifford~V.}\ \bibnamefont
  {Johnson}}, \ and\ \bibinfo {author} {\bibfnamefont {Robert~C.}\ \bibnamefont
  {Myers}},\ }\bibfield  {title} {\enquote {\bibinfo {title} {{Charged AdS
  black holes and catastrophic holography}},}\ }\href {\doibase
  10.1103/PhysRevD.60.064018} {\bibfield  {journal} {\bibinfo  {journal} {Phys.
  Rev. D}\ }\textbf {\bibinfo {volume} {60}},\ \bibinfo {pages} {064018}
  (\bibinfo {year} {1999})},\ \Eprint {http://arxiv.org/abs/hep-th/9902170}
  {arXiv:hep-th/9902170} \BibitemShut {NoStop}%
\bibitem [{\citenamefont {Kubiznak}\ and\ \citenamefont
  {Mann}(2012)}]{Kubiznak:2012wp}%
  \BibitemOpen
  \bibfield  {author} {\bibinfo {author} {\bibfnamefont {David}\ \bibnamefont
  {Kubiznak}}\ and\ \bibinfo {author} {\bibfnamefont {Robert~B.}\ \bibnamefont
  {Mann}},\ }\bibfield  {title} {\enquote {\bibinfo {title} {{P-V criticality
  of charged AdS black holes}},}\ }\href {\doibase 10.1007/JHEP07(2012)033}
  {\bibfield  {journal} {\bibinfo  {journal} {JHEP}\ }\textbf {\bibinfo
  {volume} {07}},\ \bibinfo {pages} {033} (\bibinfo {year} {2012})},\ \Eprint
  {http://arxiv.org/abs/1205.0559} {arXiv:1205.0559 [hep-th]} \BibitemShut
  {NoStop}%
\bibitem [{\citenamefont {Gregory}\ and\ \citenamefont
  {Laflamme}(1994)}]{Gregory:1994bj}%
  \BibitemOpen
  \bibfield  {author} {\bibinfo {author} {\bibfnamefont {Ruth}\ \bibnamefont
  {Gregory}}\ and\ \bibinfo {author} {\bibfnamefont {Raymond}\ \bibnamefont
  {Laflamme}},\ }\bibfield  {title} {\enquote {\bibinfo {title} {{The
  Instability of charged black strings and p-branes}},}\ }\href {\doibase
  10.1016/0550-3213(94)90206-2} {\bibfield  {journal} {\bibinfo  {journal}
  {Nucl. Phys. B}\ }\textbf {\bibinfo {volume} {428}},\ \bibinfo {pages}
  {399--434} (\bibinfo {year} {1994})},\ \Eprint
  {http://arxiv.org/abs/hep-th/9404071} {arXiv:hep-th/9404071} \BibitemShut
  {NoStop}%
\bibitem [{\citenamefont {Graham}\ and\ \citenamefont
  {Jha}(2014)}]{Graham:2014mda}%
  \BibitemOpen
  \bibfield  {author} {\bibinfo {author} {\bibfnamefont {Alexander A.~H.}\
  \bibnamefont {Graham}}\ and\ \bibinfo {author} {\bibfnamefont {Rahul}\
  \bibnamefont {Jha}},\ }\bibfield  {title} {\enquote {\bibinfo {title}
  {{Nonexistence of black holes with noncanonical scalar fields}},}\ }\href
  {\doibase 10.1103/PhysRevD.89.084056} {\bibfield  {journal} {\bibinfo
  {journal} {Phys. Rev. D}\ }\textbf {\bibinfo {volume} {89}},\ \bibinfo
  {pages} {084056} (\bibinfo {year} {2014})},\ \bibinfo {note} {[Erratum:
  Phys.Rev.D 92, 069901 (2015)]},\ \Eprint {http://arxiv.org/abs/1401.8203}
  {arXiv:1401.8203 [gr-qc]} \BibitemShut {NoStop}%
\bibitem [{\citenamefont {Luminet}(1979)}]{Luminet:1979nyg}%
  \BibitemOpen
  \bibfield  {author} {\bibinfo {author} {\bibfnamefont {J.~P.}\ \bibnamefont
  {Luminet}},\ }\bibfield  {title} {\enquote {\bibinfo {title} {{Image of a
  spherical black hole with thin accretion disk}},}\ }\href
  {https://adsabs.harvard.edu/full/1979A%26A....75..228L} {\bibfield  {journal}
  {\bibinfo  {journal} {Astron. Astrophys.}\ }\textbf {\bibinfo {volume}
  {75}},\ \bibinfo {pages} {228--235} (\bibinfo {year} {1979})}\BibitemShut
  {NoStop}%
\bibitem [{\citenamefont {Synge}(1966)}]{Synge:1966okc}%
  \BibitemOpen
  \bibfield  {author} {\bibinfo {author} {\bibfnamefont {J.~L.}\ \bibnamefont
  {Synge}},\ }\bibfield  {title} {\enquote {\bibinfo {title} {{The Escape of
  Photons from Gravitationally Intense Stars}},}\ }\href {\doibase
  10.1093/mnras/131.3.463} {\bibfield  {journal} {\bibinfo  {journal} {Mon.
  Not. Roy. Astron. Soc.}\ }\textbf {\bibinfo {volume} {131}},\ \bibinfo
  {pages} {463--466} (\bibinfo {year} {1966})}\BibitemShut {NoStop}%
\bibitem [{\citenamefont {Akiyama}\ \emph {et~al.}(2019)\citenamefont {Akiyama}
  \emph {et~al.}}]{EventHorizonTelescope:2019dse}%
  \BibitemOpen
  \bibfield  {author} {\bibinfo {author} {\bibfnamefont {Kazunori}\
  \bibnamefont {Akiyama}} \emph {et~al.} (\bibinfo {collaboration} {Event
  Horizon Telescope}),\ }\bibfield  {title} {\enquote {\bibinfo {title} {{First
  M87 Event Horizon Telescope Results. I. The Shadow of the Supermassive Black
  Hole}},}\ }\href {\doibase 10.3847/2041-8213/ab0ec7} {\bibfield  {journal}
  {\bibinfo  {journal} {Astrophys. J. Lett.}\ }\textbf {\bibinfo {volume}
  {875}},\ \bibinfo {pages} {L1} (\bibinfo {year} {2019})},\ \Eprint
  {http://arxiv.org/abs/1906.11238} {arXiv:1906.11238 [astro-ph.GA]}
  \BibitemShut {NoStop}%
\bibitem [{\citenamefont {Akiyama}\ \emph {et~al.}(2022)\citenamefont {Akiyama}
  \emph {et~al.}}]{EventHorizonTelescope:2022xnr}%
  \BibitemOpen
  \bibfield  {author} {\bibinfo {author} {\bibfnamefont {Kazunori}\
  \bibnamefont {Akiyama}} \emph {et~al.} (\bibinfo {collaboration} {Event
  Horizon Telescope}),\ }\bibfield  {title} {\enquote {\bibinfo {title} {{First
  Sagittarius A* Event Horizon Telescope Results. I. The Shadow of the
  Supermassive Black Hole in the Center of the Milky Way}},}\ }\href {\doibase
  10.3847/2041-8213/ac6674} {\bibfield  {journal} {\bibinfo  {journal}
  {Astrophys. J. Lett.}\ }\textbf {\bibinfo {volume} {930}},\ \bibinfo {pages}
  {L12} (\bibinfo {year} {2022})}\BibitemShut {NoStop}%
\bibitem [{\citenamefont {Contreras}\ \emph {et~al.}(2021)\citenamefont
  {Contreras}, \citenamefont {Rinc\'on}, \citenamefont {Panotopoulos},\ and\
  \citenamefont {Bargue\~no}}]{Contreras:2020kgy}%
  \BibitemOpen
  \bibfield  {author} {\bibinfo {author} {\bibfnamefont {E.}~\bibnamefont
  {Contreras}}, \bibinfo {author} {\bibfnamefont {\'Angel}\ \bibnamefont
  {Rinc\'on}}, \bibinfo {author} {\bibfnamefont {Grigoris}\ \bibnamefont
  {Panotopoulos}}, \ and\ \bibinfo {author} {\bibfnamefont {Pedro}\
  \bibnamefont {Bargue\~no}},\ }\bibfield  {title} {\enquote {\bibinfo {title}
  {{Geodesic analysis and black hole shadows on a general non-extremal rotating
  black hole in five-dimensional gauged supergravity}},}\ }\href {\doibase
  10.1016/j.aop.2021.168567} {\bibfield  {journal} {\bibinfo  {journal} {Annals
  Phys.}\ }\textbf {\bibinfo {volume} {432}},\ \bibinfo {pages} {168567}
  (\bibinfo {year} {2021})},\ \Eprint {http://arxiv.org/abs/2010.03734}
  {arXiv:2010.03734 [gr-qc]} \BibitemShut {NoStop}%
\bibitem [{\citenamefont {Panotopoulos}\ \emph {et~al.}(2021)\citenamefont
  {Panotopoulos}, \citenamefont {Rinc\'on},\ and\ \citenamefont
  {Lopes}}]{Panotopoulos:2021tkk}%
  \BibitemOpen
  \bibfield  {author} {\bibinfo {author} {\bibfnamefont {Grigoris}\
  \bibnamefont {Panotopoulos}}, \bibinfo {author} {\bibfnamefont {\'Angel}\
  \bibnamefont {Rinc\'on}}, \ and\ \bibinfo {author} {\bibfnamefont {Ilidio}\
  \bibnamefont {Lopes}},\ }\bibfield  {title} {\enquote {\bibinfo {title}
  {{Orbits of light rays in scale-dependent gravity: Exact analytical solutions
  to the null geodesic equations}},}\ }\href {\doibase
  10.1103/PhysRevD.103.104040} {\bibfield  {journal} {\bibinfo  {journal}
  {Phys. Rev. D}\ }\textbf {\bibinfo {volume} {103}},\ \bibinfo {pages}
  {104040} (\bibinfo {year} {2021})},\ \Eprint
  {http://arxiv.org/abs/2104.13611} {arXiv:2104.13611 [gr-qc]} \BibitemShut
  {NoStop}%
\bibitem [{\citenamefont {Panotopoulos}\ and\ \citenamefont
  {Rincon}(2022)}]{Panotopoulos:2022bky}%
  \BibitemOpen
  \bibfield  {author} {\bibinfo {author} {\bibfnamefont {Grigoris}\
  \bibnamefont {Panotopoulos}}\ and\ \bibinfo {author} {\bibfnamefont {Angel}\
  \bibnamefont {Rincon}},\ }\bibfield  {title} {\enquote {\bibinfo {title}
  {{Orbits of light rays in (1＋2)-dimensional
  Einstein\textendash{}power\textendash{}Maxwell gravity: Exact analytical
  solution to the null geodesic equations}},}\ }\href {\doibase
  10.1016/j.aop.2022.168947} {\bibfield  {journal} {\bibinfo  {journal} {Annals
  Phys.}\ }\textbf {\bibinfo {volume} {443}},\ \bibinfo {pages} {168947}
  (\bibinfo {year} {2022})},\ \Eprint {http://arxiv.org/abs/2206.03437}
  {arXiv:2206.03437 [gr-qc]} \BibitemShut {NoStop}%
\bibitem [{\citenamefont {Pantig}\ \emph {et~al.}(2022)\citenamefont {Pantig},
  \citenamefont {Mastrototaro}, \citenamefont {Lambiase},\ and\ \citenamefont
  {\"Ovg\"un}}]{Pantig:2022gih}%
  \BibitemOpen
  \bibfield  {author} {\bibinfo {author} {\bibfnamefont {Reggie~C.}\
  \bibnamefont {Pantig}}, \bibinfo {author} {\bibfnamefont {Leonardo}\
  \bibnamefont {Mastrototaro}}, \bibinfo {author} {\bibfnamefont {Gaetano}\
  \bibnamefont {Lambiase}}, \ and\ \bibinfo {author} {\bibfnamefont {Ali}\
  \bibnamefont {\"Ovg\"un}},\ }\bibfield  {title} {\enquote {\bibinfo {title}
  {{Shadow, lensing, quasinormal modes, greybody bounds and neutrino
  propagation by dyonic ModMax black holes}},}\ }\href {\doibase
  10.1140/epjc/s10052-022-11125-y} {\bibfield  {journal} {\bibinfo  {journal}
  {Eur. Phys. J. C}\ }\textbf {\bibinfo {volume} {82}},\ \bibinfo {pages}
  {1155} (\bibinfo {year} {2022})},\ \Eprint {http://arxiv.org/abs/2208.06664}
  {arXiv:2208.06664 [gr-qc]} \BibitemShut {NoStop}%
\bibitem [{\citenamefont {\"Ovg\"un}\ \emph {et~al.}(2020)\citenamefont
  {\"Ovg\"un}, \citenamefont {Sakall\i{}}, \citenamefont {Saavedra},\ and\
  \citenamefont {Leiva}}]{Ovgun:2019jdo}%
  \BibitemOpen
  \bibfield  {author} {\bibinfo {author} {\bibfnamefont {Ali}\ \bibnamefont
  {\"Ovg\"un}}, \bibinfo {author} {\bibfnamefont {\.Izzet}\ \bibnamefont
  {Sakall\i{}}}, \bibinfo {author} {\bibfnamefont {Joel}\ \bibnamefont
  {Saavedra}}, \ and\ \bibinfo {author} {\bibfnamefont {Carlos}\ \bibnamefont
  {Leiva}},\ }\bibfield  {title} {\enquote {\bibinfo {title} {{Shadow cast of
  noncommutative black holes in Rastall gravity}},}\ }\href {\doibase
  10.1142/S0217732320501631} {\bibfield  {journal} {\bibinfo  {journal} {Mod.
  Phys. Lett. A}\ }\textbf {\bibinfo {volume} {35}},\ \bibinfo {pages}
  {2050163} (\bibinfo {year} {2020})},\ \Eprint
  {http://arxiv.org/abs/1906.05954} {arXiv:1906.05954 [hep-th]} \BibitemShut
  {NoStop}%
\bibitem [{\citenamefont {\"Ovg\"un}\ and\ \citenamefont
  {Sakall\i{}}(2020)}]{Ovgun:2020gjz}%
  \BibitemOpen
  \bibfield  {author} {\bibinfo {author} {\bibfnamefont {Ali}\ \bibnamefont
  {\"Ovg\"un}}\ and\ \bibinfo {author} {\bibfnamefont {\.Izzet}\ \bibnamefont
  {Sakall\i{}}},\ }\bibfield  {title} {\enquote {\bibinfo {title} {{Testing
  generalized Einstein\textendash{}Cartan\textendash{}Kibble\textendash{}Sciama
  gravity using weak deflection angle and shadow cast}},}\ }\href {\doibase
  10.1088/1361-6382/abb579} {\bibfield  {journal} {\bibinfo  {journal} {Class.
  Quant. Grav.}\ }\textbf {\bibinfo {volume} {37}},\ \bibinfo {pages} {225003}
  (\bibinfo {year} {2020})},\ \Eprint {http://arxiv.org/abs/2005.00982}
  {arXiv:2005.00982 [gr-qc]} \BibitemShut {NoStop}%
\bibitem [{\citenamefont {Okyay}\ and\ \citenamefont
  {\"Ovg\"un}(2022)}]{Okyay:2021nnh}%
  \BibitemOpen
  \bibfield  {author} {\bibinfo {author} {\bibfnamefont {Mert}\ \bibnamefont
  {Okyay}}\ and\ \bibinfo {author} {\bibfnamefont {Ali}\ \bibnamefont
  {\"Ovg\"un}},\ }\bibfield  {title} {\enquote {\bibinfo {title} {{Nonlinear
  electrodynamics effects on the black hole shadow, deflection angle,
  quasinormal modes and greybody factors}},}\ }\href {\doibase
  10.1088/1475-7516/2022/01/009} {\bibfield  {journal} {\bibinfo  {journal}
  {JCAP}\ }\textbf {\bibinfo {volume} {01}},\ \bibinfo {pages} {009} (\bibinfo
  {year} {2022})},\ \Eprint {http://arxiv.org/abs/2108.07766} {arXiv:2108.07766
  [gr-qc]} \BibitemShut {NoStop}%
\bibitem [{\citenamefont {Javed}\ \emph {et~al.}(2021)\citenamefont {Javed},
  \citenamefont {Hamza},\ and\ \citenamefont {\"Ovg\"un}}]{Javed:2021arr}%
  \BibitemOpen
  \bibfield  {author} {\bibinfo {author} {\bibfnamefont {Wajiha}\ \bibnamefont
  {Javed}}, \bibinfo {author} {\bibfnamefont {Ali}\ \bibnamefont {Hamza}}, \
  and\ \bibinfo {author} {\bibfnamefont {Ali}\ \bibnamefont {\"Ovg\"un}},\
  }\bibfield  {title} {\enquote {\bibinfo {title} {{Weak Deflection Angle and
  Shadow by Tidal Charged Black Hole}},}\ }\href {\doibase
  10.3390/universe7100385} {\bibfield  {journal} {\bibinfo  {journal}
  {Universe}\ }\textbf {\bibinfo {volume} {7}},\ \bibinfo {pages} {385}
  (\bibinfo {year} {2021})},\ \Eprint {http://arxiv.org/abs/2110.11397}
  {arXiv:2110.11397 [gr-qc]} \BibitemShut {NoStop}%
\bibitem [{\citenamefont {\c{C}imdiker}\ \emph
  {et~al.}(2021{\natexlab{b}})\citenamefont {\c{C}imdiker}, \citenamefont
  {Demir},\ and\ \citenamefont {\"Ovg\"un}}]{Cimdiker:2021cpz}%
  \BibitemOpen
  \bibfield  {author} {\bibinfo {author} {\bibfnamefont {\.Irfan}\ \bibnamefont
  {\c{C}imdiker}}, \bibinfo {author} {\bibfnamefont {Durmu\c{s}}\ \bibnamefont
  {Demir}}, \ and\ \bibinfo {author} {\bibfnamefont {Ali}\ \bibnamefont
  {\"Ovg\"un}},\ }\bibfield  {title} {\enquote {\bibinfo {title} {{Black hole
  shadow in symmergent gravity}},}\ }\href {\doibase
  10.1016/j.dark.2021.100900} {\bibfield  {journal} {\bibinfo  {journal} {Phys.
  Dark Univ.}\ }\textbf {\bibinfo {volume} {34}},\ \bibinfo {pages} {100900}
  (\bibinfo {year} {2021}{\natexlab{b}})},\ \Eprint
  {http://arxiv.org/abs/2110.11904} {arXiv:2110.11904 [gr-qc]} \BibitemShut
  {NoStop}%
\bibitem [{\citenamefont {Uniyal}\ \emph {et~al.}(2023)\citenamefont {Uniyal},
  \citenamefont {Pantig},\ and\ \citenamefont {\"Ovg\"un}}]{Uniyal:2022vdu}%
  \BibitemOpen
  \bibfield  {author} {\bibinfo {author} {\bibfnamefont {Akhil}\ \bibnamefont
  {Uniyal}}, \bibinfo {author} {\bibfnamefont {Reggie~C.}\ \bibnamefont
  {Pantig}}, \ and\ \bibinfo {author} {\bibfnamefont {Ali}\ \bibnamefont
  {\"Ovg\"un}},\ }\bibfield  {title} {\enquote {\bibinfo {title} {{Probing a
  non-linear electrodynamics black hole with thin accretion disk, shadow, and
  deflection angle with M87* and Sgr A* from EHT}},}\ }\href {\doibase
  10.1016/j.dark.2023.101178} {\bibfield  {journal} {\bibinfo  {journal} {Phys.
  Dark Univ.}\ }\textbf {\bibinfo {volume} {40}},\ \bibinfo {pages} {101178}
  (\bibinfo {year} {2023})},\ \Eprint {http://arxiv.org/abs/2205.11072}
  {arXiv:2205.11072 [gr-qc]} \BibitemShut {NoStop}%
\bibitem [{\citenamefont {Pantig}\ and\ \citenamefont
  {\"Ovg\"un}(2023)}]{Pantig:2022ely}%
  \BibitemOpen
  \bibfield  {author} {\bibinfo {author} {\bibfnamefont {Reggie~C.}\
  \bibnamefont {Pantig}}\ and\ \bibinfo {author} {\bibfnamefont {Ali}\
  \bibnamefont {\"Ovg\"un}},\ }\bibfield  {title} {\enquote {\bibinfo {title}
  {{Testing dynamical torsion effects on the charged black
  hole\textquoteright{}s shadow, deflection angle and greybody with M87* and
  Sgr. A* from EHT}},}\ }\href {\doibase 10.1016/j.aop.2022.169197} {\bibfield
  {journal} {\bibinfo  {journal} {Annals Phys.}\ }\textbf {\bibinfo {volume}
  {448}},\ \bibinfo {pages} {169197} (\bibinfo {year} {2023})},\ \Eprint
  {http://arxiv.org/abs/2206.02161} {arXiv:2206.02161 [gr-qc]} \BibitemShut
  {NoStop}%
\bibitem [{\citenamefont {Mustafa}\ \emph {et~al.}(2022)\citenamefont
  {Mustafa}, \citenamefont {Atamurotov}, \citenamefont {Hussain}, \citenamefont
  {Shaymatov},\ and\ \citenamefont {\"Ovg\"un}}]{Mustafa:2022xod}%
  \BibitemOpen
  \bibfield  {author} {\bibinfo {author} {\bibfnamefont {Ghulam}\ \bibnamefont
  {Mustafa}}, \bibinfo {author} {\bibfnamefont {Farruh}\ \bibnamefont
  {Atamurotov}}, \bibinfo {author} {\bibfnamefont {Ibrar}\ \bibnamefont
  {Hussain}}, \bibinfo {author} {\bibfnamefont {Sanjar}\ \bibnamefont
  {Shaymatov}}, \ and\ \bibinfo {author} {\bibfnamefont {Ali}\ \bibnamefont
  {\"Ovg\"un}},\ }\bibfield  {title} {\enquote {\bibinfo {title} {{Shadows and
  gravitational weak lensing by the Schwarzschild black hole in the string
  cloud background with quintessential field*}},}\ }\href {\doibase
  10.1088/1674-1137/ac917f} {\bibfield  {journal} {\bibinfo  {journal} {Chin.
  Phys. C}\ }\textbf {\bibinfo {volume} {46}},\ \bibinfo {pages} {125107}
  (\bibinfo {year} {2022})},\ \Eprint {http://arxiv.org/abs/2207.07608}
  {arXiv:2207.07608 [gr-qc]} \BibitemShut {NoStop}%
\bibitem [{\citenamefont {Pantig}\ \emph {et~al.}(2023)\citenamefont {Pantig},
  \citenamefont {\"Ovg\"un},\ and\ \citenamefont {Demir}}]{Pantig:2022qak}%
  \BibitemOpen
  \bibfield  {author} {\bibinfo {author} {\bibfnamefont {Reggie~C.}\
  \bibnamefont {Pantig}}, \bibinfo {author} {\bibfnamefont {Ali}\ \bibnamefont
  {\"Ovg\"un}}, \ and\ \bibinfo {author} {\bibfnamefont {Durmu\c{s}}\
  \bibnamefont {Demir}},\ }\bibfield  {title} {\enquote {\bibinfo {title}
  {{Testing symmergent gravity through the shadow image and weak field photon
  deflection by a rotating black hole using the M87$^*$ and Sgr. $\hbox {A}^*$
  results}},}\ }\href {\doibase 10.1140/epjc/s10052-023-11400-6} {\bibfield
  {journal} {\bibinfo  {journal} {Eur. Phys. J. C}\ }\textbf {\bibinfo {volume}
  {83}},\ \bibinfo {pages} {250} (\bibinfo {year} {2023})},\ \Eprint
  {http://arxiv.org/abs/2208.02969} {arXiv:2208.02969 [gr-qc]} \BibitemShut
  {NoStop}%
\bibitem [{\citenamefont {Kumaran}\ and\ \citenamefont
  {\"Ovg\"un}(2022)}]{Kumaran:2022soh}%
  \BibitemOpen
  \bibfield  {author} {\bibinfo {author} {\bibfnamefont {Yashmitha}\
  \bibnamefont {Kumaran}}\ and\ \bibinfo {author} {\bibfnamefont {Ali}\
  \bibnamefont {\"Ovg\"un}},\ }\bibfield  {title} {\enquote {\bibinfo {title}
  {{Deflection Angle and Shadow of the Reissner\textendash{}Nordstr\"om Black
  Hole with Higher-Order Magnetic Correction in Einstein-Nonlinear-Maxwell
  Fields}},}\ }\href {\doibase 10.3390/sym14102054} {\bibfield  {journal}
  {\bibinfo  {journal} {Symmetry}\ }\textbf {\bibinfo {volume} {14}},\ \bibinfo
  {pages} {2054} (\bibinfo {year} {2022})},\ \Eprint
  {http://arxiv.org/abs/2210.00468} {arXiv:2210.00468 [gr-qc]} \BibitemShut
  {NoStop}%
\bibitem [{\citenamefont {Atamurotov}\ \emph {et~al.}(2023)\citenamefont
  {Atamurotov}, \citenamefont {Hussain}, \citenamefont {Mustafa},\ and\
  \citenamefont {\"Ovg\"un}}]{Atamurotov:2022knb}%
  \BibitemOpen
  \bibfield  {author} {\bibinfo {author} {\bibfnamefont {Farruh}\ \bibnamefont
  {Atamurotov}}, \bibinfo {author} {\bibfnamefont {Ibrar}\ \bibnamefont
  {Hussain}}, \bibinfo {author} {\bibfnamefont {Ghulam}\ \bibnamefont
  {Mustafa}}, \ and\ \bibinfo {author} {\bibfnamefont {Ali}\ \bibnamefont
  {\"Ovg\"un}},\ }\bibfield  {title} {\enquote {\bibinfo {title} {{Weak
  deflection angle and shadow cast by the charged-Kiselev black hole with cloud
  of strings in plasma*}},}\ }\href {\doibase 10.1088/1674-1137/ac9fbb}
  {\bibfield  {journal} {\bibinfo  {journal} {Chin. Phys. C}\ }\textbf
  {\bibinfo {volume} {47}},\ \bibinfo {pages} {025102} (\bibinfo {year}
  {2023})}\BibitemShut {NoStop}%
\bibitem [{\citenamefont {Vagnozzi}\ \emph {et~al.}(2023)\citenamefont
  {Vagnozzi}, \citenamefont {Roy}, \citenamefont {Tsai}, \citenamefont
  {Visinelli}, \citenamefont {Afrin}, \citenamefont {Allahyari}, \citenamefont
  {Bambhaniya}, \citenamefont {Dey}, \citenamefont {Ghosh}, \citenamefont
  {Joshi}, \citenamefont {Jusufi}, \citenamefont {Khodadi}, \citenamefont
  {Walia}, \citenamefont {\"Ovg\"un},\ and\ \citenamefont
  {Bambi}}]{Vagnozzi:2022moj}%
  \BibitemOpen
  \bibfield  {author} {\bibinfo {author} {\bibfnamefont {Sunny}\ \bibnamefont
  {Vagnozzi}}, \bibinfo {author} {\bibfnamefont {Rittick}\ \bibnamefont {Roy}},
  \bibinfo {author} {\bibfnamefont {Yu-Dai}\ \bibnamefont {Tsai}}, \bibinfo
  {author} {\bibfnamefont {Luca}\ \bibnamefont {Visinelli}}, \bibinfo {author}
  {\bibfnamefont {Misba}\ \bibnamefont {Afrin}}, \bibinfo {author}
  {\bibfnamefont {Alireza}\ \bibnamefont {Allahyari}}, \bibinfo {author}
  {\bibfnamefont {Parth}\ \bibnamefont {Bambhaniya}}, \bibinfo {author}
  {\bibfnamefont {Dipanjan}\ \bibnamefont {Dey}}, \bibinfo {author}
  {\bibfnamefont {Sushant~G}\ \bibnamefont {Ghosh}}, \bibinfo {author}
  {\bibfnamefont {Pankaj~S}\ \bibnamefont {Joshi}}, \bibinfo {author}
  {\bibfnamefont {Kimet}\ \bibnamefont {Jusufi}}, \bibinfo {author}
  {\bibfnamefont {Mohsen}\ \bibnamefont {Khodadi}}, \bibinfo {author}
  {\bibfnamefont {Rahul~Kumar}\ \bibnamefont {Walia}}, \bibinfo {author}
  {\bibfnamefont {Ali}\ \bibnamefont {\"Ovg\"un}}, \ and\ \bibinfo {author}
  {\bibfnamefont {Cosimo}\ \bibnamefont {Bambi}},\ }\bibfield  {title}
  {\enquote {\bibinfo {title} {{Horizon-scale tests of gravity theories and
  fundamental physics from the Event Horizon Telescope image of Sagittarius
  A$^*$}},}\ }\href {\doibase 10.1088/1361-6382/acd97b} {\bibfield  {journal}
  {\bibinfo  {journal} {Class. Quant. Grav.}\ }\textbf {\bibinfo {volume}
  {40}},\ \bibinfo {pages} {165007} (\bibinfo {year} {2023})},\ \Eprint
  {http://arxiv.org/abs/2205.07787} {arXiv:2205.07787 [gr-qc]} \BibitemShut
  {NoStop}%
\bibitem [{\citenamefont {Chen}\ \emph {et~al.}(2022)\citenamefont {Chen},
  \citenamefont {Roy}, \citenamefont {Vagnozzi},\ and\ \citenamefont
  {Visinelli}}]{Chen:2022nbb}%
  \BibitemOpen
  \bibfield  {author} {\bibinfo {author} {\bibfnamefont {Yifan}\ \bibnamefont
  {Chen}}, \bibinfo {author} {\bibfnamefont {Rittick}\ \bibnamefont {Roy}},
  \bibinfo {author} {\bibfnamefont {Sunny}\ \bibnamefont {Vagnozzi}}, \ and\
  \bibinfo {author} {\bibfnamefont {Luca}\ \bibnamefont {Visinelli}},\
  }\bibfield  {title} {\enquote {\bibinfo {title} {{Superradiant evolution of
  the shadow and photon ring of Sgr A\ensuremath{\star}}},}\ }\href {\doibase
  10.1103/PhysRevD.106.043021} {\bibfield  {journal} {\bibinfo  {journal}
  {Phys. Rev. D}\ }\textbf {\bibinfo {volume} {106}},\ \bibinfo {pages}
  {043021} (\bibinfo {year} {2022})},\ \Eprint
  {http://arxiv.org/abs/2205.06238} {arXiv:2205.06238 [astro-ph.HE]}
  \BibitemShut {NoStop}%
\bibitem [{\citenamefont {Dymnikova}\ and\ \citenamefont
  {Kraav}(2019)}]{Dymnikova2019}%
  \BibitemOpen
  \bibfield  {author} {\bibinfo {author} {\bibfnamefont {Irina}\ \bibnamefont
  {Dymnikova}}\ and\ \bibinfo {author} {\bibfnamefont {Kirill}\ \bibnamefont
  {Kraav}},\ }\bibfield  {title} {\enquote {\bibinfo {title} {{Identification
  of a regular black hole by its shadow}},}\ }\href {\doibase
  10.3390/universe5070163} {\bibfield  {journal} {\bibinfo  {journal}
  {Universe}\ }\textbf {\bibinfo {volume} {5}},\ \bibinfo {pages} {1--16}
  (\bibinfo {year} {2019})}\BibitemShut {NoStop}%
\bibitem [{\citenamefont {Kuang}\ and\ \citenamefont
  {\"Ovg\"un}(2022)}]{Kuang:2022xjp}%
  \BibitemOpen
  \bibfield  {author} {\bibinfo {author} {\bibfnamefont {Xiao-Mei}\
  \bibnamefont {Kuang}}\ and\ \bibinfo {author} {\bibfnamefont {Ali}\
  \bibnamefont {\"Ovg\"un}},\ }\bibfield  {title} {\enquote {\bibinfo {title}
  {{Strong gravitational lensing and shadow constraint from M87* of slowly
  rotating Kerr-like black hole}},}\ }\href {\doibase
  10.1016/j.aop.2022.169147} {\bibfield  {journal} {\bibinfo  {journal} {Annals
  Phys.}\ }\textbf {\bibinfo {volume} {447}},\ \bibinfo {pages} {169147}
  (\bibinfo {year} {2022})},\ \Eprint {http://arxiv.org/abs/2205.11003}
  {arXiv:2205.11003 [gr-qc]} \BibitemShut {NoStop}%
\bibitem [{\citenamefont {Kuang}\ \emph {et~al.}(2022)\citenamefont {Kuang},
  \citenamefont {Tang}, \citenamefont {Wang},\ and\ \citenamefont
  {Wang}}]{Kuang:2022ojj}%
  \BibitemOpen
  \bibfield  {author} {\bibinfo {author} {\bibfnamefont {Xiao-Mei}\
  \bibnamefont {Kuang}}, \bibinfo {author} {\bibfnamefont {Zi-Yu}\ \bibnamefont
  {Tang}}, \bibinfo {author} {\bibfnamefont {Bin}\ \bibnamefont {Wang}}, \ and\
  \bibinfo {author} {\bibfnamefont {Anzhong}\ \bibnamefont {Wang}},\ }\bibfield
   {title} {\enquote {\bibinfo {title} {{Constraining a modified gravity theory
  in strong gravitational lensing and black hole shadow observations}},}\
  }\href {\doibase 10.1103/PhysRevD.106.064012} {\bibfield  {journal} {\bibinfo
   {journal} {Phys. Rev. D}\ }\textbf {\bibinfo {volume} {106}},\ \bibinfo
  {pages} {064012} (\bibinfo {year} {2022})},\ \Eprint
  {http://arxiv.org/abs/2206.05878} {arXiv:2206.05878 [gr-qc]} \BibitemShut
  {NoStop}%
\bibitem [{\citenamefont {Wei}\ \emph {et~al.}(2019)\citenamefont {Wei},
  \citenamefont {Zou}, \citenamefont {Liu},\ and\ \citenamefont
  {Mann}}]{Wei2019}%
  \BibitemOpen
  \bibfield  {author} {\bibinfo {author} {\bibfnamefont {Shao-Wen}\
  \bibnamefont {Wei}}, \bibinfo {author} {\bibfnamefont {Yuan-Chuan}\
  \bibnamefont {Zou}}, \bibinfo {author} {\bibfnamefont {Yu-Xiao}\ \bibnamefont
  {Liu}}, \ and\ \bibinfo {author} {\bibfnamefont {Robert~B.}\ \bibnamefont
  {Mann}},\ }\bibfield  {title} {\enquote {\bibinfo {title} {{Curvature radius
  and Kerr black hole shadow}},}\ }\href {\doibase
  10.1088/1475-7516/2019/08/030} {\bibfield  {journal} {\bibinfo  {journal}
  {JCAP}\ }\textbf {\bibinfo {volume} {08}},\ \bibinfo {pages} {030} (\bibinfo
  {year} {2019})},\ \Eprint {http://arxiv.org/abs/1904.07710} {arXiv:1904.07710
  [gr-qc]} \BibitemShut {NoStop}%
\bibitem [{\citenamefont {Hou}\ \emph {et~al.}(2018)\citenamefont {Hou},
  \citenamefont {Xu},\ and\ \citenamefont {Wang}}]{Hou:2018avu}%
  \BibitemOpen
  \bibfield  {author} {\bibinfo {author} {\bibfnamefont {Xian}\ \bibnamefont
  {Hou}}, \bibinfo {author} {\bibfnamefont {Zhaoyi}\ \bibnamefont {Xu}}, \ and\
  \bibinfo {author} {\bibfnamefont {Jiancheng}\ \bibnamefont {Wang}},\
  }\bibfield  {title} {\enquote {\bibinfo {title} {{Rotating Black Hole Shadow
  in Perfect Fluid Dark Matter}},}\ }\href {\doibase
  10.1088/1475-7516/2018/12/040} {\bibfield  {journal} {\bibinfo  {journal}
  {JCAP}\ }\textbf {\bibinfo {volume} {12}},\ \bibinfo {pages} {040} (\bibinfo
  {year} {2018})},\ \Eprint {http://arxiv.org/abs/1810.06381} {arXiv:1810.06381
  [gr-qc]} \BibitemShut {NoStop}%
\bibitem [{\citenamefont {Tsukamoto}(2018)}]{Tsukamoto:2017fxq}%
  \BibitemOpen
  \bibfield  {author} {\bibinfo {author} {\bibfnamefont {Naoki}\ \bibnamefont
  {Tsukamoto}},\ }\bibfield  {title} {\enquote {\bibinfo {title} {{Black hole
  shadow in an asymptotically-flat, stationary, and axisymmetric spacetime: The
  Kerr-Newman and rotating regular black holes}},}\ }\href {\doibase
  10.1103/PhysRevD.97.064021} {\bibfield  {journal} {\bibinfo  {journal} {Phys.
  Rev. D}\ }\textbf {\bibinfo {volume} {97}},\ \bibinfo {pages} {064021}
  (\bibinfo {year} {2018})},\ \Eprint {http://arxiv.org/abs/1708.07427}
  {arXiv:1708.07427 [gr-qc]} \BibitemShut {NoStop}%
\bibitem [{\citenamefont {Kumar}\ \emph {et~al.}(2020)\citenamefont {Kumar},
  \citenamefont {Ghosh},\ and\ \citenamefont {Wang}}]{Kumar:2020hgm}%
  \BibitemOpen
  \bibfield  {author} {\bibinfo {author} {\bibfnamefont {Rahul}\ \bibnamefont
  {Kumar}}, \bibinfo {author} {\bibfnamefont {Sushant~G.}\ \bibnamefont
  {Ghosh}}, \ and\ \bibinfo {author} {\bibfnamefont {Anzhong}\ \bibnamefont
  {Wang}},\ }\bibfield  {title} {\enquote {\bibinfo {title} {{Gravitational
  deflection of light and shadow cast by rotating Kalb-Ramond black holes}},}\
  }\href {\doibase 10.1103/PhysRevD.101.104001} {\bibfield  {journal} {\bibinfo
   {journal} {Phys. Rev. D}\ }\textbf {\bibinfo {volume} {101}},\ \bibinfo
  {pages} {104001} (\bibinfo {year} {2020})},\ \Eprint
  {http://arxiv.org/abs/2001.00460} {arXiv:2001.00460 [gr-qc]} \BibitemShut
  {NoStop}%
\bibitem [{\citenamefont {Wang}\ \emph {et~al.}(2017)\citenamefont {Wang},
  \citenamefont {Chen},\ and\ \citenamefont {Jing}}]{Wang2017}%
  \BibitemOpen
  \bibfield  {author} {\bibinfo {author} {\bibfnamefont {Mingzhi}\ \bibnamefont
  {Wang}}, \bibinfo {author} {\bibfnamefont {Songbai}\ \bibnamefont {Chen}}, \
  and\ \bibinfo {author} {\bibfnamefont {Jiliang}\ \bibnamefont {Jing}},\
  }\bibfield  {title} {\enquote {\bibinfo {title} {{Shadow casted by a
  Konoplya-Zhidenko rotating non-Kerr black hole}},}\ }\href {\doibase
  10.1088/1475-7516/2017/10/051} {\bibfield  {journal} {\bibinfo  {journal} {J.
  Cosmol. Astropart. Phys.}\ }\textbf {\bibinfo {volume} {2017}},\ \bibinfo
  {pages} {1--14} (\bibinfo {year} {2017})},\ \Eprint
  {http://arxiv.org/abs/1707.09451} {arXiv:1707.09451} \BibitemShut {NoStop}%
\bibitem [{\citenamefont {Tsupko}\ \emph {et~al.}(2020)\citenamefont {Tsupko},
  \citenamefont {Fan},\ and\ \citenamefont {Bisnovatyi-Kogan}}]{Tsupko_2020}%
  \BibitemOpen
  \bibfield  {author} {\bibinfo {author} {\bibfnamefont {Oleg~Yu.}\
  \bibnamefont {Tsupko}}, \bibinfo {author} {\bibfnamefont {Zuhui}\
  \bibnamefont {Fan}}, \ and\ \bibinfo {author} {\bibfnamefont {Gennady~S.}\
  \bibnamefont {Bisnovatyi-Kogan}},\ }\bibfield  {title} {\enquote {\bibinfo
  {title} {{Black hole shadow as a standard ruler in cosmology}},}\ }\href
  {\doibase 10.1088/1361-6382/ab6f7d} {\bibfield  {journal} {\bibinfo
  {journal} {Class. Quant. Grav.}\ }\textbf {\bibinfo {volume} {37}},\ \bibinfo
  {pages} {065016} (\bibinfo {year} {2020})},\ \Eprint
  {http://arxiv.org/abs/1905.10509} {arXiv:1905.10509 [gr-qc]} \BibitemShut
  {NoStop}%
\bibitem [{\citenamefont {Konoplya}(2019)}]{Konoplya2019}%
  \BibitemOpen
  \bibfield  {author} {\bibinfo {author} {\bibfnamefont {R.~A.}\ \bibnamefont
  {Konoplya}},\ }\bibfield  {title} {\enquote {\bibinfo {title} {{Shadow of a
  black hole surrounded by dark matter}},}\ }\href {\doibase
  10.1016/j.physletb.2019.05.043} {\bibfield  {journal} {\bibinfo  {journal}
  {Phys. Lett. B}\ }\textbf {\bibinfo {volume} {795}},\ \bibinfo {pages} {1--6}
  (\bibinfo {year} {2019})},\ \Eprint {http://arxiv.org/abs/1905.00064}
  {arXiv:1905.00064 [gr-qc]} \BibitemShut {NoStop}%
\bibitem [{\citenamefont {Belhaj}\ \emph {et~al.}(2020)\citenamefont {Belhaj},
  \citenamefont {Benali}, \citenamefont {El~Balali}, \citenamefont
  {El~Moumni},\ and\ \citenamefont {Ennadifi}}]{Belhaj:2020rdb}%
  \BibitemOpen
  \bibfield  {author} {\bibinfo {author} {\bibfnamefont {A.}~\bibnamefont
  {Belhaj}}, \bibinfo {author} {\bibfnamefont {M.}~\bibnamefont {Benali}},
  \bibinfo {author} {\bibfnamefont {A.}~\bibnamefont {El~Balali}}, \bibinfo
  {author} {\bibfnamefont {H.}~\bibnamefont {El~Moumni}}, \ and\ \bibinfo
  {author} {\bibfnamefont {S.~E.}\ \bibnamefont {Ennadifi}},\ }\bibfield
  {title} {\enquote {\bibinfo {title} {{Deflection angle and shadow behaviors
  of quintessential black holes in arbitrary dimensions}},}\ }\href {\doibase
  10.1088/1361-6382/abbaa9} {\bibfield  {journal} {\bibinfo  {journal} {Class.
  Quant. Grav.}\ }\textbf {\bibinfo {volume} {37}},\ \bibinfo {pages} {215004}
  (\bibinfo {year} {2020})},\ \Eprint {http://arxiv.org/abs/2006.01078}
  {arXiv:2006.01078 [gr-qc]} \BibitemShut {NoStop}%
\bibitem [{\citenamefont {Belhaj}\ \emph {et~al.}(2021)\citenamefont {Belhaj},
  \citenamefont {Belmahi}, \citenamefont {Benali}, \citenamefont {El~Hadri},
  \citenamefont {El~Moumni},\ and\ \citenamefont
  {Torrente-Lujan}}]{Belhaj:2020okh}%
  \BibitemOpen
  \bibfield  {author} {\bibinfo {author} {\bibfnamefont {A.}~\bibnamefont
  {Belhaj}}, \bibinfo {author} {\bibfnamefont {H.}~\bibnamefont {Belmahi}},
  \bibinfo {author} {\bibfnamefont {M.}~\bibnamefont {Benali}}, \bibinfo
  {author} {\bibfnamefont {W.}~\bibnamefont {El~Hadri}}, \bibinfo {author}
  {\bibfnamefont {H.}~\bibnamefont {El~Moumni}}, \ and\ \bibinfo {author}
  {\bibfnamefont {E.}~\bibnamefont {Torrente-Lujan}},\ }\bibfield  {title}
  {\enquote {\bibinfo {title} {{Shadows of 5D black holes from string
  theory}},}\ }\href {\doibase 10.1016/j.physletb.2020.136025} {\bibfield
  {journal} {\bibinfo  {journal} {Phys. Lett. B}\ }\textbf {\bibinfo {volume}
  {812}},\ \bibinfo {pages} {136025} (\bibinfo {year} {2021})},\ \Eprint
  {http://arxiv.org/abs/2008.13478} {arXiv:2008.13478 [hep-th]} \BibitemShut
  {NoStop}%
\bibitem [{\citenamefont {Cunha}\ and\ \citenamefont
  {Herdeiro}(2018)}]{Cunha:2018acu}%
  \BibitemOpen
  \bibfield  {author} {\bibinfo {author} {\bibfnamefont {Pedro V.~P.}\
  \bibnamefont {Cunha}}\ and\ \bibinfo {author} {\bibfnamefont {Carlos A.~R.}\
  \bibnamefont {Herdeiro}},\ }\bibfield  {title} {\enquote {\bibinfo {title}
  {{Shadows and strong gravitational lensing: a brief review}},}\ }\href
  {\doibase 10.1007/s10714-018-2361-9} {\bibfield  {journal} {\bibinfo
  {journal} {Gen. Rel. Grav.}\ }\textbf {\bibinfo {volume} {50}},\ \bibinfo
  {pages} {42} (\bibinfo {year} {2018})},\ \Eprint
  {http://arxiv.org/abs/1801.00860} {arXiv:1801.00860 [gr-qc]} \BibitemShut
  {NoStop}%
\bibitem [{\citenamefont {Gralla}\ \emph {et~al.}(2019)\citenamefont {Gralla},
  \citenamefont {Holz},\ and\ \citenamefont {Wald}}]{Gralla:2019xty}%
  \BibitemOpen
  \bibfield  {author} {\bibinfo {author} {\bibfnamefont {Samuel~E.}\
  \bibnamefont {Gralla}}, \bibinfo {author} {\bibfnamefont {Daniel~E.}\
  \bibnamefont {Holz}}, \ and\ \bibinfo {author} {\bibfnamefont {Robert~M.}\
  \bibnamefont {Wald}},\ }\bibfield  {title} {\enquote {\bibinfo {title}
  {{Black Hole Shadows, Photon Rings, and Lensing Rings}},}\ }\href {\doibase
  10.1103/PhysRevD.100.024018} {\bibfield  {journal} {\bibinfo  {journal}
  {Phys. Rev. D}\ }\textbf {\bibinfo {volume} {100}},\ \bibinfo {pages}
  {024018} (\bibinfo {year} {2019})},\ \Eprint
  {http://arxiv.org/abs/1906.00873} {arXiv:1906.00873 [astro-ph.HE]}
  \BibitemShut {NoStop}%
\bibitem [{\citenamefont {Perlick}\ \emph {et~al.}(2015)\citenamefont
  {Perlick}, \citenamefont {Tsupko},\ and\ \citenamefont
  {Bisnovatyi-Kogan}}]{Perlick:2015vta}%
  \BibitemOpen
  \bibfield  {author} {\bibinfo {author} {\bibfnamefont {Volker}\ \bibnamefont
  {Perlick}}, \bibinfo {author} {\bibfnamefont {Oleg~Yu.}\ \bibnamefont
  {Tsupko}}, \ and\ \bibinfo {author} {\bibfnamefont {Gennady~S.}\ \bibnamefont
  {Bisnovatyi-Kogan}},\ }\bibfield  {title} {\enquote {\bibinfo {title}
  {{Influence of a plasma on the shadow of a spherically symmetric black
  hole}},}\ }\href {\doibase 10.1103/PhysRevD.92.104031} {\bibfield  {journal}
  {\bibinfo  {journal} {Phys. Rev. D}\ }\textbf {\bibinfo {volume} {92}},\
  \bibinfo {pages} {104031} (\bibinfo {year} {2015})},\ \Eprint
  {http://arxiv.org/abs/1507.04217} {arXiv:1507.04217 [gr-qc]} \BibitemShut
  {NoStop}%
\bibitem [{\citenamefont {Khodadi}\ \emph {et~al.}(2021)\citenamefont
  {Khodadi}, \citenamefont {Lambiase},\ and\ \citenamefont
  {Mota}}]{Khodadi:2021gbc}%
  \BibitemOpen
  \bibfield  {author} {\bibinfo {author} {\bibfnamefont {Mohsen}\ \bibnamefont
  {Khodadi}}, \bibinfo {author} {\bibfnamefont {Gaetano}\ \bibnamefont
  {Lambiase}}, \ and\ \bibinfo {author} {\bibfnamefont {David~F.}\ \bibnamefont
  {Mota}},\ }\bibfield  {title} {\enquote {\bibinfo {title} {{No-hair theorem
  in the wake of Event Horizon Telescope}},}\ }\href {\doibase
  10.1088/1475-7516/2021/09/028} {\bibfield  {journal} {\bibinfo  {journal}
  {JCAP}\ }\textbf {\bibinfo {volume} {09}},\ \bibinfo {pages} {028} (\bibinfo
  {year} {2021})},\ \Eprint {http://arxiv.org/abs/2107.00834} {arXiv:2107.00834
  [gr-qc]} \BibitemShut {NoStop}%
\bibitem [{\citenamefont {Khodadi}\ and\ \citenamefont
  {Lambiase}(2022)}]{Khodadi:2022pqh}%
  \BibitemOpen
  \bibfield  {author} {\bibinfo {author} {\bibfnamefont {Mohsen}\ \bibnamefont
  {Khodadi}}\ and\ \bibinfo {author} {\bibfnamefont {Gaetano}\ \bibnamefont
  {Lambiase}},\ }\bibfield  {title} {\enquote {\bibinfo {title} {{Probing
  Lorentz symmetry violation using the first image of Sagittarius A*:
  Constraints on standard-model extension coefficients}},}\ }\href {\doibase
  10.1103/PhysRevD.106.104050} {\bibfield  {journal} {\bibinfo  {journal}
  {Phys. Rev. D}\ }\textbf {\bibinfo {volume} {106}},\ \bibinfo {pages}
  {104050} (\bibinfo {year} {2022})},\ \Eprint
  {http://arxiv.org/abs/2206.08601} {arXiv:2206.08601 [gr-qc]} \BibitemShut
  {NoStop}%
\bibitem [{\citenamefont {Cunha}\ \emph {et~al.}(2017)\citenamefont {Cunha},
  \citenamefont {Herdeiro}, \citenamefont {Kleihaus}, \citenamefont {Kunz},\
  and\ \citenamefont {Radu}}]{Cunha:2016wzk}%
  \BibitemOpen
  \bibfield  {author} {\bibinfo {author} {\bibfnamefont {Pedro V.~P.}\
  \bibnamefont {Cunha}}, \bibinfo {author} {\bibfnamefont {Carlos A.~R.}\
  \bibnamefont {Herdeiro}}, \bibinfo {author} {\bibfnamefont {Burkhard}\
  \bibnamefont {Kleihaus}}, \bibinfo {author} {\bibfnamefont {Jutta}\
  \bibnamefont {Kunz}}, \ and\ \bibinfo {author} {\bibfnamefont {Eugen}\
  \bibnamefont {Radu}},\ }\bibfield  {title} {\enquote {\bibinfo {title}
  {{Shadows of
  Einstein\textendash{}dilaton\textendash{}Gauss\textendash{}Bonnet black
  holes}},}\ }\href {\doibase 10.1016/j.physletb.2017.03.020} {\bibfield
  {journal} {\bibinfo  {journal} {Phys. Lett. B}\ }\textbf {\bibinfo {volume}
  {768}},\ \bibinfo {pages} {373--379} (\bibinfo {year} {2017})},\ \Eprint
  {http://arxiv.org/abs/1701.00079} {arXiv:1701.00079 [gr-qc]} \BibitemShut
  {NoStop}%
\bibitem [{\citenamefont {Shaikh}(2019)}]{Shaikh:2019fpu}%
  \BibitemOpen
  \bibfield  {author} {\bibinfo {author} {\bibfnamefont {Rajibul}\ \bibnamefont
  {Shaikh}},\ }\bibfield  {title} {\enquote {\bibinfo {title} {{Black hole
  shadow in a general rotating spacetime obtained through Newman-Janis
  algorithm}},}\ }\href {\doibase 10.1103/PhysRevD.100.024028} {\bibfield
  {journal} {\bibinfo  {journal} {Phys. Rev. D}\ }\textbf {\bibinfo {volume}
  {100}},\ \bibinfo {pages} {024028} (\bibinfo {year} {2019})},\ \Eprint
  {http://arxiv.org/abs/1904.08322} {arXiv:1904.08322 [gr-qc]} \BibitemShut
  {NoStop}%
\bibitem [{\citenamefont {Allahyari}\ \emph {et~al.}(2020)\citenamefont
  {Allahyari}, \citenamefont {Khodadi}, \citenamefont {Vagnozzi},\ and\
  \citenamefont {Mota}}]{Allahyari:2019jqz}%
  \BibitemOpen
  \bibfield  {author} {\bibinfo {author} {\bibfnamefont {Alireza}\ \bibnamefont
  {Allahyari}}, \bibinfo {author} {\bibfnamefont {Mohsen}\ \bibnamefont
  {Khodadi}}, \bibinfo {author} {\bibfnamefont {Sunny}\ \bibnamefont
  {Vagnozzi}}, \ and\ \bibinfo {author} {\bibfnamefont {David~F.}\ \bibnamefont
  {Mota}},\ }\bibfield  {title} {\enquote {\bibinfo {title} {{Magnetically
  charged black holes from non-linear electrodynamics and the Event Horizon
  Telescope}},}\ }\href {\doibase 10.1088/1475-7516/2020/02/003} {\bibfield
  {journal} {\bibinfo  {journal} {JCAP}\ }\textbf {\bibinfo {volume} {02}},\
  \bibinfo {pages} {003} (\bibinfo {year} {2020})},\ \Eprint
  {http://arxiv.org/abs/1912.08231} {arXiv:1912.08231 [gr-qc]} \BibitemShut
  {NoStop}%
\bibitem [{\citenamefont {Cunha}\ \emph {et~al.}(2016)\citenamefont {Cunha},
  \citenamefont {Grover}, \citenamefont {Herdeiro}, \citenamefont {Radu},
  \citenamefont {Runarsson},\ and\ \citenamefont {Wittig}}]{Cunha:2016bjh}%
  \BibitemOpen
  \bibfield  {author} {\bibinfo {author} {\bibfnamefont {P.~V.~P.}\
  \bibnamefont {Cunha}}, \bibinfo {author} {\bibfnamefont {J.}~\bibnamefont
  {Grover}}, \bibinfo {author} {\bibfnamefont {C.}~\bibnamefont {Herdeiro}},
  \bibinfo {author} {\bibfnamefont {E.}~\bibnamefont {Radu}}, \bibinfo {author}
  {\bibfnamefont {H.}~\bibnamefont {Runarsson}}, \ and\ \bibinfo {author}
  {\bibfnamefont {A.}~\bibnamefont {Wittig}},\ }\bibfield  {title} {\enquote
  {\bibinfo {title} {{Chaotic lensing around boson stars and Kerr black holes
  with scalar hair}},}\ }\href {\doibase 10.1103/PhysRevD.94.104023} {\bibfield
   {journal} {\bibinfo  {journal} {Phys. Rev. D}\ }\textbf {\bibinfo {volume}
  {94}},\ \bibinfo {pages} {104023} (\bibinfo {year} {2016})},\ \Eprint
  {http://arxiv.org/abs/1609.01340} {arXiv:1609.01340 [gr-qc]} \BibitemShut
  {NoStop}%
\bibitem [{\citenamefont {Zakharov}(2014)}]{Zakharov:2014lqa}%
  \BibitemOpen
  \bibfield  {author} {\bibinfo {author} {\bibfnamefont {Alexander~F.}\
  \bibnamefont {Zakharov}},\ }\bibfield  {title} {\enquote {\bibinfo {title}
  {{Constraints on a charge in the Reissner-Nordstr\"om metric for the black
  hole at the Galactic Center}},}\ }\href {\doibase 10.1103/PhysRevD.90.062007}
  {\bibfield  {journal} {\bibinfo  {journal} {Phys. Rev. D}\ }\textbf {\bibinfo
  {volume} {90}},\ \bibinfo {pages} {062007} (\bibinfo {year} {2014})},\
  \Eprint {http://arxiv.org/abs/1407.7457} {arXiv:1407.7457 [gr-qc]}
  \BibitemShut {NoStop}%
\bibitem [{\citenamefont {Chakhchi}\ \emph {et~al.}(2022)\citenamefont
  {Chakhchi}, \citenamefont {El~Moumni},\ and\ \citenamefont
  {Masmar}}]{Chakhchi:2022fl}%
  \BibitemOpen
  \bibfield  {author} {\bibinfo {author} {\bibfnamefont {L.}~\bibnamefont
  {Chakhchi}}, \bibinfo {author} {\bibfnamefont {H.}~\bibnamefont {El~Moumni}},
  \ and\ \bibinfo {author} {\bibfnamefont {K.}~\bibnamefont {Masmar}},\
  }\bibfield  {title} {\enquote {\bibinfo {title} {{Shadows and optical
  appearance of a power-Yang-Mills black hole surrounded by different accretion
  disk profiles}},}\ }\href {\doibase 10.1103/PhysRevD.105.064031} {\bibfield
  {journal} {\bibinfo  {journal} {Phys. Rev. D}\ }\textbf {\bibinfo {volume}
  {105}},\ \bibinfo {pages} {064031} (\bibinfo {year} {2022})}\BibitemShut
  {NoStop}%
\bibitem [{\citenamefont {Pantig}\ and\ \citenamefont
  {Rodulfo}(2020)}]{Pantig:2020uhp}%
  \BibitemOpen
  \bibfield  {author} {\bibinfo {author} {\bibfnamefont {Reggie~C.}\
  \bibnamefont {Pantig}}\ and\ \bibinfo {author} {\bibfnamefont {Emmanuel~T.}\
  \bibnamefont {Rodulfo}},\ }\bibfield  {title} {\enquote {\bibinfo {title}
  {{Rotating dirty black hole and its shadow}},}\ }\href {\doibase
  10.1016/j.cjph.2020.08.001} {\bibfield  {journal} {\bibinfo  {journal} {Chin.
  J. Phys.}\ }\textbf {\bibinfo {volume} {68}},\ \bibinfo {pages} {236--257}
  (\bibinfo {year} {2020})},\ \Eprint {http://arxiv.org/abs/2003.06829}
  {arXiv:2003.06829 [gr-qc]} \BibitemShut {NoStop}%
\bibitem [{\citenamefont {Pantig}\ and\ \citenamefont
  {\"Ovg\"un}(2022{\natexlab{a}})}]{Pantig:2022toh}%
  \BibitemOpen
  \bibfield  {author} {\bibinfo {author} {\bibfnamefont {Reggie~C.}\
  \bibnamefont {Pantig}}\ and\ \bibinfo {author} {\bibfnamefont {Ali}\
  \bibnamefont {\"Ovg\"un}},\ }\bibfield  {title} {\enquote {\bibinfo {title}
  {{Dark matter effect on the weak deflection angle by black holes at the
  center of Milky Way and M87 galaxies}},}\ }\href {\doibase
  10.1140/epjc/s10052-022-10319-8} {\bibfield  {journal} {\bibinfo  {journal}
  {Eur. Phys. J. C}\ }\textbf {\bibinfo {volume} {82}},\ \bibinfo {pages} {391}
  (\bibinfo {year} {2022}{\natexlab{a}})},\ \Eprint
  {http://arxiv.org/abs/2201.03365} {arXiv:2201.03365 [gr-qc]} \BibitemShut
  {NoStop}%
\bibitem [{\citenamefont {Pantig}\ and\ \citenamefont
  {\"Ovg\"un}(2022{\natexlab{b}})}]{Pantig:2022whj}%
  \BibitemOpen
  \bibfield  {author} {\bibinfo {author} {\bibfnamefont {Reggie~C.}\
  \bibnamefont {Pantig}}\ and\ \bibinfo {author} {\bibfnamefont {Ali}\
  \bibnamefont {\"Ovg\"un}},\ }\bibfield  {title} {\enquote {\bibinfo {title}
  {{Dehnen halo effect on a black hole in an ultra-faint dwarf galaxy}},}\
  }\href {\doibase 10.1088/1475-7516/2022/08/056} {\bibfield  {journal}
  {\bibinfo  {journal} {JCAP}\ }\textbf {\bibinfo {volume} {08}},\ \bibinfo
  {pages} {056} (\bibinfo {year} {2022}{\natexlab{b}})},\ \Eprint
  {http://arxiv.org/abs/2202.07404} {arXiv:2202.07404 [astro-ph.GA]}
  \BibitemShut {NoStop}%
\bibitem [{\citenamefont {Xu}\ \emph {et~al.}(2020)\citenamefont {Xu},
  \citenamefont {Gong},\ and\ \citenamefont {Zhang}}]{Xu:2020jpv}%
  \BibitemOpen
  \bibfield  {author} {\bibinfo {author} {\bibfnamefont {Zhaoyi}\ \bibnamefont
  {Xu}}, \bibinfo {author} {\bibfnamefont {Xiaobo}\ \bibnamefont {Gong}}, \
  and\ \bibinfo {author} {\bibfnamefont {Shuang-Nan}\ \bibnamefont {Zhang}},\
  }\bibfield  {title} {\enquote {\bibinfo {title} {{Black hole immersed dark
  matter halo}},}\ }\href {\doibase 10.1103/PhysRevD.101.024029} {\bibfield
  {journal} {\bibinfo  {journal} {Phys. Rev. D}\ }\textbf {\bibinfo {volume}
  {101}},\ \bibinfo {pages} {024029} (\bibinfo {year} {2020})}\BibitemShut
  {NoStop}%
\bibitem [{\citenamefont {Xu}\ \emph {et~al.}(2021)\citenamefont {Xu},
  \citenamefont {Wang},\ and\ \citenamefont {Tang}}]{Xu:2021dkv}%
  \BibitemOpen
  \bibfield  {author} {\bibinfo {author} {\bibfnamefont {Zhaoyi}\ \bibnamefont
  {Xu}}, \bibinfo {author} {\bibfnamefont {Jiancheng}\ \bibnamefont {Wang}}, \
  and\ \bibinfo {author} {\bibfnamefont {Meirong}\ \bibnamefont {Tang}},\
  }\bibfield  {title} {\enquote {\bibinfo {title} {{Deformed black hole
  immersed in dark matter spike}},}\ }\href {\doibase
  10.1088/1475-7516/2021/09/007} {\bibfield  {journal} {\bibinfo  {journal}
  {JCAP}\ }\textbf {\bibinfo {volume} {09}},\ \bibinfo {pages} {007} (\bibinfo
  {year} {2021})},\ \Eprint {http://arxiv.org/abs/2104.13158} {arXiv:2104.13158
  [gr-qc]} \BibitemShut {NoStop}%
\bibitem [{\citenamefont {Konoplya}(2021)}]{Konoplya:2021ube}%
  \BibitemOpen
  \bibfield  {author} {\bibinfo {author} {\bibfnamefont {R.~A.}\ \bibnamefont
  {Konoplya}},\ }\bibfield  {title} {\enquote {\bibinfo {title} {{Black holes
  in galactic centers: Quasinormal ringing, grey-body factors and Unruh
  temperature}},}\ }\href {\doibase 10.1016/j.physletb.2021.136734} {\bibfield
  {journal} {\bibinfo  {journal} {Phys. Lett. B}\ }\textbf {\bibinfo {volume}
  {823}},\ \bibinfo {pages} {136734} (\bibinfo {year} {2021})},\ \Eprint
  {http://arxiv.org/abs/2109.01640} {arXiv:2109.01640 [gr-qc]} \BibitemShut
  {NoStop}%
\bibitem [{\citenamefont {Devi}\ \emph {et~al.}(2023)\citenamefont {Devi},
  \citenamefont {S.}, \citenamefont {Chakrabarti},\ and\ \citenamefont
  {Majhi}}]{Devi:2021ctm}%
  \BibitemOpen
  \bibfield  {author} {\bibinfo {author} {\bibfnamefont {Saraswati}\
  \bibnamefont {Devi}}, \bibinfo {author} {\bibfnamefont {Abhinove~Nagarajan}\
  \bibnamefont {S.}}, \bibinfo {author} {\bibfnamefont {Sayan}\ \bibnamefont
  {Chakrabarti}}, \ and\ \bibinfo {author} {\bibfnamefont {Bibhas~Ranjan}\
  \bibnamefont {Majhi}},\ }\bibfield  {title} {\enquote {\bibinfo {title}
  {{Shadow of quantum extended Kruskal black hole and its super-radiance
  property}},}\ }\href {\doibase 10.1016/j.dark.2023.101173} {\bibfield
  {journal} {\bibinfo  {journal} {Phys. Dark Univ.}\ }\textbf {\bibinfo
  {volume} {39}},\ \bibinfo {pages} {101173} (\bibinfo {year} {2023})},\
  \Eprint {http://arxiv.org/abs/2105.11847} {arXiv:2105.11847 [gr-qc]}
  \BibitemShut {NoStop}%
\bibitem [{\citenamefont {Xu}\ and\ \citenamefont {Tang}(2022)}]{Xu:2021xgw}%
  \BibitemOpen
  \bibfield  {author} {\bibinfo {author} {\bibfnamefont {Zhaoyi}\ \bibnamefont
  {Xu}}\ and\ \bibinfo {author} {\bibfnamefont {Meirong}\ \bibnamefont
  {Tang}},\ }\bibfield  {title} {\enquote {\bibinfo {title} {{Testing the
  quantum effects near the event horizon with respect to the black hole shadow
  *}},}\ }\href {\doibase 10.1088/1674-1137/ac6665} {\bibfield  {journal}
  {\bibinfo  {journal} {Chin. Phys. C}\ }\textbf {\bibinfo {volume} {46}},\
  \bibinfo {pages} {085101} (\bibinfo {year} {2022})},\ \Eprint
  {http://arxiv.org/abs/2109.14245} {arXiv:2109.14245 [hep-th]} \BibitemShut
  {NoStop}%
\bibitem [{\citenamefont {Lobos}\ and\ \citenamefont
  {Pantig}(2022)}]{Lobos:2022}%
  \BibitemOpen
  \bibfield  {author} {\bibinfo {author} {\bibfnamefont {Nikko John Leo~S.}\
  \bibnamefont {Lobos}}\ and\ \bibinfo {author} {\bibfnamefont {Reggie~C.}\
  \bibnamefont {Pantig}},\ }\bibfield  {title} {\enquote {\bibinfo {title}
  {Generalized extended uncertainty principle black holes: Shadow and lensing
  in the macro- and microscopic realms},}\ }\href {\doibase
  10.3390/physics4040084} {\bibfield  {journal} {\bibinfo  {journal} {Physics}\
  }\textbf {\bibinfo {volume} {4}},\ \bibinfo {pages} {1318--1330} (\bibinfo
  {year} {2022})}\BibitemShut {NoStop}%
\bibitem [{\citenamefont {Anacleto}\ \emph {et~al.}(2021)\citenamefont
  {Anacleto}, \citenamefont {Campos}, \citenamefont {Brito},\ and\
  \citenamefont {Passos}}]{Anacleto:2021qoe}%
  \BibitemOpen
  \bibfield  {author} {\bibinfo {author} {\bibfnamefont {M.~A.}\ \bibnamefont
  {Anacleto}}, \bibinfo {author} {\bibfnamefont {J.~A.~V.}\ \bibnamefont
  {Campos}}, \bibinfo {author} {\bibfnamefont {F.~A.}\ \bibnamefont {Brito}}, \
  and\ \bibinfo {author} {\bibfnamefont {E.}~\bibnamefont {Passos}},\
  }\bibfield  {title} {\enquote {\bibinfo {title} {{Quasinormal modes and
  shadow of a Schwarzschild black hole with GUP}},}\ }\href {\doibase
  10.1016/j.aop.2021.168662} {\bibfield  {journal} {\bibinfo  {journal} {Annals
  Phys.}\ }\textbf {\bibinfo {volume} {434}},\ \bibinfo {pages} {168662}
  (\bibinfo {year} {2021})},\ \Eprint {http://arxiv.org/abs/2108.04998}
  {arXiv:2108.04998 [gr-qc]} \BibitemShut {NoStop}%
\bibitem [{\citenamefont {Hu}\ \emph {et~al.}(2021)\citenamefont {Hu},
  \citenamefont {Zhong}, \citenamefont {Li}, \citenamefont {Guo},\ and\
  \citenamefont {Chen}}]{Hu:2020usx}%
  \BibitemOpen
  \bibfield  {author} {\bibinfo {author} {\bibfnamefont {Zezhou}\ \bibnamefont
  {Hu}}, \bibinfo {author} {\bibfnamefont {Zhen}\ \bibnamefont {Zhong}},
  \bibinfo {author} {\bibfnamefont {Peng-Cheng}\ \bibnamefont {Li}}, \bibinfo
  {author} {\bibfnamefont {Minyong}\ \bibnamefont {Guo}}, \ and\ \bibinfo
  {author} {\bibfnamefont {Bin}\ \bibnamefont {Chen}},\ }\bibfield  {title}
  {\enquote {\bibinfo {title} {{QED effect on a black hole shadow}},}\ }\href
  {\doibase 10.1103/PhysRevD.103.044057} {\bibfield  {journal} {\bibinfo
  {journal} {Phys. Rev. D}\ }\textbf {\bibinfo {volume} {103}},\ \bibinfo
  {pages} {044057} (\bibinfo {year} {2021})},\ \Eprint
  {http://arxiv.org/abs/2012.07022} {arXiv:2012.07022 [gr-qc]} \BibitemShut
  {NoStop}%
\bibitem [{\citenamefont {Pantig}\ and\ \citenamefont
  {\"Ovg\"un}(2022{\natexlab{c}})}]{Pantig:2022sjb}%
  \BibitemOpen
  \bibfield  {author} {\bibinfo {author} {\bibfnamefont {Reggie~C.}\
  \bibnamefont {Pantig}}\ and\ \bibinfo {author} {\bibfnamefont {Ali}\
  \bibnamefont {\"Ovg\"un}},\ }\bibfield  {title} {\enquote {\bibinfo {title}
  {{Black hole in quantum wave dark matter}},}\ }\href {\doibase
  10.1002/prop.202200164} {\bibfield  {journal} {\bibinfo  {journal} {Fortsch.
  Phys.}\ }\textbf {\bibinfo {volume} {2022}},\ \bibinfo {pages} {2200164}
  (\bibinfo {year} {2022}{\natexlab{c}})},\ \Eprint
  {http://arxiv.org/abs/2210.00523} {arXiv:2210.00523 [gr-qc]} \BibitemShut
  {NoStop}%
\bibitem [{\citenamefont {Pantig}(2023)}]{Pantig:2023yer}%
  \BibitemOpen
  \bibfield  {author} {\bibinfo {author} {\bibfnamefont {Reggie~C.}\
  \bibnamefont {Pantig}},\ }\bibfield  {title} {\enquote {\bibinfo {title}
  {{Constraining a one-dimensional wave-type gravitational wave parameter
  through the shadow of M87* via Event Horizon Telescope}},}\ }\href@noop {}
  {\bibfield  {journal} {\bibinfo  {journal} {arXiv preprint}\ } (\bibinfo
  {year} {2023})},\ \Eprint {http://arxiv.org/abs/2303.01698} {arXiv:2303.01698
  [gr-qc]} \BibitemShut {NoStop}%
\bibitem [{\citenamefont {\"Ovg\"un}\ \emph {et~al.}(2023)\citenamefont
  {\"Ovg\"un}, \citenamefont {Pantig},\ and\ \citenamefont
  {Rinc\'on}}]{Ovgun:2023ego}%
  \BibitemOpen
  \bibfield  {author} {\bibinfo {author} {\bibfnamefont {Ali}\ \bibnamefont
  {\"Ovg\"un}}, \bibinfo {author} {\bibfnamefont {Reggie~C.}\ \bibnamefont
  {Pantig}}, \ and\ \bibinfo {author} {\bibfnamefont {\'Angel}\ \bibnamefont
  {Rinc\'on}},\ }\bibfield  {title} {\enquote {\bibinfo {title} {{4D
  scale-dependent Schwarzschild-AdS/dS black holes: study of shadow and weak
  deflection angle and greybody bounding}},}\ }\href {\doibase
  10.1140/epjp/s13360-023-03793-w} {\bibfield  {journal} {\bibinfo  {journal}
  {Eur. Phys. J. Plus}\ }\textbf {\bibinfo {volume} {138}},\ \bibinfo {pages}
  {192} (\bibinfo {year} {2023})},\ \Eprint {http://arxiv.org/abs/2303.01696}
  {arXiv:2303.01696 [gr-qc]} \BibitemShut {NoStop}%
\bibitem [{\citenamefont {Virbhadra}\ and\ \citenamefont
  {Ellis}(2000)}]{Virbhadra:1999nm}%
  \BibitemOpen
  \bibfield  {author} {\bibinfo {author} {\bibfnamefont {K.~S.}\ \bibnamefont
  {Virbhadra}}\ and\ \bibinfo {author} {\bibfnamefont {George F.~R.}\
  \bibnamefont {Ellis}},\ }\bibfield  {title} {\enquote {\bibinfo {title}
  {{Schwarzschild black hole lensing}},}\ }\href {\doibase
  10.1103/PhysRevD.62.084003} {\bibfield  {journal} {\bibinfo  {journal} {Phys.
  Rev. D}\ }\textbf {\bibinfo {volume} {62}},\ \bibinfo {pages} {084003}
  (\bibinfo {year} {2000})},\ \Eprint {http://arxiv.org/abs/astro-ph/9904193}
  {arXiv:astro-ph/9904193} \BibitemShut {NoStop}%
\bibitem [{\citenamefont {Virbhadra}\ and\ \citenamefont
  {Ellis}(2002)}]{Virbhadra:2002ju}%
  \BibitemOpen
  \bibfield  {author} {\bibinfo {author} {\bibfnamefont {K.~S.}\ \bibnamefont
  {Virbhadra}}\ and\ \bibinfo {author} {\bibfnamefont {G.~F.~R.}\ \bibnamefont
  {Ellis}},\ }\bibfield  {title} {\enquote {\bibinfo {title} {{Gravitational
  lensing by naked singularities}},}\ }\href {\doibase
  10.1103/PhysRevD.65.103004} {\bibfield  {journal} {\bibinfo  {journal} {Phys.
  Rev. D}\ }\textbf {\bibinfo {volume} {65}},\ \bibinfo {pages} {103004}
  (\bibinfo {year} {2002})}\BibitemShut {NoStop}%
\bibitem [{\citenamefont {Adler}\ and\ \citenamefont
  {Virbhadra}(2022)}]{Adler:2022qtb}%
  \BibitemOpen
  \bibfield  {author} {\bibinfo {author} {\bibfnamefont {Stephen~L.}\
  \bibnamefont {Adler}}\ and\ \bibinfo {author} {\bibfnamefont {K.~S.}\
  \bibnamefont {Virbhadra}},\ }\bibfield  {title} {\enquote {\bibinfo {title}
  {{Cosmological constant corrections to the photon sphere and black hole
  shadow radii}},}\ }\href {\doibase 10.1007/s10714-022-02976-7} {\bibfield
  {journal} {\bibinfo  {journal} {Gen. Rel. Grav.}\ }\textbf {\bibinfo {volume}
  {54}},\ \bibinfo {pages} {93} (\bibinfo {year} {2022})},\ \Eprint
  {http://arxiv.org/abs/2205.04628} {arXiv:2205.04628 [gr-qc]} \BibitemShut
  {NoStop}%
\bibitem [{\citenamefont {Bozza}\ \emph {et~al.}(2001)\citenamefont {Bozza},
  \citenamefont {Capozziello}, \citenamefont {Iovane},\ and\ \citenamefont
  {Scarpetta}}]{Bozza:2001xd}%
  \BibitemOpen
  \bibfield  {author} {\bibinfo {author} {\bibfnamefont {V.}~\bibnamefont
  {Bozza}}, \bibinfo {author} {\bibfnamefont {S.}~\bibnamefont {Capozziello}},
  \bibinfo {author} {\bibfnamefont {G.}~\bibnamefont {Iovane}}, \ and\ \bibinfo
  {author} {\bibfnamefont {G.}~\bibnamefont {Scarpetta}},\ }\bibfield  {title}
  {\enquote {\bibinfo {title} {{Strong field limit of black hole gravitational
  lensing}},}\ }\href {\doibase 10.1023/A:1012292927358} {\bibfield  {journal}
  {\bibinfo  {journal} {Gen. Rel. Grav.}\ }\textbf {\bibinfo {volume} {33}},\
  \bibinfo {pages} {1535--1548} (\bibinfo {year} {2001})},\ \Eprint
  {http://arxiv.org/abs/gr-qc/0102068} {arXiv:gr-qc/0102068} \BibitemShut
  {NoStop}%
\bibitem [{\citenamefont {Bozza}(2002)}]{Bozza:2002zj}%
  \BibitemOpen
  \bibfield  {author} {\bibinfo {author} {\bibfnamefont {V.}~\bibnamefont
  {Bozza}},\ }\bibfield  {title} {\enquote {\bibinfo {title} {{Gravitational
  lensing in the strong field limit}},}\ }\href {\doibase
  10.1103/PhysRevD.66.103001} {\bibfield  {journal} {\bibinfo  {journal} {Phys.
  Rev. D}\ }\textbf {\bibinfo {volume} {66}},\ \bibinfo {pages} {103001}
  (\bibinfo {year} {2002})},\ \Eprint {http://arxiv.org/abs/gr-qc/0208075}
  {arXiv:gr-qc/0208075} \BibitemShut {NoStop}%
\bibitem [{\citenamefont {Perlick}(2004)}]{Perlick:2003vg}%
  \BibitemOpen
  \bibfield  {author} {\bibinfo {author} {\bibfnamefont {Volker}\ \bibnamefont
  {Perlick}},\ }\bibfield  {title} {\enquote {\bibinfo {title} {{On the Exact
  gravitational lens equation in spherically symmetric and static
  space-times}},}\ }\href {\doibase 10.1103/PhysRevD.69.064017} {\bibfield
  {journal} {\bibinfo  {journal} {Phys. Rev. D}\ }\textbf {\bibinfo {volume}
  {69}},\ \bibinfo {pages} {064017} (\bibinfo {year} {2004})},\ \Eprint
  {http://arxiv.org/abs/gr-qc/0307072} {arXiv:gr-qc/0307072} \BibitemShut
  {NoStop}%
\bibitem [{\citenamefont {He}\ \emph {et~al.}(2020)\citenamefont {He},
  \citenamefont {Zhou}, \citenamefont {Feng}, \citenamefont {Mu}, \citenamefont
  {Wang}, \citenamefont {Li}, \citenamefont {Pan},\ and\ \citenamefont
  {Lin}}]{He:2020eah}%
  \BibitemOpen
  \bibfield  {author} {\bibinfo {author} {\bibfnamefont {Guansheng}\
  \bibnamefont {He}}, \bibinfo {author} {\bibfnamefont {Xia}\ \bibnamefont
  {Zhou}}, \bibinfo {author} {\bibfnamefont {Zhongwen}\ \bibnamefont {Feng}},
  \bibinfo {author} {\bibfnamefont {Xueling}\ \bibnamefont {Mu}}, \bibinfo
  {author} {\bibfnamefont {Hui}\ \bibnamefont {Wang}}, \bibinfo {author}
  {\bibfnamefont {Weijun}\ \bibnamefont {Li}}, \bibinfo {author} {\bibfnamefont
  {Chaohong}\ \bibnamefont {Pan}}, \ and\ \bibinfo {author} {\bibfnamefont
  {Wenbin}\ \bibnamefont {Lin}},\ }\bibfield  {title} {\enquote {\bibinfo
  {title} {{Gravitational deflection of massive particles in Schwarzschild-de
  Sitter spacetime}},}\ }\href {\doibase 10.1140/epjc/s10052-020-8382-z}
  {\bibfield  {journal} {\bibinfo  {journal} {Eur. Phys. J. C}\ }\textbf
  {\bibinfo {volume} {80}},\ \bibinfo {pages} {835} (\bibinfo {year}
  {2020})}\BibitemShut {NoStop}%
\bibitem [{\citenamefont {Virbhadra}(2022{\natexlab{a}})}]{Virbhadra:2022ybp}%
  \BibitemOpen
  \bibfield  {author} {\bibinfo {author} {\bibfnamefont {K.~S.}\ \bibnamefont
  {Virbhadra}},\ }\bibfield  {title} {\enquote {\bibinfo {title} {{Compactness
  of supermassive dark objects at galactic centers}},}\ }\href@noop {}
  {\bibfield  {journal} {\bibinfo  {journal} {arXiv preprint}\ } (\bibinfo
  {year} {2022}{\natexlab{a}})},\ \Eprint {http://arxiv.org/abs/2204.01792}
  {arXiv:2204.01792 [gr-qc]} \BibitemShut {NoStop}%
\bibitem [{\citenamefont {Virbhadra}(2022{\natexlab{b}})}]{Virbhadra:2022iiy}%
  \BibitemOpen
  \bibfield  {author} {\bibinfo {author} {\bibfnamefont {K.~S.}\ \bibnamefont
  {Virbhadra}},\ }\bibfield  {title} {\enquote {\bibinfo {title} {{Distortions
  of images of Schwarzschild lensing}},}\ }\href {\doibase
  10.1103/PhysRevD.106.064038} {\bibfield  {journal} {\bibinfo  {journal}
  {Phys. Rev. D}\ }\textbf {\bibinfo {volume} {106}},\ \bibinfo {pages}
  {064038} (\bibinfo {year} {2022}{\natexlab{b}})},\ \Eprint
  {http://arxiv.org/abs/2204.01879} {arXiv:2204.01879 [gr-qc]} \BibitemShut
  {NoStop}%
\bibitem [{\citenamefont {Gibbons}\ and\ \citenamefont
  {Werner}(2008)}]{Gibbons:2008rj}%
  \BibitemOpen
  \bibfield  {author} {\bibinfo {author} {\bibfnamefont {G.~W.}\ \bibnamefont
  {Gibbons}}\ and\ \bibinfo {author} {\bibfnamefont {M.~C.}\ \bibnamefont
  {Werner}},\ }\bibfield  {title} {\enquote {\bibinfo {title} {{Applications of
  the Gauss-Bonnet theorem to gravitational lensing}},}\ }\href {\doibase
  10.1088/0264-9381/25/23/235009} {\bibfield  {journal} {\bibinfo  {journal}
  {Class. Quant. Grav.}\ }\textbf {\bibinfo {volume} {25}},\ \bibinfo {pages}
  {235009} (\bibinfo {year} {2008})},\ \Eprint {http://arxiv.org/abs/0807.0854}
  {arXiv:0807.0854 [gr-qc]} \BibitemShut {NoStop}%
\bibitem [{\citenamefont {\"Ovg\"un}(2018)}]{Ovgun:2018fnk}%
  \BibitemOpen
  \bibfield  {author} {\bibinfo {author} {\bibfnamefont {Ali}\ \bibnamefont
  {\"Ovg\"un}},\ }\bibfield  {title} {\enquote {\bibinfo {title} {{Light
  deflection by Damour-Solodukhin wormholes and Gauss-Bonnet theorem}},}\
  }\href {\doibase 10.1103/PhysRevD.98.044033} {\bibfield  {journal} {\bibinfo
  {journal} {Phys. Rev. D}\ }\textbf {\bibinfo {volume} {98}},\ \bibinfo
  {pages} {044033} (\bibinfo {year} {2018})},\ \Eprint
  {http://arxiv.org/abs/1805.06296} {arXiv:1805.06296 [gr-qc]} \BibitemShut
  {NoStop}%
\bibitem [{\citenamefont {\"Ovg\"un}(2019{\natexlab{a}})}]{Ovgun:2019wej}%
  \BibitemOpen
  \bibfield  {author} {\bibinfo {author} {\bibfnamefont {A.}~\bibnamefont
  {\"Ovg\"un}},\ }\bibfield  {title} {\enquote {\bibinfo {title} {{Weak field
  deflection angle by regular black holes with cosmic strings using the
  Gauss-Bonnet theorem}},}\ }\href {\doibase 10.1103/PhysRevD.99.104075}
  {\bibfield  {journal} {\bibinfo  {journal} {Phys. Rev. D}\ }\textbf {\bibinfo
  {volume} {99}},\ \bibinfo {pages} {104075} (\bibinfo {year}
  {2019}{\natexlab{a}})},\ \Eprint {http://arxiv.org/abs/1902.04411}
  {arXiv:1902.04411 [gr-qc]} \BibitemShut {NoStop}%
\bibitem [{\citenamefont {\"Ovg\"un}(2019{\natexlab{b}})}]{Ovgun:2018oxk}%
  \BibitemOpen
  \bibfield  {author} {\bibinfo {author} {\bibfnamefont {Ali}\ \bibnamefont
  {\"Ovg\"un}},\ }\bibfield  {title} {\enquote {\bibinfo {title} {{Deflection
  Angle of Photons through Dark Matter by Black Holes and Wormholes Using
  Gauss\textendash{}Bonnet Theorem}},}\ }\href {\doibase
  10.3390/universe5050115} {\bibfield  {journal} {\bibinfo  {journal}
  {Universe}\ }\textbf {\bibinfo {volume} {5}},\ \bibinfo {pages} {115}
  (\bibinfo {year} {2019}{\natexlab{b}})},\ \Eprint
  {http://arxiv.org/abs/1806.05549} {arXiv:1806.05549 [physics.gen-ph]}
  \BibitemShut {NoStop}%
\bibitem [{\citenamefont {Javed}\ \emph {et~al.}(2019)\citenamefont {Javed},
  \citenamefont {Babar},\ and\ \citenamefont {\"Ovg\"un}}]{Javed:2019ynm}%
  \BibitemOpen
  \bibfield  {author} {\bibinfo {author} {\bibfnamefont {Wajiha}\ \bibnamefont
  {Javed}}, \bibinfo {author} {\bibfnamefont {Rimsha}\ \bibnamefont {Babar}}, \
  and\ \bibinfo {author} {\bibfnamefont {Al\"\i{}}\ \bibnamefont {\"Ovg\"un}},\
  }\bibfield  {title} {\enquote {\bibinfo {title} {{Effect of the dilaton field
  and plasma medium on deflection angle by black holes in
  Einstein-Maxwell-dilaton-axion theory}},}\ }\href {\doibase
  10.1103/PhysRevD.100.104032} {\bibfield  {journal} {\bibinfo  {journal}
  {Phys. Rev. D}\ }\textbf {\bibinfo {volume} {100}},\ \bibinfo {pages}
  {104032} (\bibinfo {year} {2019})},\ \Eprint
  {http://arxiv.org/abs/1910.11697} {arXiv:1910.11697 [gr-qc]} \BibitemShut
  {NoStop}%
\bibitem [{\citenamefont {Werner}(2012)}]{Werner2012}%
  \BibitemOpen
  \bibfield  {author} {\bibinfo {author} {\bibfnamefont {M.~C.}\ \bibnamefont
  {Werner}},\ }\bibfield  {title} {\enquote {\bibinfo {title} {{Gravitational
  lensing in the Kerr-Randers optical geometry}},}\ }\href {\doibase
  10.1007/s10714-012-1458-9} {\bibfield  {journal} {\bibinfo  {journal} {Gen.
  Rel. Grav.}\ }\textbf {\bibinfo {volume} {44}},\ \bibinfo {pages}
  {3047--3057} (\bibinfo {year} {2012})},\ \Eprint
  {http://arxiv.org/abs/1205.3876} {arXiv:1205.3876 [gr-qc]} \BibitemShut
  {NoStop}%
\bibitem [{\citenamefont {Ishihara}\ \emph {et~al.}(2016)\citenamefont
  {Ishihara}, \citenamefont {Suzuki}, \citenamefont {Ono}, \citenamefont
  {Kitamura},\ and\ \citenamefont {Asada}}]{Ishihara:2016vdc}%
  \BibitemOpen
  \bibfield  {author} {\bibinfo {author} {\bibfnamefont {Asahi}\ \bibnamefont
  {Ishihara}}, \bibinfo {author} {\bibfnamefont {Yusuke}\ \bibnamefont
  {Suzuki}}, \bibinfo {author} {\bibfnamefont {Toshiaki}\ \bibnamefont {Ono}},
  \bibinfo {author} {\bibfnamefont {Takao}\ \bibnamefont {Kitamura}}, \ and\
  \bibinfo {author} {\bibfnamefont {Hideki}\ \bibnamefont {Asada}},\ }\bibfield
   {title} {\enquote {\bibinfo {title} {{Gravitational bending angle of light
  for finite distance and the Gauss-Bonnet theorem}},}\ }\href {\doibase
  10.1103/PhysRevD.94.084015} {\bibfield  {journal} {\bibinfo  {journal} {Phys.
  Rev. D}\ }\textbf {\bibinfo {volume} {94}},\ \bibinfo {pages} {084015}
  (\bibinfo {year} {2016})},\ \Eprint {http://arxiv.org/abs/1604.08308}
  {arXiv:1604.08308 [gr-qc]} \BibitemShut {NoStop}%
\bibitem [{\citenamefont {Ono}\ \emph {et~al.}(2017)\citenamefont {Ono},
  \citenamefont {Ishihara},\ and\ \citenamefont {Asada}}]{Ono:2017pie}%
  \BibitemOpen
  \bibfield  {author} {\bibinfo {author} {\bibfnamefont {Toshiaki}\
  \bibnamefont {Ono}}, \bibinfo {author} {\bibfnamefont {Asahi}\ \bibnamefont
  {Ishihara}}, \ and\ \bibinfo {author} {\bibfnamefont {Hideki}\ \bibnamefont
  {Asada}},\ }\bibfield  {title} {\enquote {\bibinfo {title} {{Gravitomagnetic
  bending angle of light with finite-distance corrections in stationary
  axisymmetric spacetimes}},}\ }\href {\doibase 10.1103/PhysRevD.96.104037}
  {\bibfield  {journal} {\bibinfo  {journal} {Phys. Rev. D}\ }\textbf {\bibinfo
  {volume} {96}},\ \bibinfo {pages} {104037} (\bibinfo {year} {2017})},\
  \Eprint {http://arxiv.org/abs/1704.05615} {arXiv:1704.05615 [gr-qc]}
  \BibitemShut {NoStop}%
\bibitem [{\citenamefont {Li}\ and\ \citenamefont
  {\"Ovg\"un}(2020)}]{Li:2020dln}%
  \BibitemOpen
  \bibfield  {author} {\bibinfo {author} {\bibfnamefont {Zonghai}\ \bibnamefont
  {Li}}\ and\ \bibinfo {author} {\bibfnamefont {Ali}\ \bibnamefont
  {\"Ovg\"un}},\ }\bibfield  {title} {\enquote {\bibinfo {title}
  {{Finite-distance gravitational deflection of massive particles by a
  Kerr-like black hole in the bumblebee gravity model}},}\ }\href {\doibase
  10.1103/PhysRevD.101.024040} {\bibfield  {journal} {\bibinfo  {journal}
  {Phys. Rev. D}\ }\textbf {\bibinfo {volume} {101}},\ \bibinfo {pages}
  {024040} (\bibinfo {year} {2020})},\ \Eprint
  {http://arxiv.org/abs/2001.02074} {arXiv:2001.02074 [gr-qc]} \BibitemShut
  {NoStop}%
\bibitem [{\citenamefont {Li}\ \emph {et~al.}(2020)\citenamefont {Li},
  \citenamefont {Zhang},\ and\ \citenamefont {\"Ovg\"un}}]{Li:2020wvn}%
  \BibitemOpen
  \bibfield  {author} {\bibinfo {author} {\bibfnamefont {Zonghai}\ \bibnamefont
  {Li}}, \bibinfo {author} {\bibfnamefont {Guodong}\ \bibnamefont {Zhang}}, \
  and\ \bibinfo {author} {\bibfnamefont {Ali}\ \bibnamefont {\"Ovg\"un}},\
  }\bibfield  {title} {\enquote {\bibinfo {title} {{Circular Orbit of a
  Particle and Weak Gravitational Lensing}},}\ }\href {\doibase
  10.1103/PhysRevD.101.124058} {\bibfield  {journal} {\bibinfo  {journal}
  {Phys. Rev. D}\ }\textbf {\bibinfo {volume} {101}},\ \bibinfo {pages}
  {124058} (\bibinfo {year} {2020})},\ \Eprint
  {http://arxiv.org/abs/2006.13047} {arXiv:2006.13047 [gr-qc]} \BibitemShut
  {NoStop}%
\bibitem [{\citenamefont {Belhaj}\ \emph {et~al.}(2022)\citenamefont {Belhaj},
  \citenamefont {Belmahi}, \citenamefont {Benali},\ and\ \citenamefont
  {Moumni~El}}]{Belhaj:2022vte}%
  \BibitemOpen
  \bibfield  {author} {\bibinfo {author} {\bibfnamefont {A.}~\bibnamefont
  {Belhaj}}, \bibinfo {author} {\bibfnamefont {H.}~\bibnamefont {Belmahi}},
  \bibinfo {author} {\bibfnamefont {M.}~\bibnamefont {Benali}}, \ and\ \bibinfo
  {author} {\bibfnamefont {H.}~\bibnamefont {Moumni~El}},\ }\bibfield  {title}
  {\enquote {\bibinfo {title} {{Light deflection by rotating regular black
  holes with a cosmological constant}},}\ }\href {\doibase
  10.1016/j.cjph.2022.04.013} {\bibfield  {journal} {\bibinfo  {journal} {Chin.
  J. Phys.}\ }\textbf {\bibinfo {volume} {80}},\ \bibinfo {pages} {229--238}
  (\bibinfo {year} {2022})},\ \Eprint {http://arxiv.org/abs/2204.10150}
  {arXiv:2204.10150 [gr-qc]} \BibitemShut {NoStop}%
\bibitem [{\citenamefont {Javed}\ \emph {et~al.}(2023)\citenamefont {Javed},
  \citenamefont {Atique}, \citenamefont {Pantig},\ and\ \citenamefont
  {\"Ovg\"un}}]{Javed:2023iih}%
  \BibitemOpen
  \bibfield  {author} {\bibinfo {author} {\bibfnamefont {Wajiha}\ \bibnamefont
  {Javed}}, \bibinfo {author} {\bibfnamefont {Mehak}\ \bibnamefont {Atique}},
  \bibinfo {author} {\bibfnamefont {Reggie~C.}\ \bibnamefont {Pantig}}, \ and\
  \bibinfo {author} {\bibfnamefont {Ali}\ \bibnamefont {\"Ovg\"un}},\
  }\bibfield  {title} {\enquote {\bibinfo {title} {{Weak Deflection Angle,
  Hawking Radiation and Greybody Bound of Reissner-Nordstr\"om Black Hole
  Corrected by Bounce Parameter}},}\ }\href {\doibase 10.3390/sym15010148}
  {\bibfield  {journal} {\bibinfo  {journal} {Symmetry}\ }\textbf {\bibinfo
  {volume} {15}},\ \bibinfo {pages} {148} (\bibinfo {year} {2023})},\ \Eprint
  {http://arxiv.org/abs/2301.01855} {arXiv:2301.01855 [gr-qc]} \BibitemShut
  {NoStop}%
\bibitem [{\citenamefont {Javed}\ \emph
  {et~al.}(2022{\natexlab{a}})\citenamefont {Javed}, \citenamefont {Atique},
  \citenamefont {Pantig},\ and\ \citenamefont {\"Ovg\"un}}]{Javed:2023IJGMMP}%
  \BibitemOpen
  \bibfield  {author} {\bibinfo {author} {\bibfnamefont {Wajiha}\ \bibnamefont
  {Javed}}, \bibinfo {author} {\bibfnamefont {Mehak}\ \bibnamefont {Atique}},
  \bibinfo {author} {\bibfnamefont {Reggie~C.}\ \bibnamefont {Pantig}}, \ and\
  \bibinfo {author} {\bibfnamefont {Ali}\ \bibnamefont {\"Ovg\"un}},\
  }\bibfield  {title} {\enquote {\bibinfo {title} {{Weak lensing, Hawking
  radiation and greybody factor bound by a charged black holes with non-linear
  electrodynamics corrections}},}\ }\href {\doibase 10.1142/s0219887823500408}
  {\bibfield  {journal} {\bibinfo  {journal} {International Journal of
  Geometric Methods in Modern Physics}\ ,\ \bibinfo {pages} {2350040}}
  (\bibinfo {year} {2022}{\natexlab{a}})}\BibitemShut {NoStop}%
\bibitem [{\citenamefont {Javed}\ \emph
  {et~al.}(2022{\natexlab{b}})\citenamefont {Javed}, \citenamefont {Riaz},
  \citenamefont {Pantig},\ and\ \citenamefont {\"Ovg\"un}}]{Javed:2022fsn}%
  \BibitemOpen
  \bibfield  {author} {\bibinfo {author} {\bibfnamefont {Wajiha}\ \bibnamefont
  {Javed}}, \bibinfo {author} {\bibfnamefont {Sibgha}\ \bibnamefont {Riaz}},
  \bibinfo {author} {\bibfnamefont {Reggie~C.}\ \bibnamefont {Pantig}}, \ and\
  \bibinfo {author} {\bibfnamefont {Ali}\ \bibnamefont {\"Ovg\"un}},\
  }\bibfield  {title} {\enquote {\bibinfo {title} {{Weak gravitational lensing
  in dark matter and plasma mediums for wormhole-like static aether
  solution}},}\ }\href {\doibase 10.1140/epjc/s10052-022-11030-4} {\bibfield
  {journal} {\bibinfo  {journal} {Eur. Phys. J. C}\ }\textbf {\bibinfo {volume}
  {82}},\ \bibinfo {pages} {1057} (\bibinfo {year} {2022}{\natexlab{b}})},\
  \Eprint {http://arxiv.org/abs/2212.00804} {arXiv:2212.00804 [gr-qc]}
  \BibitemShut {NoStop}%
\bibitem [{\citenamefont {Javed}\ \emph
  {et~al.}(2022{\natexlab{c}})\citenamefont {Javed}, \citenamefont {Irshad},
  \citenamefont {Pantig},\ and\ \citenamefont {\"Ovg\"un}}]{Javed:2022gtz}%
  \BibitemOpen
  \bibfield  {author} {\bibinfo {author} {\bibfnamefont {Wajiha}\ \bibnamefont
  {Javed}}, \bibinfo {author} {\bibfnamefont {Hafsa}\ \bibnamefont {Irshad}},
  \bibinfo {author} {\bibfnamefont {Reggie~C.}\ \bibnamefont {Pantig}}, \ and\
  \bibinfo {author} {\bibfnamefont {Ali}\ \bibnamefont {\"Ovg\"un}},\
  }\bibfield  {title} {\enquote {\bibinfo {title} {{Weak Deflection Angle by
  Kalb\textendash{}Ramond Traversable Wormhole in Plasma and Dark Matter
  Mediums}},}\ }\href {\doibase 10.3390/universe8110599} {\bibfield  {journal}
  {\bibinfo  {journal} {Universe}\ }\textbf {\bibinfo {volume} {8}},\ \bibinfo
  {pages} {599} (\bibinfo {year} {2022}{\natexlab{c}})},\ \Eprint
  {http://arxiv.org/abs/2211.07009} {arXiv:2211.07009 [gr-qc]} \BibitemShut
  {NoStop}%
\bibitem [{\citenamefont {Gibbons}(2016)}]{Gibbons:2015qja}%
  \BibitemOpen
  \bibfield  {author} {\bibinfo {author} {\bibfnamefont {G.~W.}\ \bibnamefont
  {Gibbons}},\ }\bibfield  {title} {\enquote {\bibinfo {title} {{The
  Jacobi-metric for timelike geodesics in static spacetimes}},}\ }\href
  {\doibase 10.1088/0264-9381/33/2/025004} {\bibfield  {journal} {\bibinfo
  {journal} {Class. Quant. Grav.}\ }\textbf {\bibinfo {volume} {33}},\ \bibinfo
  {pages} {025004} (\bibinfo {year} {2016})},\ \Eprint
  {http://arxiv.org/abs/1508.06755} {arXiv:1508.06755 [gr-qc]} \BibitemShut
  {NoStop}%
\bibitem [{\citenamefont {Chanda}\ \emph {et~al.}(2017)\citenamefont {Chanda},
  \citenamefont {Gibbons},\ and\ \citenamefont {Guha}}]{Chanda:2016aph}%
  \BibitemOpen
  \bibfield  {author} {\bibinfo {author} {\bibfnamefont {Sumanto}\ \bibnamefont
  {Chanda}}, \bibinfo {author} {\bibfnamefont {G.~W.}\ \bibnamefont {Gibbons}},
  \ and\ \bibinfo {author} {\bibfnamefont {Partha}\ \bibnamefont {Guha}},\
  }\bibfield  {title} {\enquote {\bibinfo {title} {{Jacobi-Maupertuis-Eisenhart
  metric and geodesic flows}},}\ }\href {\doibase 10.1063/1.4978333} {\bibfield
   {journal} {\bibinfo  {journal} {J. Math. Phys.}\ }\textbf {\bibinfo {volume}
  {58}},\ \bibinfo {pages} {032503} (\bibinfo {year} {2017})},\ \Eprint
  {http://arxiv.org/abs/1612.00375} {arXiv:1612.00375 [math-ph]} \BibitemShut
  {NoStop}%
\bibitem [{\citenamefont {Das}\ \emph {et~al.}(2017)\citenamefont {Das},
  \citenamefont {Sk},\ and\ \citenamefont {Ghosh}}]{Das:2016opi}%
  \BibitemOpen
  \bibfield  {author} {\bibinfo {author} {\bibfnamefont {Praloy}\ \bibnamefont
  {Das}}, \bibinfo {author} {\bibfnamefont {Ripon}\ \bibnamefont {Sk}}, \ and\
  \bibinfo {author} {\bibfnamefont {Subir}\ \bibnamefont {Ghosh}},\ }\bibfield
  {title} {\enquote {\bibinfo {title} {{Motion of charged particle in
  Reissner\textendash{}Nordstr\"om spacetime: a Jacobi-metric approach}},}\
  }\href {\doibase 10.1140/epjc/s10052-017-5295-6} {\bibfield  {journal}
  {\bibinfo  {journal} {Eur. Phys. J. C}\ }\textbf {\bibinfo {volume} {77}},\
  \bibinfo {pages} {735} (\bibinfo {year} {2017})},\ \Eprint
  {http://arxiv.org/abs/1609.04577} {arXiv:1609.04577 [gr-qc]} \BibitemShut
  {NoStop}%
\bibitem [{\citenamefont {Bera}\ \emph {et~al.}(2020)\citenamefont {Bera},
  \citenamefont {Ghosh},\ and\ \citenamefont {Majhi}}]{Bera:2019oxg}%
  \BibitemOpen
  \bibfield  {author} {\bibinfo {author} {\bibfnamefont {Avijit}\ \bibnamefont
  {Bera}}, \bibinfo {author} {\bibfnamefont {Subir}\ \bibnamefont {Ghosh}}, \
  and\ \bibinfo {author} {\bibfnamefont {Bibhas~Ranjan}\ \bibnamefont
  {Majhi}},\ }\bibfield  {title} {\enquote {\bibinfo {title} {{Hawking
  radiation in a non-covariant frame: the Jacobi metric approach}},}\ }\href
  {\doibase 10.1140/epjp/s13360-020-00693-1} {\bibfield  {journal} {\bibinfo
  {journal} {Eur. Phys. J. Plus}\ }\textbf {\bibinfo {volume} {135}},\ \bibinfo
  {pages} {670} (\bibinfo {year} {2020})},\ \Eprint
  {http://arxiv.org/abs/1909.12607} {arXiv:1909.12607 [gr-qc]} \BibitemShut
  {NoStop}%
\bibitem [{\citenamefont {Srinivasan}\ and\ \citenamefont
  {Padmanabhan}(1999)}]{Srinivasan:1998ty}%
  \BibitemOpen
  \bibfield  {author} {\bibinfo {author} {\bibfnamefont {K.}~\bibnamefont
  {Srinivasan}}\ and\ \bibinfo {author} {\bibfnamefont {T.}~\bibnamefont
  {Padmanabhan}},\ }\bibfield  {title} {\enquote {\bibinfo {title} {{Particle
  production and complex path analysis}},}\ }\href {\doibase
  10.1103/PhysRevD.60.024007} {\bibfield  {journal} {\bibinfo  {journal} {Phys.
  Rev. D}\ }\textbf {\bibinfo {volume} {60}},\ \bibinfo {pages} {024007}
  (\bibinfo {year} {1999})},\ \Eprint {http://arxiv.org/abs/gr-qc/9812028}
  {arXiv:gr-qc/9812028} \BibitemShut {NoStop}%
\bibitem [{\citenamefont {Iso}\ \emph {et~al.}(2006)\citenamefont {Iso},
  \citenamefont {Umetsu},\ and\ \citenamefont {Wilczek}}]{Iso:2006wa}%
  \BibitemOpen
  \bibfield  {author} {\bibinfo {author} {\bibfnamefont {Satoshi}\ \bibnamefont
  {Iso}}, \bibinfo {author} {\bibfnamefont {Hiroshi}\ \bibnamefont {Umetsu}}, \
  and\ \bibinfo {author} {\bibfnamefont {Frank}\ \bibnamefont {Wilczek}},\
  }\bibfield  {title} {\enquote {\bibinfo {title} {{Hawking radiation from
  charged black holes via gauge and gravitational anomalies}},}\ }\href
  {\doibase 10.1103/PhysRevLett.96.151302} {\bibfield  {journal} {\bibinfo
  {journal} {Phys. Rev. Lett.}\ }\textbf {\bibinfo {volume} {96}},\ \bibinfo
  {pages} {151302} (\bibinfo {year} {2006})},\ \Eprint
  {http://arxiv.org/abs/hep-th/0602146} {arXiv:hep-th/0602146} \BibitemShut
  {NoStop}%
\bibitem [{\citenamefont {Perlick}\ and\ \citenamefont
  {Tsupko}(2022)}]{Perlick:2021aok}%
  \BibitemOpen
  \bibfield  {author} {\bibinfo {author} {\bibfnamefont {Volker}\ \bibnamefont
  {Perlick}}\ and\ \bibinfo {author} {\bibfnamefont {Oleg~Yu.}\ \bibnamefont
  {Tsupko}},\ }\bibfield  {title} {\enquote {\bibinfo {title} {{Calculating
  black hole shadows: Review of analytical studies}},}\ }\href {\doibase
  10.1016/j.physrep.2021.10.004} {\bibfield  {journal} {\bibinfo  {journal}
  {Phys. Rept.}\ }\textbf {\bibinfo {volume} {947}},\ \bibinfo {pages} {1--39}
  (\bibinfo {year} {2022})},\ \Eprint {http://arxiv.org/abs/2105.07101}
  {arXiv:2105.07101 [gr-qc]} \BibitemShut {NoStop}%
\bibitem [{\citenamefont {Kocherlakota}\ \emph {et~al.}(2021)\citenamefont
  {Kocherlakota} \emph {et~al.}}]{EventHorizonTelescope:2021dqv}%
  \BibitemOpen
  \bibfield  {author} {\bibinfo {author} {\bibfnamefont {Prashant}\
  \bibnamefont {Kocherlakota}} \emph {et~al.} (\bibinfo {collaboration} {Event
  Horizon Telescope}),\ }\bibfield  {title} {\enquote {\bibinfo {title}
  {{Constraints on black-hole charges with the 2017 EHT observations of
  M87*}},}\ }\href {\doibase 10.1103/PhysRevD.103.104047} {\bibfield  {journal}
  {\bibinfo  {journal} {Phys. Rev. D}\ }\textbf {\bibinfo {volume} {103}},\
  \bibinfo {pages} {104047} (\bibinfo {year} {2021})},\ \Eprint
  {http://arxiv.org/abs/2105.09343} {arXiv:2105.09343 [gr-qc]} \BibitemShut
  {NoStop}%
\bibitem [{\citenamefont {Do~Carmo}(2016)}]{Carmo2016}%
  \BibitemOpen
  \bibfield  {author} {\bibinfo {author} {\bibfnamefont {Manfredo~P}\
  \bibnamefont {Do~Carmo}},\ }\href@noop {} {\emph {\bibinfo {title}
  {Differential geometry of curves and surfaces: revised and updated second
  edition}}}\ (\bibinfo  {publisher} {Courier Dover Publications},\ \bibinfo
  {year} {2016})\BibitemShut {NoStop}%
\bibitem [{\citenamefont {Klingenberg}(2013)}]{Klingenberg2013}%
  \BibitemOpen
  \bibfield  {author} {\bibinfo {author} {\bibfnamefont {Wilhelm}\ \bibnamefont
  {Klingenberg}},\ }\href@noop {} {\emph {\bibinfo {title} {A course in
  differential geometry}}},\ Vol.~\bibinfo {volume} {51}\ (\bibinfo
  {publisher} {Springer Science \& Business Media},\ \bibinfo {year}
  {2013})\BibitemShut {NoStop}%
\bibitem [{\citenamefont {Zaja\v{c}ek}\ \emph {et~al.}(2018)\citenamefont
  {Zaja\v{c}ek}, \citenamefont {Tursunov}, \citenamefont {Eckart},\ and\
  \citenamefont {Britzen}}]{Zajacek:2018ycb}%
  \BibitemOpen
  \bibfield  {author} {\bibinfo {author} {\bibfnamefont {Michal}\ \bibnamefont
  {Zaja\v{c}ek}}, \bibinfo {author} {\bibfnamefont {Arman}\ \bibnamefont
  {Tursunov}}, \bibinfo {author} {\bibfnamefont {Andreas}\ \bibnamefont
  {Eckart}}, \ and\ \bibinfo {author} {\bibfnamefont {Silke}\ \bibnamefont
  {Britzen}},\ }\bibfield  {title} {\enquote {\bibinfo {title} {{On the charge
  of the Galactic centre black hole}},}\ }\href {\doibase
  10.1093/mnras/sty2182} {\bibfield  {journal} {\bibinfo  {journal} {Mon. Not.
  Roy. Astron. Soc.}\ }\textbf {\bibinfo {volume} {480}},\ \bibinfo {pages}
  {4408--4423} (\bibinfo {year} {2018})},\ \Eprint
  {http://arxiv.org/abs/1808.07327} {arXiv:1808.07327 [astro-ph.GA]}
  \BibitemShut {NoStop}%
\bibitem [{\citenamefont {Liu}\ and\ \citenamefont
  {Prokopec}(2017)}]{Liu:2016nwt}%
  \BibitemOpen
  \bibfield  {author} {\bibinfo {author} {\bibfnamefont {Lei-Hua}\ \bibnamefont
  {Liu}}\ and\ \bibinfo {author} {\bibfnamefont {Tomislav}\ \bibnamefont
  {Prokopec}},\ }\bibfield  {title} {\enquote {\bibinfo {title} {{Gravitational
  microlensing in Verlinde's emergent gravity}},}\ }\href {\doibase
  10.1016/j.physletb.2017.03.061} {\bibfield  {journal} {\bibinfo  {journal}
  {Phys. Lett. B}\ }\textbf {\bibinfo {volume} {769}},\ \bibinfo {pages}
  {281--288} (\bibinfo {year} {2017})},\ \Eprint
  {http://arxiv.org/abs/1612.00861} {arXiv:1612.00861 [gr-qc]} \BibitemShut
  {NoStop}%
\bibitem [{\citenamefont {Kardashev}\ and\ \citenamefont
  {Khartov}(2013)}]{Kardashev:2013cla}%
  \BibitemOpen
  \bibfield  {author} {\bibinfo {author} {\bibfnamefont {N.~S.}\ \bibnamefont
  {Kardashev}}\ and\ \bibinfo {author} {\bibfnamefont {V.~V.}\ \bibnamefont
  {Khartov}} (\bibinfo {collaboration} {RadioAstron}),\ }\bibfield  {title}
  {\enquote {\bibinfo {title} {{RadioAstron -- a Telescope with a Size of 300
  000 km: Main Parameters and First Observational Results}},}\ }\href {\doibase
  10.1134/S1063772913030025} {\bibfield  {journal} {\bibinfo  {journal}
  {Astronomy Reports}\ }\textbf {\bibinfo {volume} {57}},\ \bibinfo {pages}
  {153--194} (\bibinfo {year} {2013})},\ \Eprint
  {http://arxiv.org/abs/1303.5013} {arXiv:1303.5013 [astro-ph.IM]} \BibitemShut
  {NoStop}%
\bibitem [{\citenamefont {Bonanno}\ and\ \citenamefont
  {Silveravalle}(2021)}]{stars}%
  \BibitemOpen
  \bibfield  {author} {\bibinfo {author} {\bibfnamefont {Alfio}\ \bibnamefont
  {Bonanno}}\ and\ \bibinfo {author} {\bibfnamefont {Samuele}\ \bibnamefont
  {Silveravalle}},\ }\bibfield  {title} {\enquote {\bibinfo {title} {{The
  gravitational field of a star in quadratic gravity}},}\ }\href {\doibase
  10.1088/1475-7516/2021/08/050} {\bibfield  {journal} {\bibinfo  {journal}
  {JCAP}\ }\textbf {\bibinfo {volume} {08}},\ \bibinfo {pages} {050} (\bibinfo
  {year} {2021})},\ \Eprint {http://arxiv.org/abs/2106.00558} {arXiv:2106.00558
  [gr-qc]} \BibitemShut {NoStop}%
\end{thebibliography}%
\end{document}